\documentclass{jfm}

\usepackage{latexsym}
\usepackage{graphicx}
\usepackage{epsfig}
\usepackage{amssymb}
\usepackage{amsmath}
\usepackage{epstopdf}
\usepackage{color}

\usepackage{graphicx,footmisc}
\usepackage{subfig,epstopdf}
\usepackage{natbib}
\usepackage{lineno,amsmath}
\ifCUPmtlplainloaded \else
  \checkfont{eurm10}
  \iffontfound
    \IfFileExists{upmath.sty}
      {\typeout{^^JFound AMS Euler Roman fonts on the system,
                   using the 'upmath' package.^^J}%
       \usepackage{upmath}}
      {\typeout{^^JFound AMS Euler Roman fonts on the system, but you
                   dont seem to have the}%
       \typeout{'upmath' package installed. JFM.cls can take advantage
                 of these fonts,^^Jif you use 'upmath' package.^^J}%
      }
  \else
  \fi
\fi

\ifCUPmtlplainloaded \else
  \checkfont{msam10}
  \iffontfound
    \IfFileExists{amssymb.sty}
      {\typeout{^^JFound AMS Symbol fonts on the system, using the
                'amssymb' package.^^J}%
       \usepackage{amssymb}%
       \let\le=\leqslant  
       \let\ge=\geqslant  
      }{}
  \fi
\fi

\ifCUPmtlplainloaded \else
  \IfFileExists{amsbsy.sty}
    {\typeout{^^JFound the 'amsbsy' package on the system, using it.^^J}%
     \usepackage{amsbsy}}
    {\providecommand\boldsymbol[1]{\mbox{\boldmath $##1$}}}
\fi

\newbox\grsign \setbox\grsign=\hbox{$>$} \newdimen\grdimen
\grdimen=\ht\grsign                
\newbox\simlessbox \newbox\simgreatbox
\setbox\simgreatbox=\hbox{\raise.5ex\hbox{$>$}\llap
     {\lower.5ex\hbox{$\sim$}}}\ht1=\grdimen\dp1=0pt
\setbox\simlessbox=\hbox{\raise.5ex\hbox{$<$}\llap 
     {\lower.5ex\hbox{$\sim$}}}\ht2=\grdimen\dp2=0pt

\newcommand{\Nu}{\mbox{\it Nu}}

\def\begineq{\begin{equation}}
\def\endeq{\end{equation}}

\def\begineqn{\begin{equation*}}
\def\endeqn{\end{equation*}}

\def\beginar{\begin{eqnarray}}
\def\endar{\end{eqnarray}}
\def\beginarn{\begin{eqnarray*}}
\def\endarn{\end{eqnarray*}}

\def\lb{\left ( }
\def\rb{\right ) }

\def\lbr{\left \langle }
\def\rbr{\right \rangle }

\def\eub{\boldsymbol{U}}

\def\ub{\boldsymbol{u}}

\def\hr{{\bf{\widehat e}_r}}

\def\hr{{\bf\widehat z}}

\newcommand\Beq{\begin{eqnarray}} 
\newcommand\Eeq{\end{eqnarray}}
\newcommand\Beqn{\begin{eqnarray}} 
\newcommand\Eeqn{\end{eqnarray}}

\newcommand{\pd}[1]{\partial_{#1}}

\newcommand{\nns}{\nonumber\\}

\newcommand{\nn}{\nonumber}

\title[
Rotating convection with Ekman pumping
]
{\bf 
A nonlinear model for rotationally constrained convection with Ekman pumping
 }
\author[K. Julien et al.]
{Keith Julien$^1$\thanks{Email address for correspondence: julien@colorado.edu},\ns 
Jonathan M. Aurnou$^2$,
Michael A. Calkins$^1$,\\
Edgar Knobloch$^3$,
Philippe Marti$^1$,
Stephan Stellmach$^4$ and \\
Geoffrey M. Vasil$^5$}

\affiliation{
$^1$Department of Applied Mathematics, University of Colorado, Boulder, CO  80309, USA\\
$^2$Department of Earth, Planetary and Space Sciences, University of California, Los Angeles, CA 90095, USA \\
$^3$Department of Physics, University of California, Berkeley, CA  94720, USA\\
$^4$Institut f\"ur Geophysik, Westf\"alische Wilhelms Universit\"at M\"unster, Germany\\
$^5$School of Mathematics and Statistics, University of Sydney, Australia
}

\pubyear{2015}
\volume{650}
\pagerange{119--126}
\date{?; revised ?; accepted ?. - To be entered by editorial office}

\begin{document}
\maketitle

\begin{abstract}

It is a well established result of linear theory that the influence of differing mechanical boundary conditions, i.e., 
stress-free or no-slip,  on the primary instability in rotating convection becomes asymptotically small
in the limit of rapid rotation \citep{sC61}. This is accounted for by the diminishing impact of the viscous stresses 
exerted within Ekman boundary layers and the associated vertical momentum transport by  Ekman pumping \citep{pN65,wH71}.  
By contrast, in the nonlinear regime recent laboratory experiments and supporting numerical simulations are now providing evidence 
that the efficiency of heat transport remains strongly
influenced by Ekman pumping in the rapidly rotating limit  \citep{sS14,jS15}. 
In this paper, a reduced model is developed for the case of low Rossby number convection in a plane layer 
geometry with no-slip upper and lower boundaries held at fixed temperatures. A complete description of the dynamics requires the existence of three distinct regions within the fluid layer: a geostrophically balanced interior where fluid motions are predominately aligned with the axis of rotation, Ekman boundary layers immediately adjacent to the bounding plates, and thermal wind layers driven by Ekman pumping in between. The reduced model uses a classical Ekman pumping parameterization to alleviate the need for spatially resolving the Ekman boundary layers.  Results are presented for both linear stability theory and a special class of nonlinear solutions described by a single horizontal spatial wavenumber.  
{It is shown that Ekman pumping (which correlates positively with interior convection) allows for significant enhancement in the heat transport relative to that observed in simulations with stress-free boundaries. Without the intermediate thermal wind layer the nonlinear feedback from Ekman pumping would be able to generate a heat transport that diverges to infinity. This layer arrests this blowup resulting in finite heat transport at a significantly enhanced value.} {With increasing buoyancy forcing the heat transport transitions to a more efficient regime, a transition that is always achieved within the regime of asymptotic validity of the theory, suggesting this behavior may be prevalent in geophysical and astrophysical settings. As the rotation rate increases the slope of the heat transport curve below this transition steepens, a result that is in agreement with observations from laboratory experiments and direct numerical simulations.}
\end{abstract}

\section{Introduction}

Rotating Rayleigh-B\'enard convection (RBC),  i.e., a rotating horizontal fluid layer heated from
below and cooled from above,  provides a canonical framework  for the study of fluid
phenomena influenced by rotation and thermal forcing. It has proven to be an
indispensable framework for understanding fluid motions in geophysical and
astrophysical systems including planetary atmospheres and interiors \citep{jmA15}, rapidly
rotating stars \citep{mM05}, and open ocean deep convection \citep{jM99}. In many of these examples the
dominant influence of rotation results in geostrophy where the Coriolis force
is balanced by the pressure gradient force. Small departures from this dominant
force balance, referred to as quasigeostrophy, are known to be capable of
producing turbulent fluid motions characterized by anisotropic eddies elongated
along the rotation axis \citep{sS97}. The effect of spatial anisotropy on fluid mixing and
global transport properties such as heat and energy  transport remains an important and
largely unanswered question.  

In the context of the Boussinesq approximation for incompressible motions the
rotating RBC problem is completely specified via three nondimensional
parameters, namely, the Rayleigh number $Ra$, the Ekman number $E$ and the
Prandtl number $\sigma$, defined by
\beginar
Ra = \frac{g\alpha\Delta T H^3}{\nu\kappa}, \quad
E = \frac{\nu}{2\Omega H^2},\quad
\sigma = \frac{\nu}{\kappa}.
\endar
Here $H$ is the layer depth, $\Delta T>0$ is the destabilizing temperature difference,
$\Omega$ is the rotation rate of the system,
$\nu$ is the kinematic viscosity, $\kappa$ is the thermal diffusivity, $g$ is
the gravitational acceleration and $\alpha$ is the coefficient of thermal
expansion.  The Rayleigh number measures the magnitude of the thermal forcing
and the Ekman number measures the importance of viscous forces relative to the
Coriolis force.  The Prandtl number is the ratio of the thermal and viscous
diffusion timescales in the system and describes the thermophysical properties
of the working fluid.  \textcolor{black}{Another dimensionless parameter of  importance} 
%An alternative, but critical dependent, 
 is
the convective Rossby number $Ro_{cv}=\sqrt{Ra/\sigma} E$ that measures the relative
importance of thermal forcing and the Coriolis force.  Specifically,
rotationally constrained convection is characterized by $(E, Ro_{cv}) \ll1$. 
Importantly, it is known that rotation imparts rigidity to the fluid in the RBC problem that delays the onset
of convection until a critical Rayleigh number $Ra_c\sim E^{-4/3}$ \citep{sC61}  is reached
with associated $Ro_{cv}\sim E^{1/3}$. At onset, motions are columnar with a horizontal scale $L\sim E^{1/3} H$.
By definition of the convective Rossby number, the rotationally constrained branch is defined as $E^{1/3} \lesssim Ro_{cv} \ll 1$ and characterized by
$E^{-4/3} \lesssim Ra \ll E^{-5/3}$, or equivalently, $1 \le \widetilde{Ra} \ll E^{-1/3}$ where $\widetilde{Ra}=Ra E^{4/3}$ is the reduced Rayleigh number
\citep{kJ12b}. The range of permissible $\widetilde{Ra}$ can thus be vast covering as much as five decades in geophysical and astrophysical settings where $E=\mathcal{O}(10^{-15})$. Rich dynamics are observed along this rotationally constrained branch, ranging from coherent laminar to highly turbulent states
% It is thus to be expected that observed dynamics on rotationally constrained branch is extremely rich, ranging from laminar to highly turbulent 
\citep{mS06,kJ12,aR14}.

\begin{figure}
  \begin{center}
        \includegraphics[height=7.5cm]{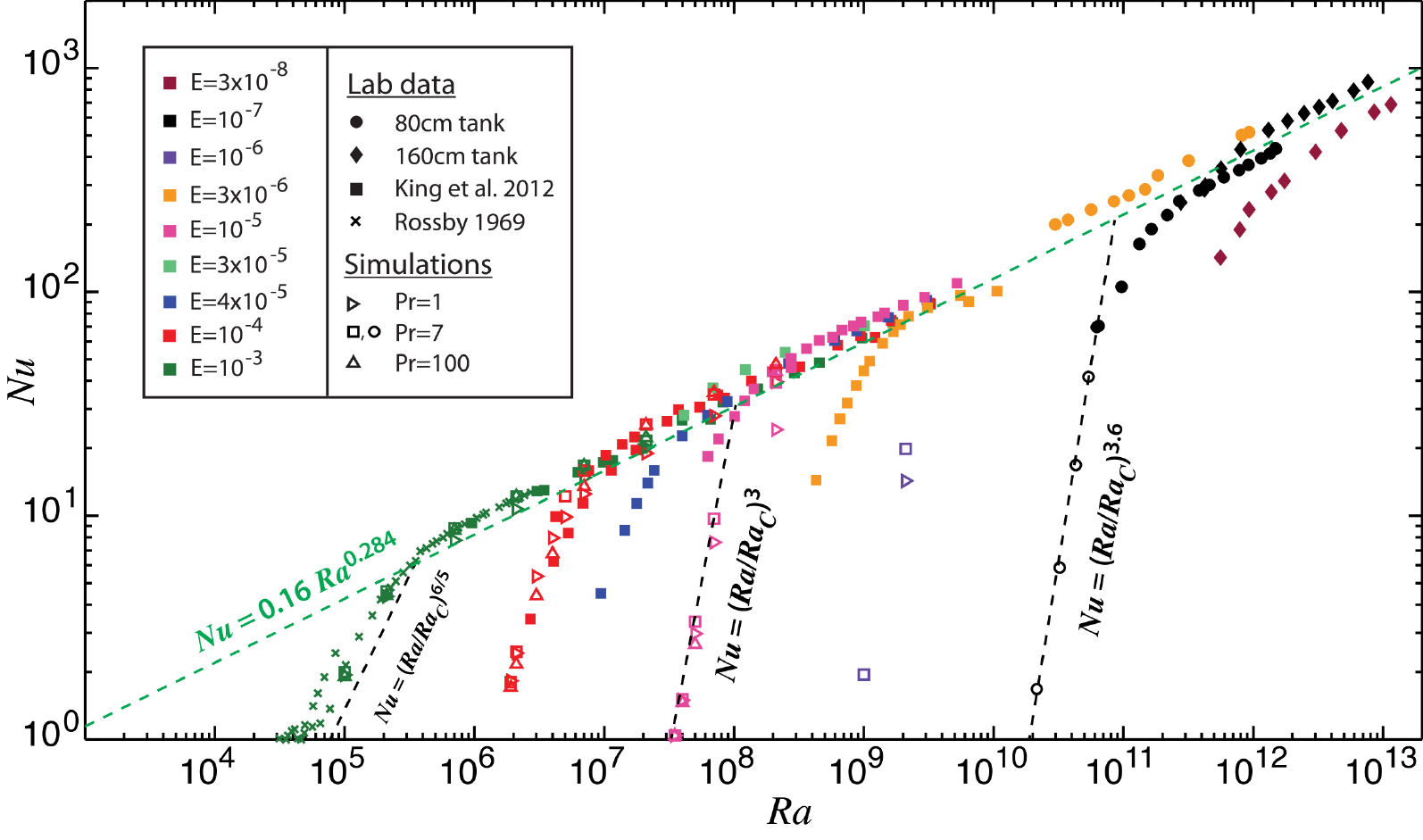}
  \end{center}
  \vspace{-2ex}
\caption{\small{Laboratory ($\sigma\approx 7$) and DNS ($\sigma=7$) rotating convection heat transfer data  (adapted from \cite{jS15}, \cite{hR69} and \cite{eK12b}). The best-fitting heat transfer trend of $Nu\propto(Ra/Ra_c)^{3.6}$ is plotted for $E=10^{-7}$. For comparison, $Nu\propto(Ra/Ra_c)^{3}$ \citep{eK12b} is plotted for $E=10^{-5}$ and $Nu\propto(Ra/Ra_c)^{6/5}$ \citep{eK09, eK10} for $E=10^{-3}$. Note that with each study at lower $E$, the scaling exponent becomes larger. This implies that the behavior of rotating convection  has not reached an asymptotic regime in the currently accessible range of $Nu$, $Ra$ and $E$.}} 
\label{fig:NuRa}
\end{figure}

The efficiency of heat transport, as measured by the nondimensional Nusselt
number $Nu = q H/\rho_0 c_p \kappa \Delta T$, is perhaps the most common result
reported in the literature given that the functional dependence $Nu = f(Ra, E, \sigma)$
is tied to the underlying dynamics. Here $q$ is the heat flux and $\rho_0 c_p$
is the volumetric heat capacity.  \textcolor{black}{Attempts have been made to characterize the rotationally constrained regime by a heat transport scaling law of the form $Nu \propto
(Ra/Ra_c)^{\beta_{rot}}$ with ${\beta_{rot}}>1$}
\citep{hR69,lY97,eK09,rE14}. At fixed $E\ll1$, combined investigations of laboratory
experiments and DNS with no-slip boundaries have reported ${\beta^{NS}_{rot}}$
values in the range $6/5<\beta^{NS}_{rot}< 3.6$ with a trend towards the upper
bound occurring at the lowest Ekman number $E=10^{-7}$ (Figure~\ref{fig:NuRa})
\citep{jS15}.  At sufficiently large $Ra$ a transition to the weakly rotating or non-rotating regime 
occurs, characterized by $Nu \propto (Ra/Ra_c)^{\beta_{no rot}}$, where ${\beta_{no rot}}\in(2/7, 1/3)$ 
(see Figure~\ref{fig:NuRa}). Comparative studies in the
presence of stress-free boundaries using direct numerical simulations
(DNS) have reported similar findings to those seen in Figure~\ref{fig:NuRa} but
with substantially smaller exponents $\beta^{SF}_{rot}\in(3/2,11/5)$ at $E=10^{-7}$ \citep{eK09,sS14}. 
\textcolor{black}{Irrespective of the boundary conditions, \cite{eK09} have established that the
transition to the nonrotating scaling law occurs at smaller and smaller $Ro_{cv}$ as the rotation rate increases, i.e., $\lim_{E\rightarrow0} Ro^{trans}_{cv}\sim E^\gamma\rightarrow0$ where $\gamma > 0$. This result is attributed to the loss of geostrophic balance in the thermal boundary layer where the local Rossby number reaches unity while the Rossby number in the bulk remains small \citep{kJ12b}.}

The difference in the heat transport scaling exponents ${\beta^{NS}_{rot}}$ and
${\beta^{SF}_{rot}}$ has been associated with Ekman pumping \citep{sS14} -- the
vertical momentum transport that results from the transition from an interior
geostrophic balance to a dominant boundary layer force balance between the
Coriolis and viscous force in the presence of no-slip boundaries. In the linear
regime, Ekman pumping promotes the destabilization of the fluid layer, an effect
quantified by the positive difference $Ra_{SF}$ - $Ra_{NS} = \mathcal{O}(E^{1/6})$ in the
critical Rayleigh numbers \citep{pN65,wH71}.  This difference is
asymptotically small in the limit $E\rightarrow 0$ indicating that differing
mechanical boundaries become asymptotically indistinguishable.  By contrast,
laboratory experiments and supporting DNS indicate that the discrepancy due to pumping 
in the fully nonlinear regime appears to remain finite as $E\rightarrow 0$ \citep{sS14,jS15}. 
Unfortunately, surveying the high $Ra$--low $(Ro_{cv}, E)$ regime remains a prohibitive challenge for both laboratory experiments and DNS. 
\textcolor{black}{Laboratory experiments are constrained by their inability to access the
rotationally constrained branch at sufficiently low $(Ro_{cv},E)$ 
owing to the decreasing accuracy of global heat transport measurements resulting from
unknown heat leaks \citep{eK09,rE14}.}
%owing to a lower bound on the achievable $Ra$, and by extension, $Ro_{cv}$  \citep{eK09,rE14}. 
DNS studies
are restricted by spatiotemporal resolution constraints imposed by {the} presence of
$\mathcal{O}(E^{1/2} H)$ Ekman boundary layers and fast inertial waves
propagating on an \textcolor{black}{$\mathcal{O}(\Omega^{-1})$} timescale \citep{dN14,sS14}.

Although an impediment to DNS, the stiff character of the governing fluid
equations in the limit $(Ro_{cv}, E)\rightarrow0$ provides a possible path forward for
simplifying, or reducing, the governing equations. Indeed, a system of reduced
NonHydrostatic Quasigeostrophic Equations (NH-QGE) appropriate for rotating  RBC
in the presence of stress-free boundaries have been successfully derived and utilized by Julien and
collaborators \citep{kJ98a,mS06,kJ06,kJ07,iG10,kJ12b,kJ12,aR14,dN14}. The NH-QGE, which filter fast
inertial waves while retaining inertial waves propagating on slow advective
timescales, enable parameter space explorations in the high $Ra$--low $Ro_{cv}$ limit. 
The NH-QGE have been used to identify various flow regimes and heat
transfer scaling behavior as a function of $Ra$, thus providing a valuable
roadmap for both DNS and laboratory experiments. Importantly, comparisons with
DNS for stress-free boundaries \citep{sS14} have established good quantitative agreement in both heat
transport and flow morphology at $E=10^{-7}$. 
For $E \ll 10^{-7}$ and large $Ra$  this approach has established an ultimate exponent $\beta^{SF}_{rot} = 3/2$, a result that 
corresponds to a dissipation-free scaling law in the rotationally constrained turbulence regime \citep{kJ12b,kJ12}.

\begin{figure}
  \begin{center}
  \vspace{+1ex}
      \includegraphics[height=7cm]{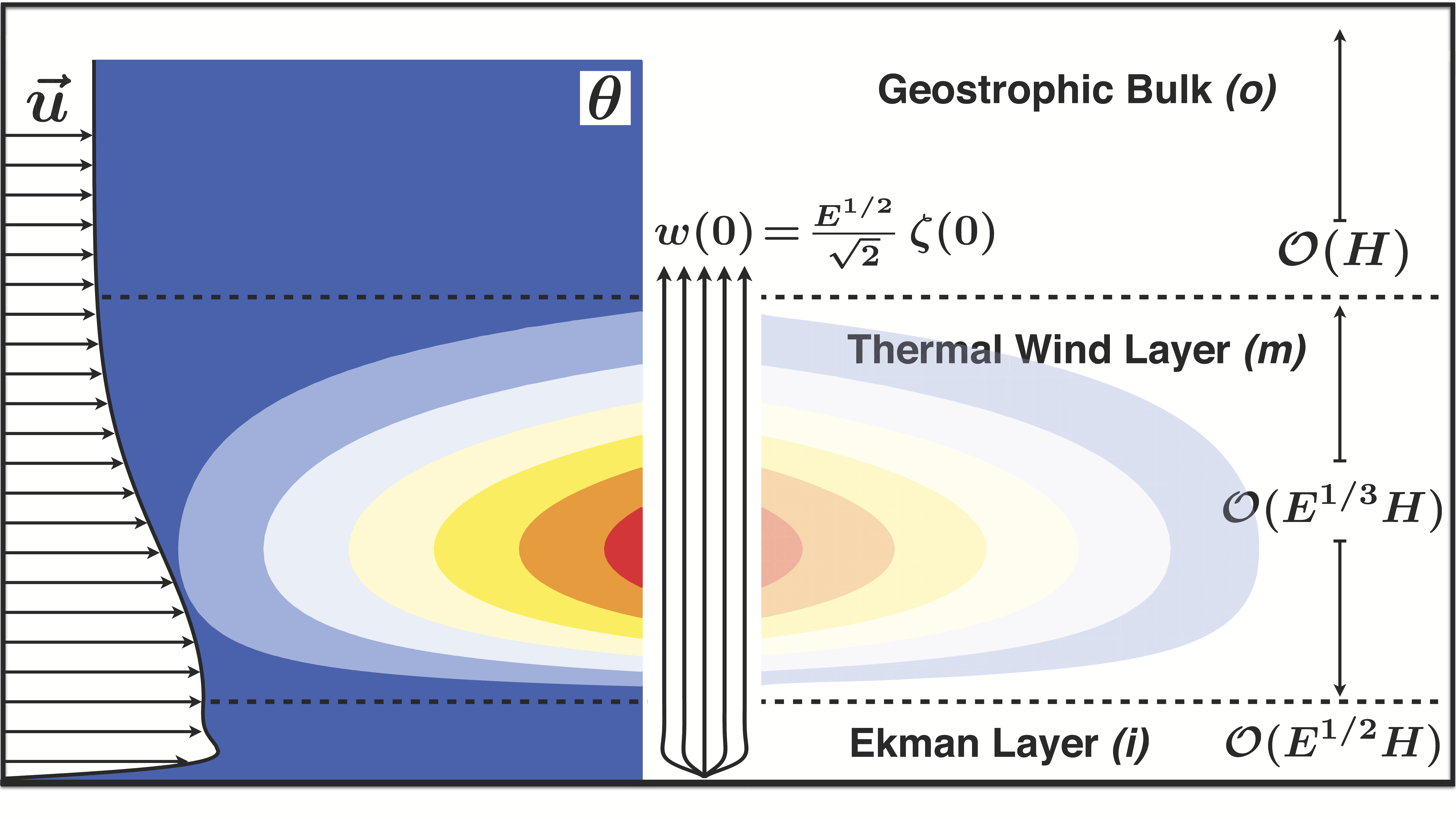}
  \end{center}
  \vspace{+0ex}
\caption{\small{A schematic diagram of the boundary layer structure for the enhanced-thermal-transport regime of rapidly rotating convection. Three distinct layers characterise the dynamics. The outer layer \textit{(o)} is geostrophically balanced with dynamically significant flows and thermal perturbations. The middle layer \textit{(m)} contains the largest thermal perturbations, $\theta$, of the entire system. The horizontal flow, $\ub_\perp$, responds with a thermal wind driven shear. The inner layer \textit{(i)} produces a frictional response that causes the flow to vanish at the physical bottom boundary. This layer contains no significant thermal perturbations. The Ekman response produces a vertical momentum flux out of the inner and middle layers and into the bulk fluid. The details of the dynamical pathway ensures that the Ekman flux correlates with the largest bulk vertical flows, thermal anomalies, and vorticity. This correlation is the means by which no-slip boundary layers can so significantly enhance the heat flux through the entire system. }}
\label{fig:config}
\end{figure}

In the present work we extend the asymptotic theory resulting in the NH-QGE to
the case of no-slip boundary conditions in which Ekman boundary layers are
present. This is the situation pertinent to the laboratory and of relevance to geophysical 
scenarios such as convection in the Earth's liquid iron outer core which is bounded from below
by the solid iron inner core and above by a rocky mantle. In agreement with the linear investigation of \cite{wH71}, our
nonlinear analysis shows that the presence of no-slip boundaries requires the
existence of three distinct regions each characterized by a different dominant physical balance:
and  hereafter denoted as the outer $(o)$, middle $(m)$ and inner $(i)$ regions (see Figure~\ref{fig:config}).  

The outer region $(o)$ corresponds to the fluid interior (i.e., the bulk region of depth $\mathcal{O}(H)$).  Within this region fluid motions are horizontally
non-divergent and in pointwise geostrophic balance.  The dynamics are asymptotically described by the NH-QGE investigated by Julien \& collaborators
(1998, 2006, 2007, 2012a,b, 2014). Similar to the classical quasigeostrophic equations \citep{jC48,jC71,Eady49,jP87,gv06} toroidal or vortical fluid motions aligned with the axis of rotation are forced solely by the vortex stretching associated with the linear Coriolis force.
%system or planetary rotation. 
%This necessarily requires the existence of vertical gradients in the vertical velocity field. 
The NH-QGE differ from the classical quasigeostrophic equations  in that vertical
motions required by vortex stretching are  now comparable in magnitude to horizontal motions and thus the nonhydrostatic 
inertial acceleration force must be retained in the vertical momentum balance. This results in a dynamic, as opposed to a diagnostic, evolution of the vertical velocity field. In the presence of stress-free boundaries the NH-QGE are sufficient to completely describe all the `\textit{slow}' dynamics occurring within the fluid layer. 

The inner $(i)$ regions  are the $\mathcal{O}(E^{1/2}H)$ Ekman boundary layers
immediately adjacent to the horizontal boundaries. These layers are required
to attenuate the interior geostrophic velocity fields to zero. Within them,
the geostrophic balance of the interior is relaxed and as a consequence fluid
motions become horizontally divergent with cross-isobaric flow.  Mass
conservation requires that vertical motions are induced, a process
referred to as Ekman pumping.  The dynamics within the Ekman layers are
described by a classical set of reduced linear equations \citep{hG69}, where
as a consequence of the $L/H= \mathcal{O}(E^{1/3})$ spatial anisotropy and the 
%$=\mathcal{O}(U/L)$ 
vortical magnitude $\zeta^*$ observed in rotating RBC \citep{sC61}, vertical pumping velocities of magnitude 
$w^*_E=\mathcal{O}(E^{1/2} \zeta^* H)=
\mathcal{O}(E^{1/6}\zeta^* L)$
%\mathcal{O}(E^{1/6}U)$
 are found \citep{pN65,wH71}. As done in large-scale oceanic and atmospheric applications,
this linear property of the dynamics can be successfully utilized and the
effects of this layer on the fluid interior can be parameterized by the application of simple pumping boundary conditions \citep{jP87}. This parameterized approach has been applied successfully in numerous previous works ranging from the investigation of Stewartson layer instabilities \citep{nS05}, spherical convection \citep{jA03,mC12b}, rotating RBC \citep{sS14}, and is also employed in the present work.

As shown by \cite{wH71} Ekman pumping induces thermal fluctuations that cannot be regulated within the Ekman boundary layer to satisfy the thermal boundary conditions. Middle layers $(m)$ of depth $\mathcal{O}(E^{1/3}H)$ arise that separate the Ekman boundary layers from the geostrophically  balanced fluid interior. Importantly,  the requirement that thermal fluctuations vanish at the fixed temperature bounding plates necessitates the introduction of vertical diffusion for temperature fluctuations. This is an insignificant process in the interior in the rapidly rotating limit
\citep{mS06,kJ12b}. However, as shown in this paper, the middle region is characterized by a thermal wind balance, i.e., a geostrophic and hydrostatic diagnostic balance, and the dynamics of this region evolve according to an advection-diffusion equation for the temperature fluctuations forced by Ekman pumping. Specifically, the magnitude of the vertical advection of the mean temperature field within the middle layers increases with Rayleigh number until it becomes a dominant source of buoyancy production that is comparable to that produced in the geostrophic interior. This enhancement is due to the intensification of vortical motions, and hence vertical transport, in the vicinity of the boundaries.  Convective fluxes driven by Ekman transport are also comparable to that produced in the geostrophic interior. Therefore, $\mathcal{O}(1)$ changes to the heat transport are to be expected, as observed in laboratory experiments and simulations \citep{sS14,jS15}.

We show that the asymptotic results for the different regions may be combined
into a single composite system of reduced equations, which we refer to as the
CNH-QGE (Composite NonHydrostatic QuasiGeostrophic Equations).
% As with the NH-QGE, appropriate for the application of stress-free boundaries, 
On the rotationally constrained branch of RBC, the CNH-QGE 
are valid in the interval $\mathcal{O}(1) \lesssim \widetilde{Ra}\lesssim \mathcal{O}(E^{-1/3})$. The linear stability properties of these equations are shown to be
consistent with earlier work \citep{pN65,wH71}. Moreover, comparisons with the
NH-QGE for single-mode (or single horizontal wavenumber) solutions show that significant departures in
heat transport occur in the interval $\mathcal{O}(E^{-1/9}) \lesssim \widetilde{Ra} \lesssim
\mathcal{O}(E^{-1/3})$.

The remainder of the paper is organized as follows. In section 2, we present the formulation 
of the rotating RBC problem in terms of
the incompressible Navier-Stokes equations.    
In section 3, we present the asymptotic
development in the presence of no-slip bounding plates using the `Method of Composite Expansions' (see section 4.2, \cite{nayfeh2008perturbation}). 
In particular, the Ekman and thermal wind boundary layers are identified and analyzed. The significance of Ekman pumping is also deduced by determining the Rayleigh number threshold at which the resulting vertical pumping velocity induces order one changes in the heat transport. It is also established that such a transition always occurs on the rotationally constrained branch of RBC.
In section 4, we present the composite reduced model, i.e., the CNH-QGE where the dynamics of each layer are combined into a unified description  (see Equations (\ref{eqn:mean_comp})-(\ref{eqn:fluc_comp})).  Results from the model are presented in section 5,  and establish that this unified model captures the physics associated with no-slip boundary conditions observed in laboratory experiments and DNS. Concluding remarks are given in section 6. 

\section{Governing Equations}
We consider thermally driven fluid flows that are characterized by the dimensional scales of length $[L]$, velocity $[U]$, time $[L/U]$, pressure $[P]$ and destabilizing temperature jump $[\Delta T]$.  Assuming a Cartesian coordinate system $\boldsymbol{x} =(x, y, z),$ we adopt the Rayleigh-B\'enard configuration of a plane-parallel geometry rotating about the $z$-axis with constant angular velocity $\Omega$ in the presence of constant gravity $\boldsymbol{g} =-g \hat{\boldsymbol{z}}$. The nondimensional equations of motion are given by the Boussinesq equations
\beginar
\label{eqn:gmtm}
D_t \ub + \frac{1}{Ro}\hr\times\ub &=& - Eu \nabla p + \Gamma \theta \hr + \frac{1}{Re} \nabla^2 \ub, 
%\\ 
\endar
\beginar
\label{eqn:gtemp}
D_t \theta &=& \frac{1}{Pe} \nabla^2 \theta, 
\\ 
\label{eqn:gmass}
%\endar
%\beginar
\nabla\cdot\ub&=&0
\endar
for the velocity field $\ub\equiv (u,v,w)$, temperature $\theta$ and the pressure $p$, where the
material derivative $D_t \equiv \partial_t + \ub\cdot\nabla$.  The nondimensional parameters that appear
are defined as 
\Beq
\label{eqn:nondim1}
Ro=\frac{U}{2\Omega L},\quad
Eu=\frac{P}{\rho_o U^2},\quad
\Gamma = \frac{g\alpha \Delta T L}{U^2},\quad
Re=\frac{U L}{\nu},\quad
Pe=\frac{U L}{\kappa},
\Eeq
respectively denoting the Rossby, Euler, buoyancy, Reynolds and P\'eclet
numbers. In the present work we are interested in the rotationally constrained regime
characterized by $Ro\ll1$ and aspect ratio  $A\equiv H/L=Ro^{-1}\gg1$ for
columnar structures of depth $H$ \citep{kJ98a,kJ06,mS06,kJ12}.  

For fluid motions in a statistically stationary state, the nondimensional vertical heat transport, i.e., the Nusselt number $Nu$, is given by 
\Beq
\label{eqn:Nuss_orig}
Nu = \frac{H}{L} \lb -\pd{z}\overline{\theta}^{\mathcal{A,T}} +  Pe\;  \overline{w \theta}^{\mathcal{A,T}}\rb   
\Eeq
obtained upon averaging equation (\ref{eqn:gtemp}) over time and the horizontal $(x,y)$ cross-section. Here
%\break
$\overline{f}^{\mathcal{A,T}}=\lim_{\mathcal{T}\rightarrow\infty}
\frac{1}{\mathcal{A T}} \int_{\mathcal{A,T}} f dx dy dt$, where $\mathcal{A}$ is the horizontal cross-sectional area.  This result indicates that the heat flux $Nu$ through the layer is constant at every vertical level.

\section{Asymptotic Development}
\label{sec:asymptotic}
It is well established that in the  geophysically and astrophysically relevant regimes $Ro\ll1$ 
the presence of fast inertial waves, propagating on $\mathcal{O}(Ro^{-1})$ timescales, poses a severe restriction on DNS.
\footnote{The discretized equations resulting from the Boussinesq equations will, in general, be coupled through the Coriolis force
${Ro^{-1}}\hr\times\ub$. This coupling is routinely eliminated by an explicit treatment in many timestepping algorithms,
i.e., by its relegation and evaluation at previous steps. This favorable numerical situation occurs at the expense of imposing severe timestepping restrictions. \textcolor{black}{Implicit treatment circumvents this issue. However, prohibitive timestepping restrictions persist owing to the Ekman-dependent CFL time constraint
%$\propto E^{2/3}$ 
associated with advective nonlinearites.}} 
The evolution of turbulent eddies is insensitive to these waves which can be filtered from the governing equations by asymptotic reduction methods in much the same manner as done in atmospheric and oceanic sciences for stably-stratified layers. Indeed, Julien and collaborators \citep{kJ06, mS06, kJ07} have established that an asymptotic reduction of the governing equations (\ref{eqn:gmtm})-(\ref{eqn:gmass}) can be deduced upon using $Ro\equiv\epsilon$ as a small parameter and introducing the distinguished limits
\Beq
\label{eqn:dl1}
A=\epsilon^{-1},\quad Eu= \epsilon^{-2}, \quad \Gamma= \mathcal{O}(\epsilon^{-1}), \quad Re=Pe=\mathcal{O}(1).
\Eeq
For the appropriate choice of the horizontal diffusive velocity scale, $U=\nu/L$, it follows that
\Beq
\label{eqn:dl2}
E=\epsilon^{3},\quad
\Gamma =\frac{Ra}{\sigma} \epsilon^3,\quad
Re=1,\quad
Pe=\sigma,
\Eeq
such that  $\Gamma =\widetilde{\Gamma} \epsilon^{-1}$.  
Here $\sigma\widetilde{\Gamma} = Ra \epsilon^{4}=\mathcal{O}(1)$ corresponds to the reduced Rayleigh number $\widetilde{Ra} = \sigma\widetilde{\Gamma}$ \textcolor{black}{involving the Prandtl number  $\sigma$.}

In accord with Figure~\ref{fig:config}, we anticipate the existence of three distinct regions: the bulk and middle regions, and the Ekman layers, each with their respective nondimensional depths $\mathcal{O}(\epsilon^{-1}), \mathcal{O}(1)$ and $\mathcal{O}(\epsilon^{1/2})$. We therefore employ a multiple scales expansion in the vertical direction and time,
\Beq
\label{eqn:mscalezt}
\pd{z}\rightarrow \epsilon^{-1/2}\pd{\mu} + \pd{z} + \epsilon \pd{Z},\quad  \pd{t}\rightarrow\pd{t}+ \epsilon^2 \pd{\tau},
\Eeq
where the slow vertical coordinate of the bulk is defined by $Z=\epsilon z$, the fast coordinate of the Ekman layer defined by $\mu =\epsilon^{-1/2} z$, and
the slow time by $\tau =\epsilon^2 t$.
We find that this setup necessitates the decomposition of  the fluid variables into mean (horizontally averaged) and fluctuating (horizontally varying) components, respectively denoted by overbars and primes, e.g.,
\Beq
\ub = \overline{\ub}+\ub^\prime,\quad \overline{\ub} =\frac{1}{\mathcal{A}}  \int_{\mathcal{A}} \ub dxdy, \quad
\overline{\ub^\prime}=0.
\Eeq
Averaging over the fast time $t$ is also required,
\Beq
\label{eqn:timeave}
\overline{\ub}^\mathcal{T}  = \lim_{\mathcal{T} \rightarrow\infty} \frac{1}{\mathcal{T}} \int_{\mathcal{T}} \ub  dt.
\Eeq
We note \textit{a posteriori} that, unlike derivations of the reduced dynamics in the presence of stress-free boundaries \citep{mS06}, the consideration of no-slip boundaries and Ekman layers requires the separation of spatial and time-averaging operations as performed here.

To proceed, all fluid variables are decomposed into outer $(o)$, middle $(m,\pm)$ and inner $(i,\pm)$ components representing the fluid bulk, middle regions, and Ekman layers.  For example, 
\Beq
\label{eqn:MOCE0}
  \ub &=& \eub^{(o)}(x,y,Z,t,\tau)+\eub^{(m,-)}(x,y,z,t,\tau)+\eub^{(m,+)}(x,y,z,t,\tau) \\
  &+&\eub^{(i,-)}(x,y,\mu,t,\tau)+\eub^{(i,+)}(x,y,\mu,t,\tau).\nn
 \Eeq
Here, capitalizations are used to identify the individual contributions of the fluid variables to each region and the notation $\pm$ denotes the upper and lower boundaries, respectively.
The boundary layer coordinates $\mu, z$ are assumed to increase away from the physical boundaries at $Z=0,1.$
Each region of the fluid layer may be accessed by the following actions for the outer, middle, and inner limits
\beginar
\label{eqn:climits_a}
\lim (\ub)^o &\equiv& \lim_{\mu\rightarrow\infty \atop z\rightarrow\infty}(\ub)= \eub^{(o)} \\
  &\Rightarrow& \quad   \lim (\eub^{(o)})^o = \eub^{(o)},\  \lim (\eub^{(m)},\eub^{(i)})^o =0,\nns
\label{eqn:climits_b}
\lim (\ub)^m &\equiv&  \lim_{\mu\rightarrow\infty \atop Z\rightarrow 0}(\ub)=\eub^{(o)}(0) +\eub^{(m)} \\
&\Rightarrow& \quad  \lim (\eub^{(o)})^m= \eub^{(o)}(0),\ \lim (\eub^{(m)})^m= \eub^{(m)},\  \lim (\eub^{(i)})^m =0, \nns
\label{eqn:climits_c}
\lim (\ub)^i &\equiv&  \lim_{z\rightarrow0 \atop Z\rightarrow 0}(\ub)=\eub^{(o)}(0) +\eub^{(m)}(0) +\eub^{(i)}  \\
& \Rightarrow & \quad  
\lim (\eub^{(o)}+\eub^{(m)})^i = \eub^{(o)}(0)+\eub^{(m)}(0),\  \lim (\eub^{(i)})^i =\eub^{(i)}.  \nn
\endar
\textcolor{black}{Similar expressions hold for the upper inner and middle layers located at  $Z=1$ upon replacing $(0)$ with $(1)$. By definition, the middle variables are identically zero in the outer region, while the inner variables are identically zero in both the middle and outer regions.}
\textcolor{black}{Contributions to regions $(i)$ or $(m)$ involving outer variables, indicated in (\ref{eqn:climits_b}a) and (\ref{eqn:climits_c}a),
are obtained by Taylor-expanding and then taking the appropriate limit}.
Hereinafter, for notational convenience, contractions $(m,\pm)\rightarrow(m)$ and $(i,\pm)\rightarrow(i)$ are used when referring to both upper and 
lower regions. Furthermore, contractions such as $(0)$ omit reference to the dependence on other variables (i.e., $x,y,t,\tau$) and refer to the vertical coordinate. 

Asymptotic series in powers of $\epsilon$ are now introduced for all fluid variables and substituted into the governing equations
(\ref{eqn:gmtm})-(\ref{eqn:gmass}). An order by order analysis is then performed. The asymptotic procedure using 
decompositions of the form (\ref{eqn:MOCE0}) with
(\ref{eqn:climits_a})-(\ref{eqn:climits_c}) is referred to as the `Method of Composite Expansions'
(see section 4.2, \cite{nayfeh2008perturbation}). Specifically, rather than using
the `Method of Matched Asymptotic Expansions' -- {i.e.,} first determining the inner and outer
expansions analytically, matching them, and then forming the composite expansion
\citep{van1964perturbation} --  here variables of the form (\ref{eqn:MOCE0}) are automatically valid
everywhere provided they satisfy the physical boundary conditions at the bounding plates.

\subsection{The Outer Region: Nonhydrostatic Quasigeostrophic Equations}
\label{sec:outer}

Within the fluid interior, inner $(i)$ and middle $(m)$ variables are identically zero. We introduce expansions in powers of $\epsilon$ of the form
\beginar
\label{eqn:asym00}
\ub^{(o)}=\lim(\ub)^o = \eub^{(o)}_0 + \epsilon \eub^{(o)}_1 + \epsilon^2 \eub^{(o)}_2 + \cdots,
\endar
where $Ro\equiv \epsilon$ \citep{mS06, kJ06, kJ07}. The leading order mean component satisfies the motionless hydrostatic balance
\beginar
\label{eqn:mhydro0}
\overline{\eub}^{(o)}_0 = 0, \qquad 
\pd{Z} \overline{P}^{(o)}_0  =\frac{\widetilde{Ra}}{\sigma} \overline{\Theta}^{(o)}_0,
\endar
together with $P^{\prime(o)}_0=\Theta^{\prime(o)}_0=0$.
% and $\pd{t,z}  \lb \overline{\Theta}^{(o)}_0, \overline{P}^{(o)}_0\rb=0$.  {\bf this is confusing - you already assumed that the outer variables are independent of $z$}
The leading order convective dynamics are found to be incompressible and geostrophically balanced, i.e.,  
\beginar
\label{eqn:Geoa}
\begin{array}{ccc}
\hr\times\eub^{\prime(o)}_{0} + \nabla P^{\prime(o)}_1 &=& 0,\\
\nabla\cdot\eub^{\prime(o)}_0 &=& 0,
\end{array}
\qquad
\mbox{or}
\qquad
\mathcal{L}_{geo}
\lb 
\begin{array}{c}
\eub^{\prime(o)}_0 \\ P^{\prime(o)}_1 
\end{array}
\rb = \boldsymbol{0} ,
\endar
where $\mathcal{L}_{geo}$ denotes the geostrophic operator. By definition, all outer variables are independent of
$z$ and so $\nabla=\nabla_\perp\equiv(\pd{x},\pd{y},0)$.\footnote{
If the small-scale $z$ dependence were retained, geostrophy would automatically imply the Proudman-Taylor (PT) constraint 
$\pd{z} \lb \eub^{\prime(o)}_0, P^{\prime(o)}_1\rb \equiv 0$ on the small vertical scale $z$ \citep{jP16,gT23}.}
%(\ref{eqn:MOCE0}).} 
The diagnostic balance given by (\ref{eqn:Geoa}) is solved by 
\Beq
\eub^{\prime(o)}_0= \nabla^\perp \Psi^{(o)}_0 + W^{(o)}_0\hr, \quad P^{\prime(o)}_1=\Psi^{(o)}_0,
\Eeq
for the streamfunction $\Psi^{(o)}_0(x,y,Z,t,\tau)$ and vertical velocity
$W^{(o)}_0(x,y,Z,t,\tau)$.  Here we adopt the definition $\nabla^\perp \Psi^{(o)}_0 \equiv
-\nabla_\perp\times\Psi^{(o)}_0\hr$ with $\nabla^\perp=(-\pd{y},\pd{x},0)$, so that the pressure is now identified as the
geostrophic streamfunction.  It follows that leading order motions are horizontally
nondivergent with $\nabla_\perp\cdot\eub^{\prime(o)}_{0\perp}=0$.  Three-dimensional
incompressiblity is captured at the next order, $\mathcal{O}(\epsilon)$, where 
\Beq
\nabla_\perp\cdot\eub^{\prime(o)}_{1\perp} + \pd{Z}W^{(o)}_0 =0.
\Eeq
This results in the production of subdominant ageostrophic motions
$\eub^{\prime(o)}_{1\perp}$ driven by vertical gradients in $W_0$.
\textcolor{black}{The prognostic evolution of these variables is obtained from balances at next order,
\beginar
\label{eqn:Geob}
\mathcal{L}_{geo}
\lb
\begin{array}{c}
\eub^{\prime(o)}_1 \\ P^{\prime(o)}_2
\end{array}
\rb &=& \boldsymbol{RHS} \\
&\equiv&
\lb
\begin{array}{c}
-
D^{\perp}_{0t} 
%\lb\pd{t} + \eub^{\prime(o)}_0\cdot\nabla_\perp\rb 
\eub^{\prime(o)}_0 - \pd{Z} P_1\hr + \displaystyle{\frac{\widetilde{Ra}}{\sigma}} {\Theta}^{(o)}_1\hr
+\nabla^2_\perp \eub^{\prime(o)}_0  \\
-\pd{Z} W_0 
\end{array}
\rb,
\nonumber
\endar
obtained by projecting $\boldsymbol{RHS}$ onto the null space of $\mathcal{L}_{geo}$ \citep{mS06,mC13}.
%(\cite{mS06}, sec. 2.2 and (\cite{mC13}, sec. 3.3).
This amounts to performing the  projections $\hr\cdot$ and $\nabla^\perp\cdot$ on (\ref{eqn:Geob}).
On noting that $D^{\perp}_{0t} \equiv \pd{t}+\eub^{\prime(o)}_{0\perp}\cdot\nabla_\perp$ this procedure results in asymptotically reduced equations for the vertical vorticity $\zeta^{(o)}_0=\nabla_\perp^2\Psi^{(o)}_0$, vertical velocity $W^{(o)}_0$, and thermal anomaly $\Theta_1^{\prime(o)}$:}
\beginar
\label{eqn:rpsi0}
D^{\perp}_{0t} \zeta^{(o)}_{0}  -\pd{Z}  W^{(o)}_0 &=& \nabla_\perp^2 \zeta^{(o)}_{0}, 
\\ 
%\nn \\
\label{eqn:rw0}
D^{\perp}_{0t} W^{(o)}_0 + \pd{Z} \Psi^{(o)}_0   &=&\frac{\widetilde{Ra}}{\sigma} \Theta_1^{\prime(o)}  + \nabla_\perp^2 W^{(o)}_0, 
\\ 
%\nn \\
\label{eqn:rtempf0}
D^{\perp}_{0t}\Theta_1^{\prime(o)}  + W^{(o)}_0 \pd{Z} \overline{\Theta}^{(o)}_0  &=& \frac{1}{\sigma} \nabla^2_\perp   \Theta_1^{\prime(o)}.
\endar
%where $D^{\perp}_{0t} \equiv \pd{t}+\eub^{\prime(o)}_{0\perp}\cdot\nabla_\perp$. 
The evolution of the 
mean temperature field $\overline{\Theta}^{(o)}_0$ is deduced at $\mathcal{O}(\epsilon^2)$ upon 
averaging over  the fast scales $x,y,z,t$:
\beginar
\label{eqn:rmeant0}
\pd{\tau}\overline{\Theta}^{(o)}_0 + \pd{Z} \lb \overline{\overline{ W^{(o)}_0   \Theta_1^{\prime(o)} }}^\mathcal{T}\rb 
&=& \frac{1}{\sigma} \pd{ZZ}\overline{\Theta}^{(o)}_0.
\endar
In a statistically stationary state, it follows that 
\Beq
Nu = \sigma  \lb \overline{\overline{ W^{(o)}_0    \Theta_1^{\prime(o)} }}^\mathcal{T}\rb  - \pd{Z}\overline{\Theta}^{(o)}_0.
\Eeq
 Equations (\ref{eqn:rpsi0})-(\ref{eqn:rtempf0}),  (\ref{eqn:mhydro0}) and
(\ref{eqn:rmeant0}) constitute the asymptotically reduced system referred
to as the NonHydrostatic QuasiGeostrophic Equations (NH-QGE). A notable 
feature in the NH-QGE is the absence of (higher order) vertical advection.
This is a hallmark characteristic of quasigeostrophic theory.
 Equation (\ref{eqn:rpsi0}) states that vertical vorticity, or toroidal motions, are affected by horizontal
 advection, vortex stretching arising from  the linear Coriolis force, and horizontal diffusion,  while equation (\ref{eqn:rw0}) shows that vertical motions are affected by horizontal
 advection, unbalanced pressure gradient, horizontal diffusion and buoyancy.  The buoyancy
 forces are captured by the fluctuating and mean temperature equations (\ref{eqn:rtempf0}, \ref{eqn:rmeant0}).

The system (\ref{eqn:rpsi0})-(\ref{eqn:rmeant0}) is accompanied by impenetrable boundary conditions
\Beq
\label{eqn:impen0}
W_0^{(o)}(0)=W_0^{(o)}(1)=0,
\Eeq
together with thermal conditions, hereafter taken to be the fixed temperature conditions
\Beq
\label{eqn:meantbc}
\overline{\Theta}^{(o)}_0(0)=1,\quad 
\overline{\Theta}^{(o)}_0(1)=0.
\Eeq

In the presence of impenetrable boundaries, the boundary limits $Z\rightarrow
0$ or $Z\rightarrow 1$ of equation (\ref{eqn:rtempf0}) for the temperature
fluctuations $\Theta^{(o)}_1$ reduce to the advection-diffusion
equation  
\beginar
D_{0t}   \Theta^{\prime(o)}_1   = \frac{1}{\sigma} \nabla_\perp^2   \Theta^{\prime(o)}_1.
\endar
The horizontally-averaged variance $\overline{\Theta^{\prime(o)2}_1}$ for such an
equation evolves according to 
\beginar
\pd{t} \overline{\Theta^{\prime(o)2}_1}  =  -\frac{1}{\sigma} 
\overline{\vert \nabla_\perp  \Theta^{\prime(o)}_1 \vert^2}.
\endar
Therefore $\overline{\Theta^{\prime(o)2}_1}$ decreases monotonically to
zero in time implying the implicit thermal boundary condition $
\Theta^{\prime(o)}_1(0)  = \Theta^{\prime(o)}_1(1) =0$.  
Together with  the impenetrability condition (\ref{eqn:impen0}), the vertical momentum equation (\ref{eqn:rw0}) implies that motions along the horizontal boundaries are implicitly stress-free with 
\Beq
\label{eqn:SF0}
\pd{Z}\Psi^{(o)}_0(0)=\pd{Z}\Psi^{(o)}_0(1)=0.
\Eeq
If rapidly rotating RBC in the presence of stress-free boundary conditions is
the primary objective, a well-posed closed system is obtained from 
(\ref{eqn:rpsi0})-(\ref{eqn:rtempf0}), (\ref{eqn:mhydro0}) and
(\ref{eqn:rmeant0}) together with impenetrable boundary conditions
(\ref{eqn:impen0}) and fixed mean temperature boundary conditions
(\ref{eqn:meantbc}).

\begin{figure}
  \begin{center}
   \includegraphics[height=4.5cm]{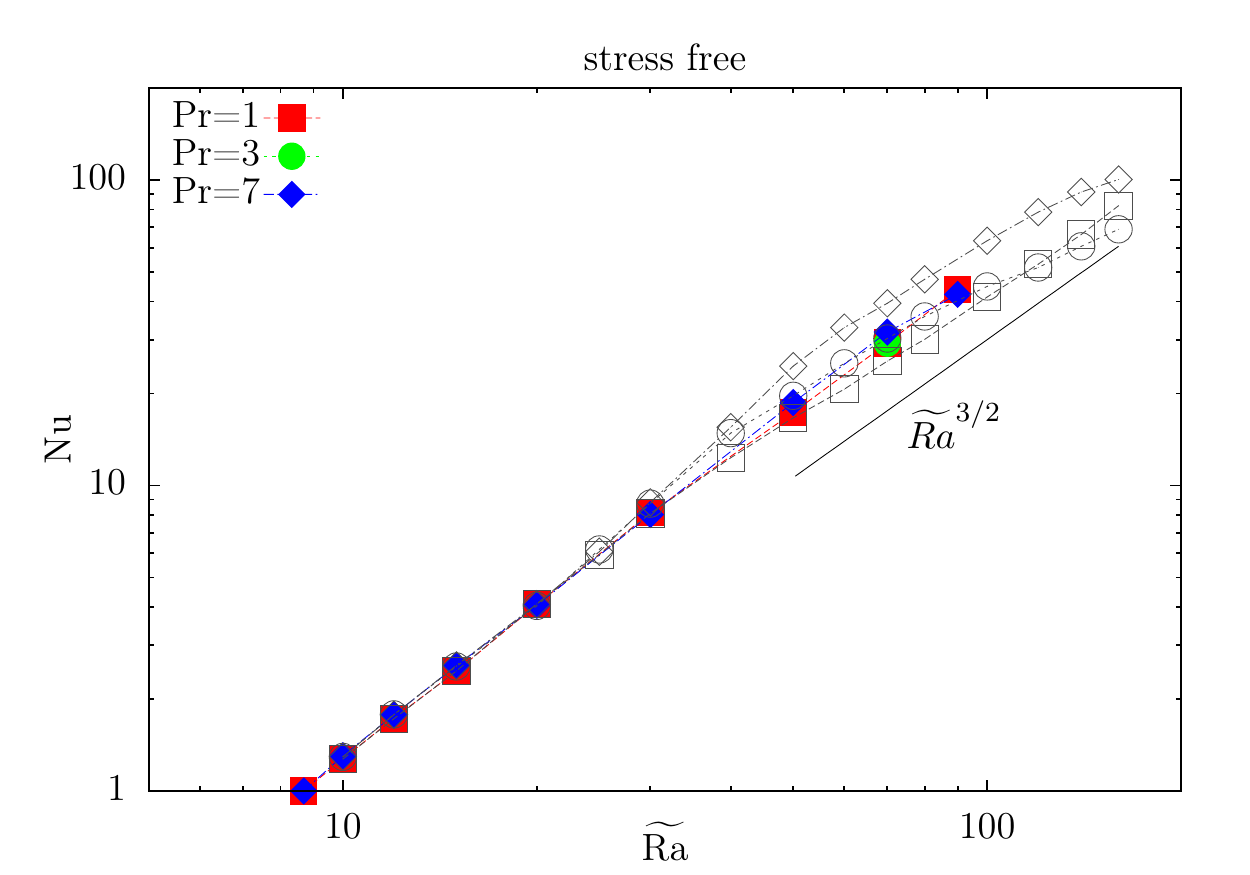}
    \includegraphics[height=4.5cm]{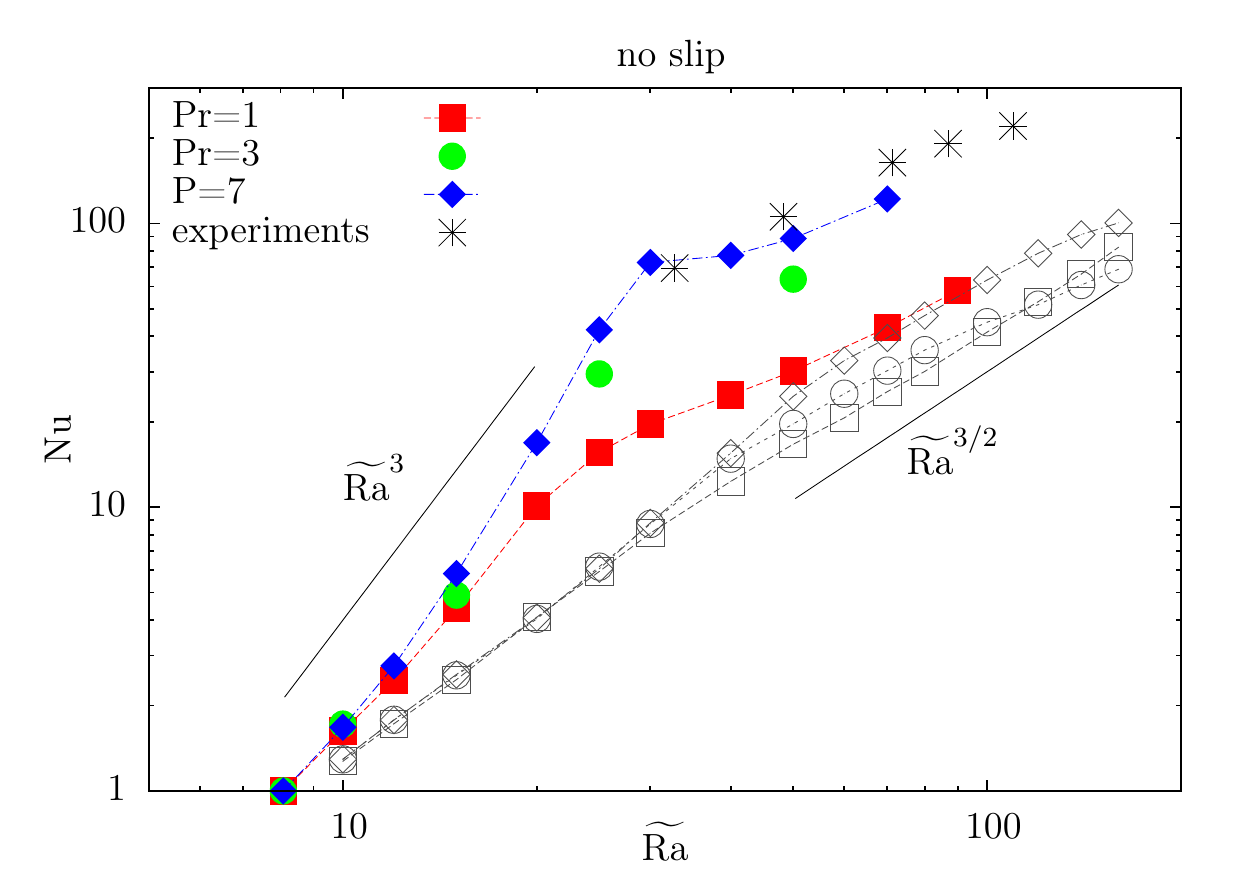}
  \end{center}
\caption{\small{Comparison of the Nusselt number $Nu$ as a function of the reduced 
Rayleigh number $\widetilde{Ra}=Ra E^{4/3}$ for NH-QGE and DNS with (a) stress-free and (b) no-slip boundary
conditions. The filled symbols illustrate DNS data obtained at $E=10^{-7}$ while open symbols are 
from the reduced NH-QGE. Data courtesy of \cite{kJ12} and \cite{sS14}.}}
\label{fig:NuRa_DNS-NH-QGE}
\end{figure}

\subsubsection{Validity of the NH-QGE}
\label{sec:validity}
In the presence of stress-free boundary conditions the NH-QGE remain valid throughout the entire flow domain provided geostrophy, Eq.~(\ref{eqn:Geoa}), holds. This remains so provided the local Rossby number $Ro_l \ll 1$. Given $\eub_0^* = \eub^{(o)}_0\nu/L$, one finds\footnote{The Landau notation little-$o$ denotes a function that is of lower order of magnitude than a given function, that is, the function $o(\epsilon^{-1})$ is of a lower order than the function $\epsilon^{-1}$.}
\Beq
\label{eqn:vorthres}
Ro_l = \frac{\vert \eub^*_0\vert}{2\Omega L} = {\vert\eub^{(o)}_0\vert} E\lb 
\frac{H}{L}\rb^2 =  {\vert\eub^{(o)}_0\vert} E^{1/3}
\quad\Rightarrow\quad 
\vert\eub^{(o)}_0\vert\sim \vert\zeta^{(o)}_0\vert=o(\epsilon^{-1}).
\Eeq
%\footnote{The Landau notation little-$o$ denotes a function that is of lower order of magnitude than a given function, that is the function $o(\epsilon^{-1})$ is of a lower order than the function $\epsilon^{-1}$.}
In a detailed investigation of the NH-QGE, \cite{kJ12b} have shown that this criterion is violated at the transitional value
\Beq
\widetilde{Ra}_{tr} =\mathcal{O}(\epsilon^{-4/5}), \quad
Ro_{tr} = \mathcal{O}(\epsilon^{3/5}) \quad\mbox{as}\quad \epsilon\rightarrow 0.
\Eeq
\begin{figure}
  \begin{center}
      \includegraphics[height=17cm]{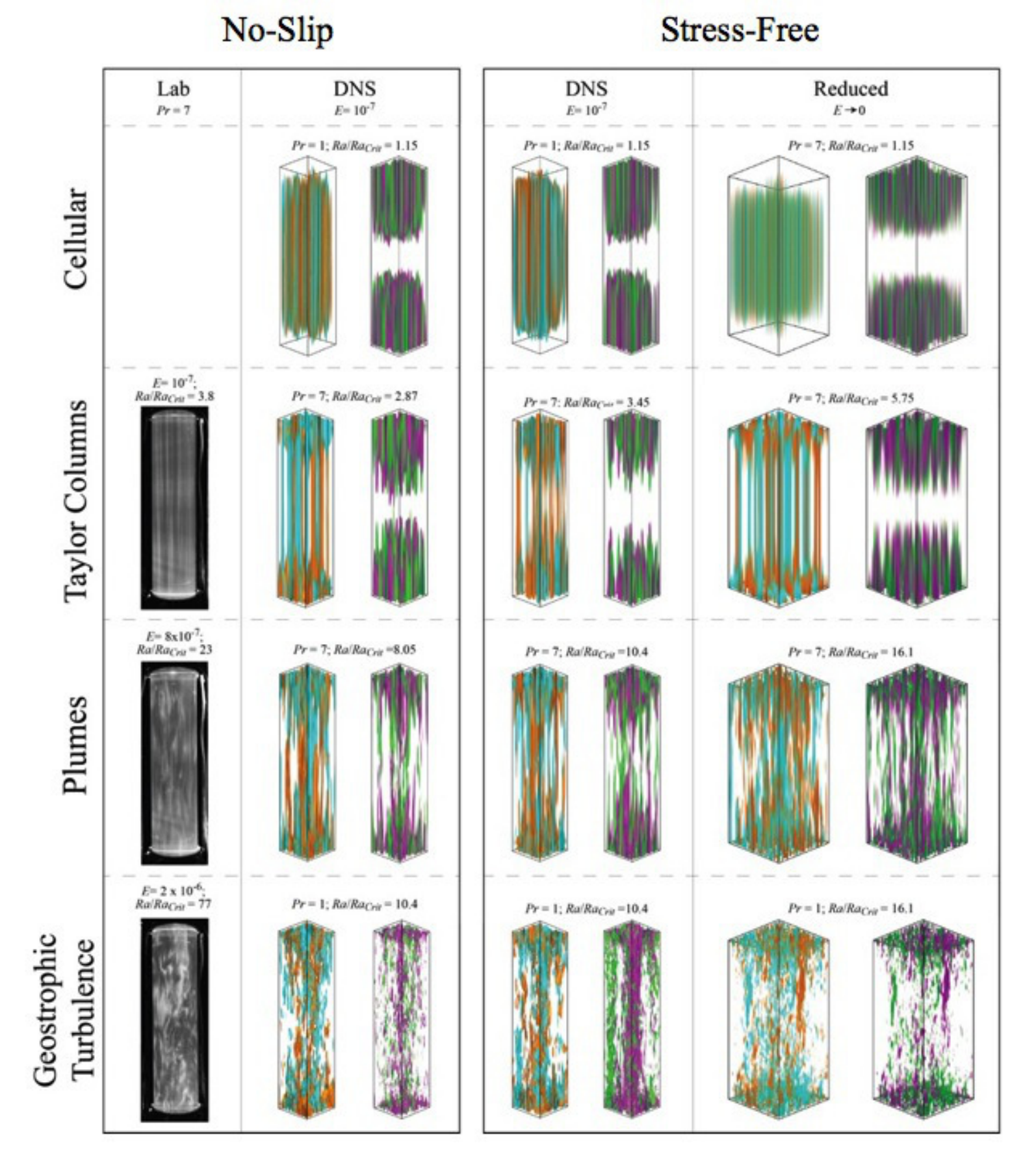}
  \end{center}
\caption{\small{Comparisons between laboratory experiments, DNS and NH-QGE of
flow morphologies of rotationally constrained Rayleigh-B\'enard convection.  As
$\widetilde{Ra}$ increases the flow transitions  between cellular, convective
Taylor columns, plumes and geostrophic turbulence regimes. 
}}
\label{fig:Vis_DNS-NH-QGE}
\end{figure}

In this regime the thermal boundary layers experience a loss of geostrophic balance. 
This provides an upper bound for comparisons  of NH-QGE with DNS with stress-free boundary
conditions. Indeed, as illustrated in Figure~\ref{fig:NuRa_DNS-NH-QGE}(a) for $E=10^{-7}$, 
within the regime of validity, $\widetilde{Ra}\lesssim 70$, good quantitative agreement in the heat 
transport measurements is observed  \citep{sS14}.  Simulations \citep{kJ12, dN14} of the
NH-QGE prior to this transition have revealed four different flow morphologies
subsequently confirmed by both DNS \citep{sS14} and laboratory experiments
\citep{jS15} as $\widetilde{Ra}$ is increased (see
Figure~\ref{fig:Vis_DNS-NH-QGE}): a cellular regime, a convective Taylor column
(CTC) regime consisting of weakly interacting shielded columns, a plume regime
where CTCs have lost stability, and finally a geostrophic turbulence regime that
is also associated with an inverse energy cascade that produces a
depth-independent large-scale dipole vortex pair. All regimes are identifiable
by changes in the heat transport exponent: the CTC regime exhibits a steep
heat transport scaling law where $Nu\propto \widetilde{Ra}^{2.1}$, whereas the
geostrophic turbulence regime is characterized by a dissipation-free scaling
law $Nu\propto \sigma^{-1/2} \widetilde{Ra}^{3/2}$ \citep{kJ12b} and an inverse turbulent energy cascade \citep{kJ12b,kJ12,aR14,bF14,cG14,sS14}.
\textcolor{black}{A rigorous upper bound heat transport result for the NH-QGE, $Nu< C \widetilde{Ra}^{3}$, where $C$ is a constant, has also been reported \citep{iG2015, iG2015b}.}

On the other hand, comparison between the NH-QGE and DNS with no-slip boundaries and laboratory
experiments \citep{sS14} reveals substantial differences (Figure~\ref{fig:NuRa_DNS-NH-QGE}(b)). 
Specifically, a steep scaling law in the heat transport
is observed in the DNS study. Moreover, DNS and laboratory results both suggest that the
steep scaling continues as $E\rightarrow 0$ (Figure~\ref{fig:NuRa}).  It is now
evident from the implicitly enforced stress-free boundary condition (\ref{eqn:SF0}) that
the reduced NH-QGE system and its solutions cannot be uniformly continued to
impenetrable no-slip boundaries, where
\Beq
\label{eqn:mechbc}
\ub_0(0)=\ub_0(1)=\boldsymbol{0}. 
\Eeq
In the presence of no-slip boundaries, it is well-known that the viscous
boundary layers are Ekman layers of depth ${\mathcal O}(E^{1/2} H)$
\citep{hG69}. Within these layers the geostrophic velocity field $\eub^{(o)}_{0\perp}$
in the bulk must be reduced to zero.  In the following, we proceed with an
analysis of the rotationally constrained regime with the intent of
extending the NH-QGE to the case of no-slip boundaries.

\subsection{Inner Region: Ekman Boundary Layers}
\label{sec:Ekman}

To avoid duplication, we focus on the lower Ekman boundary layer at $Z=0$ with nondimensional 
depth ${\mathcal O}(\epsilon^{1/2})$ (an identical analysis applies for the upper boundary layer 
at $Z=1$). \textcolor{black}{This depth arises as a result of the 
spatially anisotropic structure of rapidly rotating convection \citep{wH71}. In dimensional 
terms the Ekman layer depth $\epsilon^{1/2} L \equiv E^{1/2} H$ since $L/H=\epsilon=E^{1/3}$. }

\begin{figure}
  \begin{center}
            \includegraphics[height=10cm]{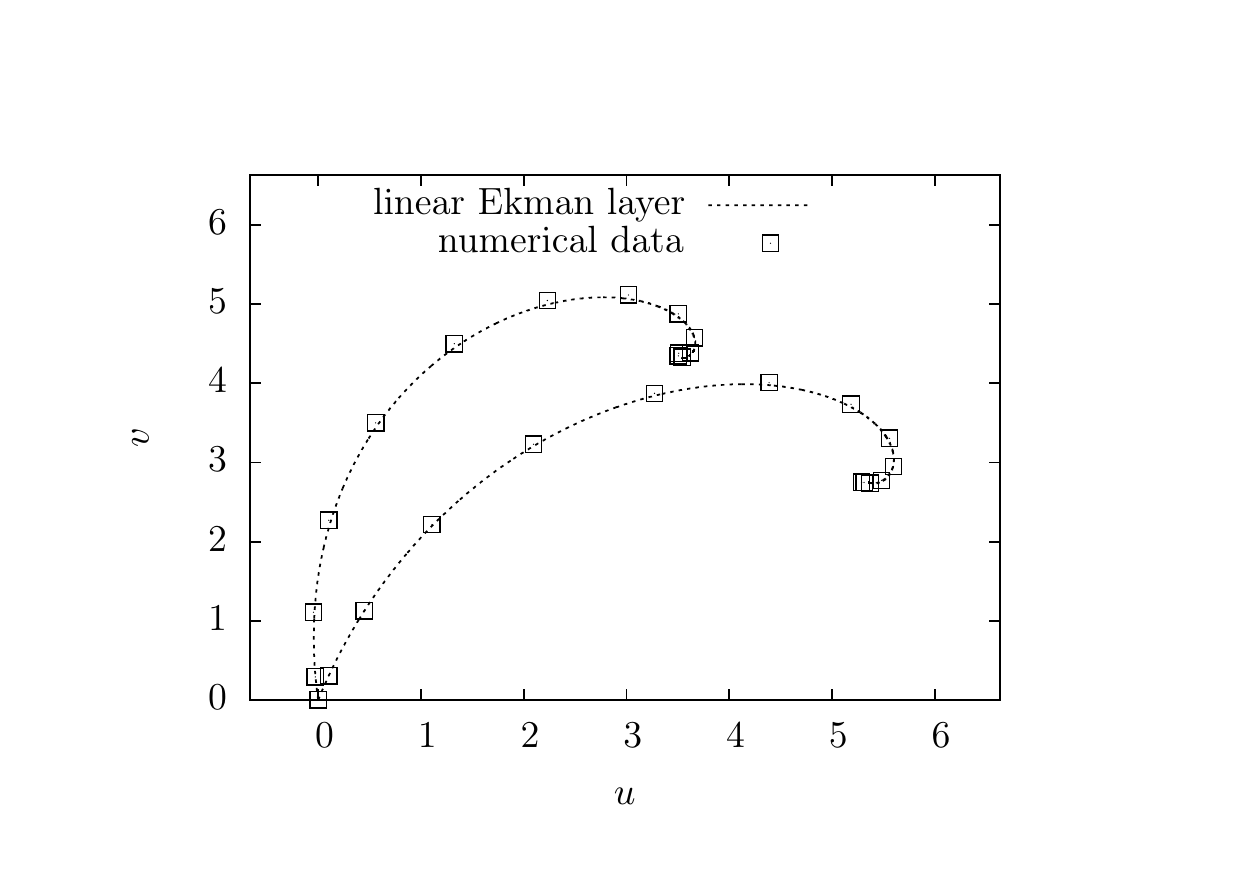}
  \end{center}
  \vspace{-6ex}
\caption{\small{Sample Ekman spiral profiles, i.e., projections of the vertical profiles of the horizontal velocities $u(z),v(z)$ onto the $u,v$ plane, at two fixed horizontal location in the lower viscous layer of the DNS at $Ra E^{4/3}=20$, $E=10^{-7}$, $\sigma=7$. The dashed line illustrates the analytic solution $(U_0, V_0)$ obtained from (\ref{eqn:ek1}), (\ref{eqn:ek2}). Open squares correspond to different vertical locations in the DNS boundary layer, computed using a vertical Chebyshev discretization with $385$ grid points.}} 
\label{fig:Ekspiral} 
\end{figure}

\textcolor{black}{The equations that capture the Ekman layer dynamics are obtained by taking the
inner limit of the governing equations (\ref{eqn:gmtm})-(\ref{eqn:gmass}) about $Z=0$ using 
(\ref{eqn:climits_c}).
We pose an asymptotic inner expansion in powers of $\epsilon^{1/2}$ of the form
\beginar
\label{eqn:asym1}
\ub^{(i)}=\lim(\ub)^i  = \eub^{(i)}_0 + \epsilon^{1/2} \eub^{(i)}_{1/2} + \epsilon \eub^{(i)}_1 + \cdots, 
\endar
and utilize, {\it a posteriori}, knowledge that the middle layer variables have the asymptotic form
%{\bf EK: wrong superscript?}
\beginar
\label{eqn:asymm}
\ub^{(m)}=\epsilon \eub^{(m)}_{1} + \cdots, \quad
p^{(m)}=\epsilon^2 P^{(m)}_{2} + \cdots 
\endar
and therefore do not contribute to leading order. Subtracting the contributions of the outer region then yields}   
\beginar
\hr\times\eub^{(i)}_{0\perp} &=& - \nabla_\perp P^{(i)}_1 + \pd{\mu\mu} \eub^{(i)}_{0\perp}, \\
                              0  &=& -\pd{\mu} P^{(i)}_1, \label{E:eblvert} \\
\nabla_\perp \cdot \eub^{(i)}_{0\perp} + \pd{\mu}  W^{(i)}_{1/2} &=& 0.
\endar
Given that the mean components are identically zero, the primed notation is omitted.
It follows from equation \eqref{E:eblvert} that the pressure within the Ekman
boundary layer is the same as that outside it and we thus take $P^{(i)}_1\equiv
0$.  The classical linear Ekman equations are therefore
\beginar
\hr\times\eub^{(i)}_{0\perp} &=&  \pd{\mu\mu} \eub^{(i)}_{0\perp}, \\
\label{eqn:clekman}
\nabla_\perp \cdot \eub^{(i)}_{0\perp} + \pd{\mu}  W^{(i)}_{1/2} &=& 0.
\endar
After a simple reformulation and the introduction of no-slip boundary conditions, 
we have 
\beginar
\left ( \partial^4_{\mu}  + 1 \right ) \eub^{(i)}_{0\perp} = 0, \quad \eub^{(i)}_{0\perp}(0)+ \eub^{(o)}_{0\perp}(0)  =0, \quad \eub^{(i)}_{0\perp}(\mu \rightarrow\infty)=0,
\endar
where we have utilized in advance that the leading order middle layer variables $(\eub_0^{(m)},P_1^{(m)})$ $\equiv0$.
Since the flow within the Ekman layer is horizontally divergent, equation (\ref{eqn:clekman}) 
implies the presence of vertical motions with velocity $w = \epsilon^{1/2} W^{(i)}_{1/2}$. 

\textcolor{black}{The classical solution \citep{hG69} is found within the Ekman layer at $Z=0$ and is given by}
\beginar
\label{eqn:ek1}
U^{(i)}_0(x,y,\mu,t) &=& - e^{-\frac{\mu}{\sqrt{2}}} \left (  U^{(o)}_0(x,y,0,t) \cos \frac{\mu}{\sqrt{2}} + V^{(o)}_0(x,y,0,t) \sin \frac{\mu}{\sqrt{2}}\right ),
\\
\label{eqn:ek2}
V^{(i)}_0(x,y,\mu,t) &=&  - e^{-\frac{\mu}{\sqrt{2}}} \left (  V^{(o)}_0(x,y,0,t) \cos \frac{\mu}{\sqrt{2}}- U^{(o)}_0(x,y,0,t) \sin \frac{\mu}{\sqrt{2}} \right ),
\\
\label{eqn:ek3}
P^{(i)}_1(x,y,\mu,t) &=& 0.
\endar
Figure~\ref{fig:Ekspiral} illustrates sample boundary layer profiles obtained
from DNS. The projected horizontal  velocities are in perfect agreement with 
(\ref{eqn:ek1}) and (\ref{eqn:ek2}). These structures are robust throughout the boundary layer, thus
providing confirmation of the existence of a linear Ekman layer \citep{sS14}.  
Application of mass conservation (\ref{eqn:clekman}) yields the inner and outer vertical velocities
\beginar
\label{eqn:ekpumpT}
w &=& W^{(o)}_0(x,y,0,t) +\epsilon^{1/2} W_{1/2}^{(i)}(x,y,\mu,t) \\
&=&\epsilon^{1/2} \frac{1}{\sqrt{2}} \zeta^{(o)}_0(x,y,0,t)  - \epsilon^{1/2} \frac{1}{\sqrt{2}} \zeta^{(o)}_0(x,y,0,t) e^{-\frac{\mu}{\sqrt{2}}}  \left [ \cos \frac{\mu}{\sqrt{2}}+\sin \frac{\mu}{\sqrt{2}} \right]. \nn
\endar
As $\mu\rightarrow\infty$, we see that $\lim\lb W_{1/2}^{(i)}\rb^o= 0$ and hence that
\Beq
\label{eqn:ekpump0}
%\lim (w(x,y,0,t))^o =
 W_0^{(o)}(x,y,0,t)  
= \epsilon^{1/2} \frac{1}{\sqrt{2}} \zeta^{(o)}_0(x,y,0,t).
\Eeq
\textcolor{black}{This relation, often referred to as the Ekman pumping boundary condition,
constitutes an exact parameterization of the linear Ekman layer in rotating RBC and represents
Ekman pumping when $W_0^{(o)}>0$ and Ekman suction when $W_0^{(o)}<0$. Hereafter, we do not distinguish between these two cases and refer to this phenomenon generically as Ekman pumping.  
%The relation (\ref{eqn:ekpump0}), often referred to as the Ekman pumping boundary condition,
%constitutes an exact parameterization of the linear Ekman layer in rotating RBC. 
Use of the Ekman pumping boundary condition alleviates the need to resolve the velocity 
dynamics that occur within the Ekman layer.} A similar analysis of the Ekman layer at $Z=1$ gives
\Beq
\label{eqn:ekpump1}
%\lim (w)^o (x,y,1,t)=
  W_0^{(o)}(x,y,1,t) 
= -\epsilon^{1/2}  \frac{1}{\sqrt{2}} \zeta^{(o)} _0(x,y,1,t).
%W_{1/2}^{(o)}(x,y,1,t,\tau) = - \frac{1}{\sqrt{2}} \zeta^{(o)} _0(x,y,1,t,\tau).
\Eeq

Equations (\ref{eqn:ekpump0})-(\ref{eqn:ekpump1}) exemplify the following important distinction between the asymptotic procedure performed here and the linear analyses of \cite{pN65} and \cite{wH71} which implicitly assume that $W_{1/2}^{(o)}=\pm \frac{1}{\sqrt{2}} \zeta^{(o)}_0$ 
\textcolor{black}{as opposed to $W_{0}^{(o)}=\pm \epsilon^{1/2} \frac{1}{\sqrt{2}} \zeta^{(o)}_0$}
(parameterized boundary conditions were not uncovered in these articles). The latter approach
yields a perturbative analysis that captures a range of $\widetilde{Ra}$ for which the effect of Ekman pumping
remains asymptotically close to the stress-free problem. 
However, for a sufficiently large $\widetilde{Ra}$, determined below in \S\ref{sec:rathres}, the asymptotic expansion 
breaks down and becomes nonuniform. For completeness, this result is summarized in the  Appendix.  We find that maintaining asymptotic uniformity requires that pumping be elevated to contribute to the leading order vertical velocity, as pursued here.
This approach enables a complete exploration of Ekman pumping for all $\widetilde{Ra}$ for which the NH-QGE remain valid.

\textcolor{black}{Several studies \citep{barcilon1965stability,faller1966numerical,dudis1971energy} have established that solutions to the classical linear Ekman layer are unstable to small $\mathcal{O}(E^{1/2} H)$ horizontal scale disturbances that evolve on rapid timescales that are  filtered  from the reduced dynamics. This occurs when the boundary layer Reynolds number $R_E = \vert \eub^{(o)}_{0\perp} \vert E^{1/6}$ $\sim 55$. Although this is within the realm of possibility for the rotationally constrained regime according to (\ref{eqn:vorthres}), to alter the pumping parameterization such an instability must also reach amplitudes comparable to the bulk vorticity. No evidence of this is presently seen in DNS.}

The Ekman layer solutions (\ref{eqn:ek1})-(\ref{eqn:ek2}) together with the
application of the exact parameterizations
(\ref{eqn:ekpump0})-(\ref{eqn:ekpump1}) indicate that the outer velocity fields
$(\eub^{(o)}_{0\perp},W_0^{(o)})$ can be \textit{continued uniformly} to the
computational boundaries {at $Z=0,1$. Evidence for this two-layer structure is
provided in Figure~\ref{fig:vbl} which illustrates the RMS velocity profiles at various magnifications 
obtained from DNS at $\widetilde{Ra}=20, E=10^{-7}, \sigma=7$.
 Specifically, the figure demonstrates that the velocity structure is unaffected by the presence of the thermal boundary layer, defined in terms the maxima of the RMS of $\theta$ (purple line, plot (b)).
%of a thermal boundary layer.  \textcolor{red}{WHICH FIGURE HERE?}.
\begin{figure}
  \begin{center}
       \includegraphics[height=4.5cm]{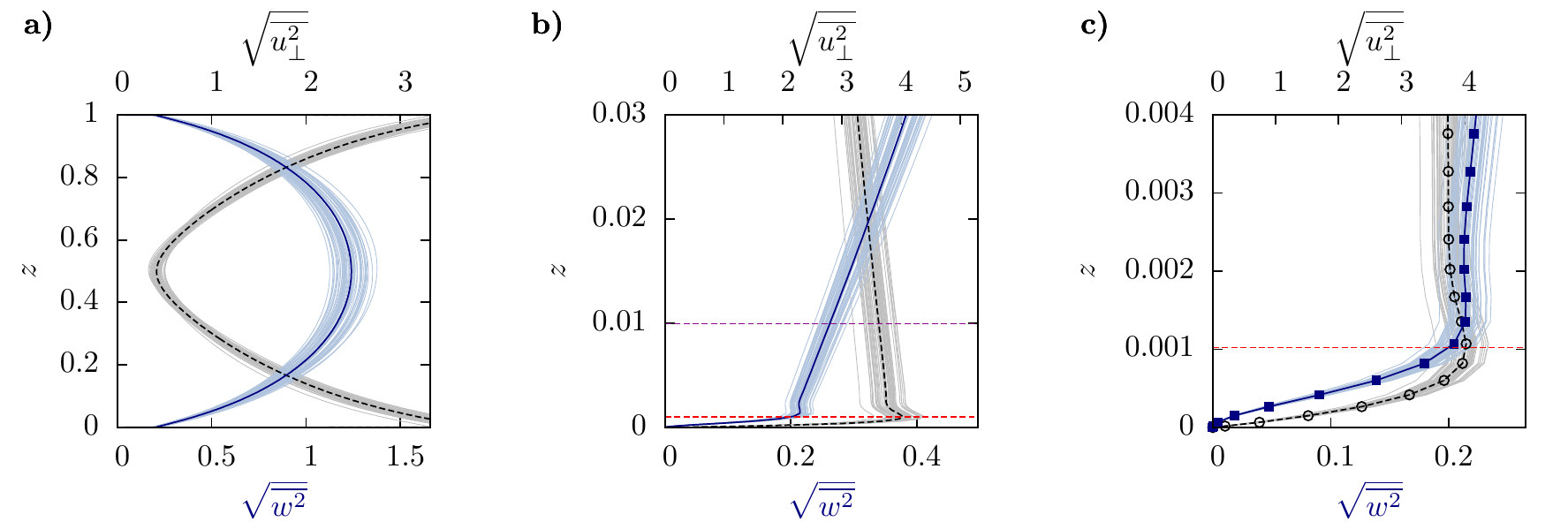}
  \end{center}
  \caption{\small{RMS velocity profiles obtained via DNS with no-slip boundaries at $\widetilde{Ra}=20$, $E=10^{-7}$, $\sigma=7$.  Blue, solid vertical line:  RMS horizontal velocity $\sqrt{\overline{u^2_\perp}}$;  black, solid vertical line: RMS vertical velocity $\sqrt{\overline{w^2_\perp}}$. Shaded regions denote the variance obtained from a time series. (a) \textcolor{black}{Entire layer}, (b) magnification of the $\mathcal{O}(E^{1/3} H)$ thermal boundary layer scale, and (c) magnification of the $\mathcal{O}(E^{1/2} H)$ lower Ekman boundary layer. The Ekman boundary layer is delineated by the red, horizontal dashed line, while the fluctuating thermal boundary layer (that has no visible effect on the velocity profiles) is delineated by the purple, horizontal dashed line. The numerical grid points are marked by square and circular symbols in plot (c). The \textcolor{black}{bulk} velocity profile plotted in (a) transitions at the Ekman layer illustrated in (c).}}
\label{fig:vbl}
\end{figure}
These results suggests that Ekman layers in DNS 
of rotating RBC may be replaced by the parameterized pumping boundary conditions
\Beq
w(x,y,0,t) = \epsilon^{1/2} \frac{1}{\sqrt{2}} \zeta (x,y,0,t),
\quad
w(x,y,1,t) = -\epsilon^{1/2} \frac{1}{\sqrt{2}} \zeta (x,y,1,t).
\Eeq
\cite{sS14} have demonstrated the accuracy of this boundary layer parameterization via
comparison of the heat transport obtained from DNS with no-slip boundaries and DNS 
where the Ekman pumping conditions are used. With all other
details being identical in the two DNS studies, excellent quantitative agreement is reached.
This result establishes that  it is the presence of Ekman layers that is responsible for the 
strong differences in heat transport observed between stress-free and no-slip boundaries 
(Figure~\ref{fig:NuRa_DNS-NH-QGE}). 

The inner component of the pumping velocity $W_{1/2}^{(i)}$ defined in
(\ref{eqn:ekpumpT}) gives rise to an inner temperature fluctuation
$\Theta^{\prime(i)}$ satisfying 
\Beq
%\epsilon^{5/2} 
W_{1/2}^{(i)} \pd{Z} \overline{\Theta}^{(o)}_0 (0)  = \frac{1}{\sigma} \pd{\mu\mu} \Theta_{5/2}^{\prime(i)}.
\Eeq
Utilizing (\ref{eqn:ekpumpT}), the solution to this equation is given by 
%$\theta^{(i)}\equiv\epsilon^{5/2} \theta_{5/2}^{(i)}$ where
\Beq
 \Theta_{5/2}^{\prime(i)} (x,y,\mu,t) =  -\frac{1}{\sqrt{2}} e^{-\frac{\mu}{\sqrt{2}}}  \left [ \cos \frac{\mu}{\sqrt{2}}-\sin \frac{\mu}{\sqrt{2}} \right]\zeta^{(o)}_0(0) 
 \pd{Z} \overline{\Theta}^{(o)}_0 (0).
\Eeq
The limiting values as a function of $\mu\rightarrow 0$  and $\mu\rightarrow\infty$ are %{\bf I think you mean $\mu\rightarrow 0$, correct?}
\Beq
\label{eqn:ebltht}
 \Theta^{\prime(i)}( 0 ) = -\epsilon^{5/2}\frac{1}{\sqrt{2}} \zeta^{(o)}_0(0)\,\pd{Z} \overline{\Theta}_0^{(o)} (0), \qquad  \Theta^{(i)}\lb {\mu\rightarrow\infty} \rb =0.
\Eeq 
It thus follows that  temperature fluctuations  within the Ekman layer are of
magnitude $\Theta^{\prime(i)} = \mathcal{O}\lb \epsilon^{5/2}\zeta^{(o)}_0(0)\,\pd{Z}
\overline{\Theta}_0^{(o)} (0)\rb$. This observation yields an estimate of the convective heat
transport,
\Beq
 \epsilon^{3} W_{1/2}^{(i)}( 0 ) \Theta_{5/2}^{\prime(i)}( 0 )\sim 
\epsilon^{3}\lb\zeta_0^{(o)}(0)\rb^2\,\pd{Z} \overline{\Theta}_0^{(o)} (0),
\Eeq
which is smaller in magnitude by a factor of $\mathcal{O}(\epsilon^2)$ when
compared to that occurring in the convective interior, namely $\epsilon \lb
W_0^{(o)} \Theta_1^{\prime(o)}\rb$. We can thus conclude that the observed
enhancement of heat flux as measured by $Nu$ (Figure~\ref{fig:NuRa_DNS-NH-QGE}) cannot occur
directly within the Ekman layer given the vorticity bound (\ref{eqn:vorthres}).

\begin{figure}
  \begin{center}
       \includegraphics[height=4.5cm]{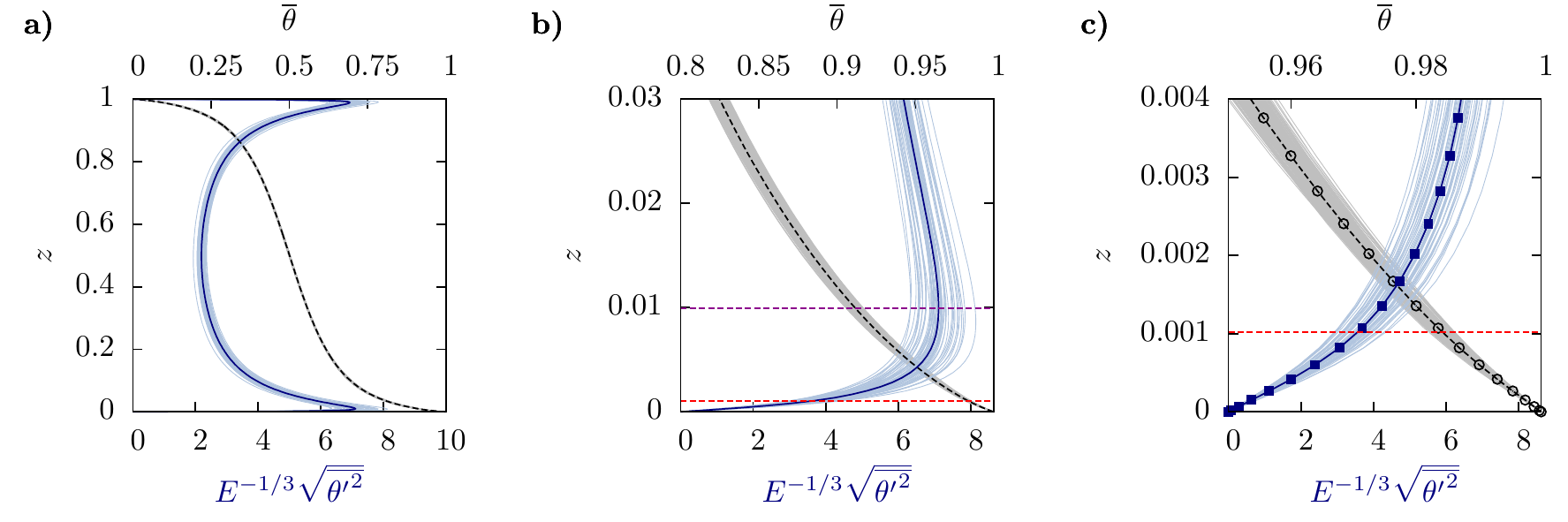}
  \end{center}
  \caption{\small{RMS temperature profiles obtained via DNS with no-slip boundaries at $\widetilde{Ra}=Ra E^{4/3}=20$, $E=10^{-7}$, $\sigma=7$. Blue, solid vertical line: RMS temperature $E^{-1/3} \sqrt{\overline{\theta^{\prime 2}}}$;  black, solid vertical line: RMS mean temperature  ${\overline{\theta}}$. Shaded regions denote the variance obtained from a time series. (a) Entire layer, (b) magnification of the $\mathcal{O}(E^{1/3} H)$ thermal boundary layer scale, and (c) magnification of the Ekman boundary layer.  The fluctuating thermal boundary layer is delineated by the purple, horizontal dashed line, while the Ekman boundary layer is delineated by the red, horizontal dashed line. The numerical grid points are marked by square and circular symbols in plot (c). The thermal profile exhibits no visible boundary layer structure on the Ekman layer scale (plot (c)).}}
%  \caption{\small{RMS velocity profiles obtained via DNS with no-slip boundaries at $\widetilde{Ra}=20,
%E=10^{-7}, \sigma=7$.  Blue, solid vertical line: RMS horizontal  velocity $\sqrt{\overline{u^2_\perp}}$;  black, solid vertical line
%RMS vertical  velocity $\sqrt{\overline{w^2_\perp}}$. Shaded regions denote the variance obtained from a time series. (a) Entire interior region, (b) magnification of the $\mathcal{O}(E^{1/3} H)$ thermal boundary layer scale, and (c) magnification of the $\mathcal{O}(E^{1/2} H)$ lower Ekman boundary layer. The Ekman boundary layer is delineated by the red, horizontal dashed line, whilst the fluctuating thermal boundary layer (that has no visible effect
%on the velocity profiles) is delineated by the purple, horizontal dashed line. The numerical grid points are  marked by square and circular symbols in plot (c).   The interior velocity profile plotted in (a) transitions at the Ekman layer illustrated in (c).}}
%  
\label{fig:bl}
\end{figure}

Inspection of the RMS thermal profiles at increasing magnification obtained at
$E=10^{-7}$ shows no visible boundary layer structure in the
vicinity of the Ekman layer (cf.~Figure~\ref{fig:bl}, plots (b) and (c)).  The RMS profiles reveal
that the thermal boundary layer extends much  farther into the interior than the
Ekman layer observed in Figure \ref{fig:vbl}.  Indeed, we recall that the mean
temperature $\overline{\Theta}_0^{(o)}(Z,\tau)$ is an outer variable
independent of the fast spatial variables $\mu$ and $z$ \citep{mS06}.

\subsubsection{The Significance of Ekman Pumping} 
\label{sec:eksig}

It now remains to determine the nature of the thermal
response in the immediate vicinity of the Ekman layer. Focusing again on the lower boundary, this requires
consideration of how the reduced outer equation for the temperature
fluctuations (\ref{eqn:rtempf0}) can be continued to  the physical boundaries.  On
Taylor-expanding all fluid variables within the Ekman layer, the  outer
component of the pumping velocity $W_0^{(o)}(0)$ induces outer temperature
fluctuations $\Theta_{1}^{\prime(o)}(0)$ that satisfy
\Beq
\label{eqn:stateq}
\lb \pd{t}+ \eub^{(o)}_{0\perp}(0) \cdot\nabla_\perp\rb \Theta_{1}^{\prime(o)}(0) 
+ \frac{\epsilon^{1/2}}{\sqrt{2}} \zeta^{(o)}_0(0)  \pd{Z} \overline{\Theta}_0^{(o)} (0) =\frac{1}{\sigma}
 \nabla^2_\perp \Theta_{1}^{\prime(o)}(0).
\Eeq
Here, we have set $W_0^{(o)}(0)= \frac{\epsilon^{1/2}}{\sqrt{2}} \zeta^{(o)}_0(0)$.

In a statistically stationary state, averaging the equation for the thermal
variance obtained from (\ref{eqn:stateq}) gives the balance
\Beq
 \frac{\epsilon^{1/2}}{\sqrt{2}}\overline{\overline{\Theta_{1}^{\prime(o)}(0)  \zeta^{(o)}_0(0)}}^\mathcal{T}  
\pd{Z} \overline{\Theta}_0^{(o)} (0) =- \frac{1}{\sigma}
\overline{\overline{\vert \nabla_\perp \Theta_{1}^{\prime(o)}(0)\vert^2 }}^\mathcal{T}.
\Eeq
Clearly, in contrast to stress-free boundaries where $\Theta_{1}^{\prime(o)}(0)\equiv 0$,
pumping induces the enhanced thermal response 
\Beq
\label{eqn:thtbc}
\Theta_{1}^{\prime(o)}(0)=\mathcal{O}\lb \epsilon^{1/2} \zeta^{(o)}_0(0)  \pd{Z} \overline{\Theta}_0^{(o)} (0) \rb
\Eeq
immediately outside the Ekman layer.  The associated enhancement in the  convective flux 
is estimated as
\Beq
\label{eqn:convflux}
 \overline{W_{0}^{(o)}(0)  \Theta_{1}^{\prime(o)}(0)} = 
\mathcal{O}\lb  \epsilon\lb \zeta^{(o)}_0(0)\rb^2  \pd{Z} \overline{\Theta}_0^{(o)} (0) \rb.
\Eeq
This equation shows that the convective flux  induced by Ekman pumping becomes as important as the conductive transport $- \pd{Z} \overline{\Theta}_0^{(o)} (0)$ when
\Beq
\zeta^{(o)}_0(0) = \mathcal{O}\lb \epsilon^{-1/2} \rb.
\Eeq
This threshold is \textit{always} achieved within the regime of asymptotic validity of the NH-QGE  that demands $\zeta^{(o)}_0(0) = o(\epsilon^{-1})$
(see \S\ref{sec:validity}). This result also indicates that the differences between stress-free and no-slip boundary conditions are asymptotically $\mathcal{O}(\epsilon^{1/2})$-small for rotationally constrained RBC when $\zeta^{(o)}_0(0) =o\lb \epsilon^{-1/2} \rb$ (see Appendix). Strong, $\mathcal{O}(1)$ departures are predicted to occur in the range
\Beq
\label{eqn:ekpregme}
\mathcal{O}\lb \epsilon^{-1/2} \rb \le  \zeta^{(o)}_0(0) < \mathcal{O}\lb \epsilon^{-1} \rb.
\Eeq
As we show below, the vast majority of laboratory experiments and DNS studies fall within this range.

\subsection{The Middle Region: Thermal Wind Layer} 

It is evident from (\ref{eqn:thtbc}) that $\Theta_{1}^{\prime(o)}(0)$ cannot
satisfy the thermal boundary condition $\Theta^{(o)}(0)=0$, implying that the thermal
response to Ekman pumping in the NH-QGE requires a boundary layer regularization.  However, temperature
fluctuations $ \Theta^{(i)}(0) =\mathcal{O}(\epsilon^{5/2} \Theta^{\prime(0)}_{1}(0))$ within
the Ekman layer are too small to provide compensation (see Eq.~(\ref{eqn:ebltht})). This indicates the
existence of a middle $(m)$ region $\mathcal{O}(\epsilon)$ in depth \citep{wH71}.  

Within the middle region, the inner $(i)$ variables are identically zero and the interior variables achieve their boundary values. \textcolor{black}{We introduce an expansion of the form
\beginar
\label{eqn:asym01}
\ub^{(m)}=\lim(\ub)^m = \eub^{(m)}_0 + \epsilon \eub^{(m)}_1 + \epsilon^2 \eub^{(m)}_2 + \cdots
\endar
and retain geostrophic balance as the leading order fluctuating balance in the middle region, as in
Eq.~(\ref{eqn:Geoa}).}
%  geostrophic balance is retained as the leading order fluctuating balance in the middle region. 
Thus the Proudman-Taylor constraint $\pd{z}(\eub^{(m)}_0, P^{(m)}_1)=0$ implies $(\eub^{(m)}_0, P^{(m)}_1)=0$. 
Departure from the NH-QGE  (\ref{eqn:rpsi0})-(\ref{eqn:rtempf0}),  (\ref{eqn:mhydro0}) and (\ref{eqn:rmeant0}) due to Ekman
pumping can be deduced from the  following prognostic equations 
\beginar
\label{eqn:mtm0}
D^{\perp}_{0t} \eub^{\prime(o)}_{0\perp}  +
 \hr\times\eub_{1} &=& -  \nabla_\perp P^{\prime}_2 + \nabla_\perp^2 \eub^{\prime(o)}_{0\perp}, \\
D^{\perp}_{0t} W^{\prime(o)}_0 &=&-\pd{z} P^{(m)}_2 - \pd{Z} P^{(o)}_1  +\frac{\widetilde{Ra}}{\sigma}\Theta_1+ \nabla_\perp^2 W^{\prime(o)}_0,\hspace{2em} \\ 
\label{eqn:theta0}
D^\perp_{0t} \Theta_1  + W^{\prime(o)}_0\lb \pd{z} \Theta^{(m)}_1 + \pd{Z} \overline{\Theta}^{(o)}_0 \rb &=& \frac{1}{\sigma} \nabla^2  \Theta_1, \\ 
\label{eqn:incomp0}
\nabla\cdot\eub_{1} +  \pd{Z} W^{\prime(o)}_0&=&0.
\endar
Variables without superscripts contribute to both the outer $(o)$ and middle $(m)$ regions and can be separated by taking the limits
(\ref{eqn:climits_a})-(\ref{eqn:climits_c}). Note that Eq.~(\ref{eqn:theta0}) contains the three-dimensional (3D) Laplacian $\nabla^2\equiv\partial^2_x+\partial^2_y+\partial^2_z$.

Inspection of the momentum equation (\ref{eqn:mtm0}) and the continuity equation (\ref{eqn:incomp0}) in the middle region yields the hydrostatic thermal wind balance 
 \beginar
 \hr\times\eub^{\prime(m)}_1 &=& - \nabla_\perp P_2^{\prime(m)},\\
 \pd{z} P_2^{\prime(m)} &=& \frac{\widetilde{Ra}}{\sigma} \Theta^{\prime(m)}_1,\\
 \nabla_\perp\cdot\eub^{\prime(m)}_{1} = 0, \quad & \& &\quad  W^{\prime(m)}_1 \equiv 0,
 \endar
from which we find
\Beq
\label{eqn:thwind}
\pd{z} \eub_{1\perp}^{\prime(m)} = \frac{\widetilde{Ra}}{\sigma} \nabla^\perp \Theta^{\prime(m)}_1,\quad\mbox{s.t.}\quad
\eub_{1\perp}^{\prime(m)} = \nabla^\perp  \Psi^{(m)}_1, \quad
P^{\prime(m)}_2 = \Psi^{(m)}_1.   
\Eeq
The upper and lower middle layer convective dynamics are thus completely reduced to the determination of $\Theta^{\prime(m)}_1$ which from (\ref{eqn:theta0}) evolves according to 
\beginar
\label{eqn:mtempf}
D^\perp_{0t} \Theta^{\prime(m)}_1  + W^{\prime(o)}_0 \pd{z} \lb\overline{\Theta}^{(m)}_1 + \Theta^{\prime(m)}_1 \rb- \pd{z}\lb \overline{ W^{\prime^{(o)}}_0 \Theta^{\prime^{(m)}}_1}\rb  &=& \frac{1}{\sigma} \nabla^2  \Theta^{\prime(m)}_1.
\endar
This couples to the mean state via
\beginar
\pd{z} {\overline{P}}^{(m)}_2&=& \frac{\widetilde{Ra}}{\sigma}  {\overline{\Theta}}^{(m)}_1,\\
\label{eqn:mtempm}
\pd{t}  {\overline{\Theta}}^{(m)}_1+
\pd{z}\lb \overline{ W^{\prime^{(o)}}_0 \Theta^{\prime^{(m)}}_1} \rb &=&\frac{1}{\sigma} \pd{zz} {\overline{\Theta}}^{(m)}_1.
\endar
Equations   (\ref{eqn:mtempf})  and  (\ref{eqn:mtempm}) yield the thermal variance relation
\Beq
\frac{1}{2} \pd{t} \lbr \lb {\overline{\Theta}}^{(m)}_1\rb^2 + \overline{\lb \Theta^{\prime(m)}_1\rb^2}\rbr =
-\frac{1}{\sigma}\lbr \lb \pd{z} {\overline{\Theta}}^{(m)}_1\rb^2 + \overline{\lb \pd{z}  \Theta^{\prime(m)}_1\rb^2} \rbr.
\Eeq
It follows that 
nonzero values of ${\overline{\Theta}}^{(m)}_1, \Theta^{\prime(m)}_1$ only exist in the middle region if they are sustained through the regularizing boundary conditions:
 \beginar
 \label{eqn:mbct}
 \Theta^{\prime(o)}_1 +\Theta^{\prime(m)}_1  =0, \ \   \Theta^{\prime(m)}_1(z\rightarrow\infty) =0,\\ %\quad \mbox{and}\quad
 \label{eqn:mbctb}
 {\overline{\Theta}}^{(o)}_1 + {\overline{\Theta}}^{(m)}_1  =0,\ \ \  {\overline{\Theta}}^{(m)}_1(z\rightarrow\infty) =0%\hspace{2em}
 \endar
 at $Z=0$ or $1$.
 Importantly, we conclude from (\ref{eqn:mbct}, \ref{eqn:mbctb})  that equations (\ref{eqn:mtempf}, \ref{eqn:mtempm}) are fully coupled to the interior dynamics. Integrating (\ref{eqn:mtempm}) over $z$ and $t$  generates the heat transport relation
  \Beq
 \label{eqn:nuss0}
 \sigma \lb\overline{\overline{W^{\prime(o)}_0 \Theta^{\prime(m)}_1}}^\mathcal{T} \rb - \pd{z}\overline{ \overline{\Theta}^{(m)}_1}^\mathcal{T}= 0.
 \Eeq
{By application of the outer limit to this equation,  the constant of integration must be identically zero}. This relation indicates that there is no net heat flux associated with the middle layer dynamics.
 
 We now see that  within the upper and lower middle regions relation (\ref{eqn:Nuss_orig}) yields
\Beq
Nu=\sigma \lb
\overline{\overline{ W^{\prime(o)}_0 \lb  {\Theta^{\prime(o)}_1} + {\Theta^{\prime(m)}_1} \rb } }^\mathcal{T}  
\rb -\lb \pd{Z} \overline{\Theta}_0^{(o)} + \pd{z} \overline{\overline{\Theta}^{(m)}_1}^\mathcal{T} \rb.
\Eeq
Given (\ref{eqn:nuss0}), we find
\Beq
\label{eqn:heat}
Nu=\sigma\lb \overline{\overline{W^{\prime(o)}_0 \Theta^{\prime(o)}_1}}^\mathcal{T} \rb - \pd{Z} \overline{\Theta}_0^{(o)},
\Eeq
valid at every vertical level. This result states that the heat transport within the fluid layer is determined entirely within the \textcolor{black}{bulk}. Moreover, equations  (\ref{eqn:stateq}), (\ref{eqn:thtbc}) and (\ref{eqn:convflux}) now imply that any enhancement in heat transport is entirely due to buoyancy production in $\Theta^{\prime(o)}_1$ arising through the nonlinear advection of the mean temperature gradient $\epsilon^{1/2} \zeta^{\prime(o)}_0\pd{Z} \overline{\Theta}^{(o)}_0/\sqrt{2}$ generated by Ekman pumping.

\begin{figure}
  \begin{center}
       \includegraphics[height=4.5cm]{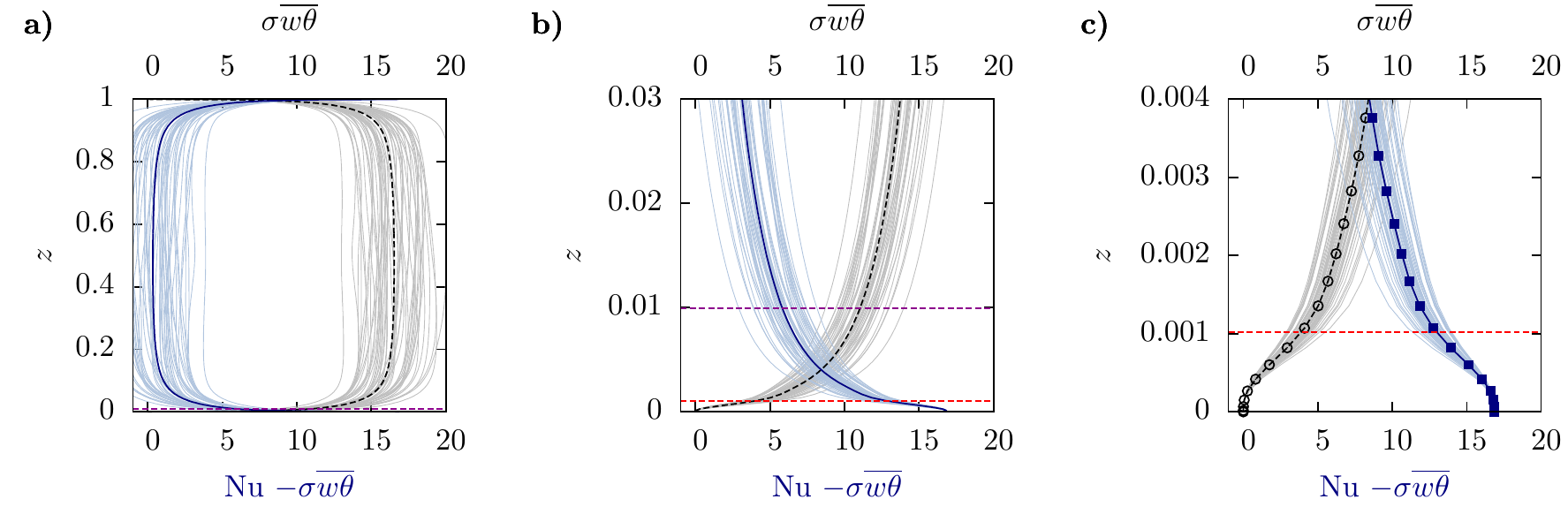}  
       \end{center}
  \caption{\small{Convective flux $\sigma \overline{w\theta}$ (vertical, dashed line) and mean temperature gradient $-\pd{z}\overline{\theta}=Nu-\sigma \overline{w\theta}$
  (vertical, solid line) obtained for DNS with no-slip boundaries at $\widetilde{Ra}=20, E=10^{-7}, \sigma=7$ and average $Nu=17$. Shaded regions denote the variance obtained from a time series.  (a) Entire layer, (b) magnification of the $\mathcal{O}(E^{1/3} H)$ thermal boundary layer scale, and (c) magnification of the Ekman boundary layer. The fluctuating thermal boundary layer is delineated by the purple, horizontal dashed line, while the Ekman boundary layer is delineated by the red, horizontal dashed line. The numerical grid points are marked by square and circular symbols. Equipartition is reached within the thermal wind layer (plot (b)).}}
\label{fig:ttbl}
\end{figure}
DNS results indicate that equipartition between convective and conductive heat transport is achieved within the middle region (see Figure~\ref{fig:ttbl}). For comparison, for stress-free boundary conditions equipartition occurs at a vertical depth well outside that associated with the middle layer.

\subsubsection{Estimation of the transition threshold, $\widetilde{Ra}_{thres}$}
\label{sec:rathres}

In the following, we determine the threshold Rayleigh number $\widetilde{Ra}_{thres}$ at which Ekman pumping gains significance according to the  criterion $\zeta^{(o)}_0(0)\sim\epsilon^{-1/2}$ in (\ref{eqn:ekpregme}). This is achieved by assessing the $\widetilde{Ra}$ dependence of the outer fluid variables prior to the transition threshold where stress-free and no-slip boundary
conditions are presumed to be asymptotically indistinguishable at leading order. DNS at $E= 10^{-7}$ (Figure~\ref{fig:NuRa_DNS-NH-QGE}, \cite{sS14}) have established that Ekman pumping has a significant effect on the heat transport within the laminar CTC regime. We therefore make an {\it a priori} assumption that the transition occurs within this regime.
Simulations of the NH-QGE for rotating Rayleigh-B\'enard convection \citep{kJ12} have established that the dynamics within the CTC regime  
exhibit power law scalings with respect to $\widetilde{Ra}$ both in the bulk and the thermal boundary layer. Hence, we pose the following scaling relations
\beginar
\label{eqn:exp}
& W^{(o)}_0 = \widetilde{Ra}^{\hat w} \widehat{W}^{(o)}_0,\
\Psi^{(o)}_0 = \widetilde{Ra}^{\hat \psi} \widehat{\Psi}^{(o)}_0,\
\zeta^{(o)}_0 = \widetilde{Ra}^{\hat \zeta} \widehat{\zeta}^{(o)}_0,\
\Theta^{\prime(o)}_1 = \widetilde{Ra}^{\hat \theta} \widehat{\Theta}^{(o)}_1,\\
&\pd{Z} \overline{\Theta}^{(o)}_0 =  \widetilde{Ra}^{\hat {dt}} \widehat{\pd{Z} \overline{\Theta}}^{(o)}_0,\
Nu =  \widetilde{Ra}^{\hat{\beta}} \widehat{Nu}. \nn
\endar
On noting that $\pd{t},\nabla_\perp,\pd{Z}=\mathcal{O}(1)$ in the core region and that the CTC structures are known to be axisymmetric to leading order \citep{iG10}, the following balances hold in the NH-QGE 
\beginar
&\pd{t} \zeta_0^{(o)}\sim -\pd{Z}  W_0^{(o)} \sim  \nabla_\perp^2 \zeta^{(o)}_{0}, \\ 
& \pd{t} W_0^{(o)}\sim \pd{Z} \Psi_1^{(o)}   \sim \widetilde{Ra} \Theta_1^{\prime(o)}\sim \nabla^2_\perp W_0^{(o)}, \\
&  \pd{t} \Theta_1^{\prime(o)}\sim  W_0^{(o)} \pd{Z} \overline{\Theta}_0^{(o)}  \sim\displaystyle{ \frac{1}{\sigma}} \nabla_\perp^2  \Theta_1^{\prime(o)},\\
 & \sigma \overline{\overline{W_0^{(o)}  \Theta_1^{\prime(o)}}}^\mathcal{T} \sim Nu. 
\endar
The algebraic equations satisfied by the exponents defined in (\ref{eqn:exp}) are given by
\Beq
{\hat w}={\hat \psi} = {\hat \zeta} = 1 + {\hat \theta},\quad 
{\hat w} + \hat {dt} = {\hat \theta},\quad
{\hat w} + {\hat \theta}  = \hat{\beta}.
\Eeq
On assuming $\hat{\beta}$ is known empirically, we obtain 
\Beq
\label{eqn:exp1}
{\hat w}={\hat \psi} ={\hat \zeta} = \frac{ \hat{\beta}+1}{2},\quad
{\hat \theta}=\frac{ \hat{\beta}-1}{2},\quad
\hat {dt} =-1.
\Eeq
This result indicates that as the amplitude of convection intensifies with increasing $\widetilde{Ra}$ the mean bulk temperature gradient approaches an increasingly well-mixed interior according to $\widetilde{Ra}^{-1}$. This is a well-established result of the CTC regime \citep{mS06, kJ12}  that is confirmed by the reduced simulations \citep{kJ12} which yield $Nu\sim \widetilde{Ra}^{2.1}$ together with the bulk scalings %{\bf replaced $T$ by ${\Theta}_0$}
\beginar
\pd{Z} \overline{\Theta}_0^{(o)} \sim \widetilde{Ra}^{-0.96},\
W_0^{(o)}=\widetilde{Ra}^{1.53},\
\zeta^{(o)}_0 \sim \widetilde{Ra}^{1.55},\
\Theta_1^{\prime(o)}\sim \widetilde{Ra}^{0.62}
\endar
 evaluated at $Z=1/2$ for all variables except vorticity which is evaluated at $Z=3/4$  owing to its antisymmetry. The empirically measured scalings are in good quantitative agreement with the choice $\hat{\beta}\approx 2$, giving 
\Beq
Nu\sim \widetilde{Ra}^{2},\quad W_0^{(o)}=\zeta^{(o)}_0 \sim \widetilde{Ra}^{3/2},\quad
\Theta_1^{(o)} \sim \widetilde{Ra}^{1/2}.
\Eeq

In the thermal boundary layers, where it is once again assumed that $\pd{t},\nabla_\perp=\mathcal{O}(1)$, but now $\pd{Z}=\widetilde{Ra}^{\hat\eta}\gg1$, the following balances hold
\beginar
&\pd{t} \zeta_0^{(o)}\sim -\pd{Z}  W_0^{(o)} \sim  \nabla_\perp^2 \zeta^{(o)}_{0}, \\ 
& \pd{Z} \Psi_1^{(o)}   \sim \widetilde{Ra} \Theta_1^{\prime(o)}, \\
&  \pd{t} \Theta_1^{\prime(o)}\sim  W_0^{(o)} \pd{Z} \overline{\Theta}_0^{(o)}  \sim\displaystyle{ \frac{1}{\sigma}} \nabla_\perp^2  \Theta_1^{\prime(o)},\\
 & \sigma \overline{\overline{W_0^{(o)}  \Theta_1^{\prime(o)}}}^\mathcal{T} \sim \pd{Z} \overline{\Theta}^{(o)}_0 \sim Nu. 
\endar
The algebraic equations satisfied by the exponents defined in (\ref{eqn:exp}) now satisfy
\Beq
{\hat\eta} + {\hat w}={\hat \psi} = {\hat \zeta},\quad 
{\hat\eta} + {\hat \psi} =  1 + {\hat \theta},\quad 
{\hat w} + \hat {dt} = {\hat \theta},\quad
{\hat w} + {\hat \theta} = \hat {dt}  = \hat{\beta}
\Eeq
with the solution
\Beq
\label{eqn:exp2}
{\hat \eta}={\hat \psi} ={\hat \zeta} = \frac{ \hat{\beta}+1}{2},\quad
{\hat w}=0,\quad
{\hat \theta}=\hat {dt} =\hat{\beta}.
\Eeq
%Although simulations have established good order of magnitude estimates for the fields and their gradients, empirical confirmation of the predicted scalings with $\widetilde{Ra}$ is challenging because the boundary layer dynamics are complex and contain structures other than fully developed CTCs \citep{kJ12}.

From (\ref{eqn:exp1}) and (\ref{eqn:exp2}) it now follows that the vorticity satisfies
\Beq
\label{eqn:vscaling}
\zeta^{(o)}_0 \sim \widetilde{Ra}^{\frac{\hat{\beta}+1}{2}} \widehat{\zeta}^{(o)}_0.
\Eeq
and hence that (see Eq.~(\ref{eqn:ekpregme}))
\Beq
\zeta^{(o)}_0(0)\sim\epsilon^{-1/2}\quad\Rightarrow\quad \widetilde{Ra}_{thres} = \mathcal{O}\lb\epsilon^{-\frac{1}{\hat{\beta}+1}}\rb =  \mathcal{O}\lb E^{-\frac{1}{3(\hat{\beta}+1)}}\rb.
\Eeq
Specifically, for the empirically observed CTC value of $\hat{\beta}\approx 2$, 
\Beq
\label{eqn:rthres}
\widetilde{Ra}_{thres} = \mathcal{O}\lb\epsilon^{-{1}/{3}}\rb =  \mathcal{O}\lb E^{-{1}/{9}}\rb.
\Eeq
At $E=10^{-7}$ this gives a value $\widetilde{Ra}_{thres} \sim 6.0$. This is of the same magnitude as the critical Rayleigh number $\widetilde{Ra}_c =8.05$ \citep{wH71} indicating an immediate departure from the stress-free case.  This is a conclusion borne out by the recent DNS (Figure~\ref{fig:NuRa_DNS-NH-QGE}(b)).  \textcolor{black}{At lower values of $E$, it is found that $\widetilde{Ra}_{thres}>\widetilde{Ra}_c$}. From the interior and boundary layer scaling exponents (\ref{eqn:exp1}) and (\ref{eqn:exp2}), we also estimate the magnitude of Ekman pumping normalized by the midplane velocity as
\Beq
S=\frac{E^{1/6}\zeta^{(o)}_0(0)/\sqrt{2}}{w^{(o)}_0(\frac{1}{2})}= \mathcal{O}\lb E^{{1}/{6}}\rb,
\Eeq
thus confirming empirical DNS results (see Figure 5 of \citet{sS14}).

\section{Composite Reduced NH-QGE}

The findings of section~\ref{sec:asymptotic} are now combined to deduce a composite reduced model capable of capturing the thermal effect of Ekman pumping in a single domain $Z=[0,1]$. This is accomplished by reconstituting the fluid variables in each region as defined in (\ref{eqn:MOCE0}) with the exception that the Ekman layer is now parameterized. For convenience, we first summarize the reduced asymptotic equations of the outer and middle regions.

In  the presence of no-slip boundary conditions we have established that the outer, i.e., bulk, region is described by the reduced NH-QGE system:
\begin{center}\underline{Outer Region:}\end{center}
\Beq
  \overline{\eub}^{(o)}_0 \equiv 0, \ 
  \eub^{\prime(o)}_0= \nabla^\perp \Psi^{(o)}_0 + W^{(o)}_0 \hr,\ 
P^{(o)} = \overline{P}^{(o)}_0+ \epsilon  \Psi^{(o)}_0, \ 
 \Theta^{(o)} =  \overline{\Theta}^{(o)}_0 + \epsilon   \Theta^{\prime(o)}_1,
\Eeq
with a hydrostatically evolving mean component
\beginar
\label{eqn:mean_redc}
\left.
\begin{array}{c}
  \pd{Z} \overline{P}_0  =\displaystyle{\frac{\widetilde{Ra}}{\sigma}} \overline{\Theta}_0 \\ \\
 \pd{\tau}\overline{\Theta}^{(o)}_0 + 
\pd{Z} \lb \overline{{\overline{ W^{(o)}_0  \Theta^{\prime(o)}_1 }}}^\mathcal{T}\rb = \displaystyle{\frac{1}{\sigma}} \pd{ZZ}\overline{\Theta}^{(o)}_0 \\ \\
\overline{\Theta}_0^{(o)}(0) = 1,\  \ 
\overline{\Theta}_0^{(o)}(1) = 0
\end{array}
\right\}
\endar
with fixed mean temperature boundaries.
The quasigeostrophically evolving fluctuating components are given by
\beginar
  \label{eqn:ekpumpc}
\left.
\begin{array}{c}
D^{\perp}_{0t} \zeta^{(o)}_{0}  -\pd{Z}  W^{(o)}_0 = \nabla_\perp^2 \zeta^{(o)}_{0}    \\ \\
 D^{\perp}_{0t} W^{(o)}_0 + \pd{Z} \Psi^{(o)}_0   =\displaystyle{\frac{\widetilde{Ra}}{\sigma}} \Theta^{\prime(o)}_1 + \nabla_\perp^2 W^{(o)}_0   \\ \\
 D^\perp_{0t} \Theta^{\prime(o)}_1   + W^{(o)}_0 \pd{Z} \overline{\Theta}^{(o)}_0  = \displaystyle{\frac{1}{\sigma}}  \nabla^2_\perp   \Theta^{\prime(o)}_1 \\ \\
  W_0^{(o)}(0) = \displaystyle{\frac{\epsilon^{1/2}}{\sqrt{2}}} \zeta^{(o)}_{0}(0),\quad
W_0^{(o)}(1) = -\displaystyle{\frac{\epsilon^{1/2}}{\sqrt{2}}}  \zeta^{(o)}_{0}(1)
\end{array}
\right\}
\endar
with parameterized pumping boundary conditions.
The requirement of zero thermal fluctuations on the boundaries was shown to require the introduction of  a pair of middle boundary layer regions.
\begin{center}{\underline{Middle Regions at $Z=0$ or $1$:}}\end{center}
\Beq
\eub^{\prime(m)}=\epsilon \nabla^\perp \Psi^{(m)}_1,\ 
% +\epsilon^2  W^{(m)}_2 \hr,\ 
P^{(m)}=\epsilon \lb \overline{P}^{(m)}_1 + \Psi^{(m)}_1 \rb, \ 
\Theta^{(m)} =\epsilon \lb  \overline{\Theta}^{(m)}_1 +   \Theta^{\prime(m)}_1\rb   \hspace{1.5em}
\Eeq
with a hydrostatically evolving mean component
\beginar
\label{eqn:allmiddlem}
\left.
\begin{array}{c}
\pd{z} \overline{P}^{(m)}_1   =\displaystyle{\frac{\widetilde{Ra}}{\sigma}} \overline{\Theta}^{(m)}_1   \\ \\
 \pd{t} \overline{\Theta}^{(m)}_1+  \pd{z} \lb\overline{W^{(o)}_0 \Theta^{\prime(m)}_1} \rb = \displaystyle{\frac{1}{\sigma}} \pd{zz} \overline{\Theta}^{(m)}_1 \\ \\
% \overline{T}^{(m)}_1 (Z_b)=0,\quad 
 \overline{\Theta}^{(m)}_1 (\infty)=0,
 \end{array}
\right\}
\endar
and a geostrophically evolving fluctuating component, $\hr\times\eub^{\prime(m)}_1 = -\nabla \Psi^{(m)}_1$ in thermal wind balance:
\beginar
\label{eqn:allmiddlef}
\left.
\begin{array}{c}
\pd{z} \Psi^{(m)}_1   =\displaystyle{\frac{\widetilde{Ra}}{\sigma}} \Theta^{\prime(m)}_1  \\ \\
   D^{\perp}_{0t} \Theta^{\prime(m)}_1 + W^{(o)}_0  \pd{z} \overline{\Theta}^{(m)}_1  
+  \pd{z} \lb   W^{(o)}_0 \Theta^{\prime(m)}_1 - \overline{W^{(o)}_0 \Theta^{\prime(m)}_1} \rb
    =\displaystyle{ \frac{1}{\sigma}} \nabla^2   \Theta^{\prime(m)}_1 \\ \\
    \Theta^{\prime(m)}_1 (Z_b)+\Theta^{\prime(o)}_1(Z_b)=0, \quad \Theta^{\prime(m)}_1 (\infty)=0.
\end{array}
\right\}
\hspace{2em}
\endar
Notably, the reduced dynamics within the middle layer is captured solely by the evolution of $\Theta^{\prime(m)}_1$ in (\ref{eqn:allmiddlef}b) which is coupled to the leading order outer dynamics through the regularizing boundary condition (\ref{eqn:allmiddlef}c). It is now observed that Ekman pumping gives rise to two additional physical effects in (\ref{eqn:allmiddlef}b)  that are absent in the bulk:  nonlinear vertical advection and vertical diffusion of thermal fluctuations, each of which becomes important when $\zeta^{(o)}_0 = \mathcal{O}\lb\epsilon^{-1/2}\rb$.
%$\widetilde{Ra}=\mathcal{O}\lb\epsilon^{-1/2}\rb$.
  
Following  (\ref{eqn:MOCE0}), the outer and middle regions may now be combined into a single composite system capturing dominant contributions upon defining the following composite variables:
\beginar
\overline{\Theta}^{(c)}  &=&   \overline{\Theta}^{(o)}_0(\tau, Z) +   
\epsilon\lb  \overline{\overline{\Theta}^{(m,-)}_1(t,\tau, Z/\epsilon)  +    \overline{\Theta}^{(m,+)}_1(t,\tau, 1-Z/\epsilon)}^\mathcal{T}\rb 
\\
\overline{P}^{(c)}  &=&   \overline{P}^{(o)}_0(\tau, Z) +   \epsilon\lb  \overline{\overline{P}^{(m,-)}_1(t,\tau, Z/\epsilon)  +    \overline{P}^{(m,+)}_1(t,\tau, 1-Z/\epsilon)}^\mathcal{T}\rb 
\\
\Theta^{\prime(c)} &=&   \Theta^{\prime(o)}_1(x,y,t;\tau, Z) +  \lb \Theta^{\prime(m,-)}_1(x,y,t;\tau,Z/\epsilon)  +   \Theta^{\prime(m,+)}_1(x,y,t;\tau,1-Z/\epsilon)\rb \hspace{1em}
\\
\Psi^{(c)} &=&   \Psi^{(o)}_0(x,y,t;\tau, Z) + \epsilon  \lb \Psi^{(m,-)}_1(x,y,t;\tau,Z/\epsilon)  +   \Psi^{(m,+)}_1(x,y,t;\tau,1-Z/\epsilon)\rb
 \\
W^{(c)} &=& W_0^{(o)}(x,y,t;\tau, Z) 
\hspace{2em}
\endar
and reverting to a single vertical coordinate $Z$.
We note that no middle layer corrections to the vertical velocity are required.
%to $\mathcal{O}(\epsilon)$ 
%that the first middle correction to the vertical velocity field occurs at $\mathcal{O}(\epsilon^2)$ and is subdominant {\bf I do not understand this remark}. 
\newpage
We thus arrive at  
\begin{center} \underline{Composite System or CNH-QGE:}\end{center}
\Beq
\label{eqn:var_comp}
  \overline{\eub}^{(c)} \equiv 0, \ 
  \eub^{\prime(c)}= \nabla^\perp \Psi^{(c)} + W^{(c)} \hr,\ 
P^{(c)} = \overline{P}^{(c)}+ \epsilon  \Psi^{(c)}, \ 
 \Theta^{(c)} =  \overline{\Theta}^{(c)} + \epsilon   \Theta^{\prime(c)}
\Eeq
with
\beginar
\label{eqn:mean_comp}
&\left.
\begin{array}{c}
  \pd{Z} \overline{P}^{(c)} =\displaystyle{\frac{\widetilde{Ra}}{\sigma}} \overline{\Theta}^{(c)}  \\ \\
 \pd{\tau}\overline{\Theta}^{(c)} + 
\pd{Z} \lb \overline{{\overline{ W^{(c)} \Theta^{\prime(c)} }}}^\mathcal{T}\rb = \displaystyle{\frac{1}{\sigma}} \pd{ZZ}\overline{\Theta}^{(c)} \\ \\
\overline{\Theta}^{(c)} (0) = 1,\  \ 
\overline{\Theta}^{(c)} (1) = 0,
\end{array}
\right\}
 \\ \nn \\
\label{eqn:fluc_comp}
&\left.
\begin{array}{c}
D^{\perp}_{ct} \zeta^{(c)}  -\pd{Z}  W^{(c)} = \nabla_\perp^2 \zeta^{(c)}    \\ \\
 D^{\perp}_{ct} W^{(c)} + \pd{Z} \Psi^{(c)}   =\displaystyle{\frac{\widetilde{Ra}}{\sigma}} \Theta^{\prime(c)} + \nabla_\perp^2 W^{(c)}  \\ \\
D^\perp_{ct} \Theta^{\prime(c)}  + W^{(c)} \pd{Z}  \overline{\Theta}^{(c)}
+ \epsilon \nabla_\perp\cdot\lb\eub^{(c)}_{1\perp}  \Theta^{\prime(c)} \rb
%\underline
+{\epsilon  \pd{Z} \lb  W^{(c)}  \Theta^{\prime(c)}  -  \overline{W^{(c)} \Theta^{\prime(c)}} \rb }
 = \\ \\
\hspace{20em} \displaystyle{ \frac{1}{\sigma} }\lb \nabla^2_\perp + { \epsilon^2 \pd{ZZ}} \rb   \Theta^{\prime(c)} \\ \\
W^{(c)}(0) =\displaystyle{\frac{\epsilon^{1/2}}{\sqrt{2}}} \zeta^{(c)}(0),\quad W^{(c)}(1) = -\displaystyle{\frac{\epsilon^{1/2}}{\sqrt{2}}} \zeta^{(c)}(1)
 \\ \\
 \Theta^{\prime(c)} (0) = \Theta^{\prime(c)} (1)=0.
\end{array}
\right\} 
%\nn
\endar
Here $D_{ct}^\perp \equiv \pd{t} + \eub^{(c)}_{0\perp}\cdot \nabla_\perp$
\textcolor{black}{and $\eub^{(c)}_{1\perp} $ denotes the ageostrophic field determined through the three-dimensional incompressibility condition
\beginar
\label{eqn:cont2}
\nabla_\perp\cdot \eub^{(c)}_{1\perp}  + \pd{Z} W^{(c)} =0.
\endar}
Descriptively, the composite system captures the geostrophically balanced domain (\ref{eqn:var_comp}), where Ekman layers are parameterized by (\ref{eqn:fluc_comp}d). The fluctuating dynamics (\ref{eqn:fluc_comp}) indicate that vortical dynamics are driven by vortex stretching associated with the linear Coriolis force while vertical motions are driven by buoyancy and unbalanced pressure gradients. The buoyancy source term in (\ref{eqn:fluc_comp}b) is controlled by the evolution of the fluctuating temperature  (\ref{eqn:fluc_comp}c). Here, nonlinear vertical advection and linear vertical diffusion appear as new physical terms in the composite model.  The latter is required in order to enforce fixed temperature boundary conditions (\ref{eqn:fluc_comp}d). We note, however, that it is the vertical advection of a strong mean temperature gradient that gives rise to new near boundary source terms  activated by Ekman pumping when  $  W^{(c)} = \mathcal{O}(1)$. The resulting adjustments to the convective fluxes give rise to significant changes in the mean background state that remains in hydrostatic balance, Eq.~(\ref{eqn:mean_comp}a).
\textcolor{black}{The ageostrophic advective nonlinearity in (\ref{eqn:fluc_comp}c) is retained together with the continuity condition (\ref{eqn:cont2}) in order to maintain asymptotic consistency with known power integrals for the kinetic and thermal energy dissipation, namely 
\beginar
\label{eqn:powering}
&\mathcal{E}_{\boldsymbol{u}} \equiv\left  \langle \overline{\lb\zeta^{(c)}\rb^2} + \overline{\vert\nabla_{\perp}W^{(c)}\vert^2} \right \rangle +
\displaystyle{\frac{\epsilon^{1/2}}{\sqrt{2}}} \lb \overline{\vert \nabla_\perp \Psi^{(c)}(0) \vert^2} + \overline{\vert \nabla_\perp \Psi^{(c)}(1) \vert^2}  \rb =
\displaystyle{ \frac{\widetilde{Ra}}{\sigma^2}} \lb Nu -1 \rb,  \nn
\\
&\mathcal{E}_{\Theta} \equiv\left  \langle  \lb  \pd{Z}   \overline{\Theta}^{(c)} \rb^2 \right \rangle +
 \left  \langle \overline{ \vert \nabla_\perp   \Theta^{\prime (c)} \vert^2}
 + \epsilon^2 \overline{( \pd{Z}   \Theta^{\prime (c)} )^2} \right \rangle   = Nu.
\endar 
From the analysis of section~\ref{sec:rathres} and (\ref{eqn:powering}a) we deduce that Ekman friction remains a subdominant contributor to the kinetic energy dissipation and therefore to the Nusselt number $Nu$ throughout the entire rotationally constrained regime. However, vertical thermal dissipation becomes dominant in the thermal dissipation rate and $Nu$ within the thermal wind layer once  the critical threshold $\widetilde{Ra} = \mathcal{O}(E^{-1/9})$ is reached.}

Several additional and slightly technical comments about the composite system are in order. 
The mean temperature equation (\ref{eqn:mean_comp}b) is obtained upon time filtering the evolution equations (\ref{eqn:allmiddlem}b) over the fast time $t$ to obtain $\overline{\overline{\Theta}_1^{(m)}}^\mathcal{T}$ prior to composition with (\ref{eqn:mean_redc}b). As a consequence of this filtering, evaluation of the advection of the mean temperature in (\ref{eqn:fluc_comp}c) 
incurs the  asymptotically small $\mathcal{O}(\epsilon)$ error
\Beq
\label{eqn:therrcorr}
\epsilon W^{(o)}_0 \lb \pd{Z} \overline{\overline{\Theta}^{(m)}_0}^\mathcal{T} - \pd{Z}  \overline{\Theta}^{(m)}_0
\rb
\Eeq
in the vicinity of the upper and lower bounding plates.
We also note that the composite formulation produces $\mathcal{O}(\epsilon)$ errors in horizontal advection terms of the form 
$\epsilon\eub^{(c)}_{1\perp}\cdot \nabla_\perp \lb \zeta^{(c)}, W^{(c)}%, \Theta^{\prime(c)}
\rb$.
%where $\nabla_\perp \cdot \eub^{(c)}_{1\perp} + \pd{Z} W^{(c)}=0$.
%$\epsilon \nabla^\perp\Psi^{(m)}_1\cdot \nabla_\perp \lb \zeta^{(c)}, W^{(c)}, \Theta^{\prime(c)}\rb$.

\section{Results}

\begin{figure}
  \begin{center}
      \includegraphics[height=4.5cm]{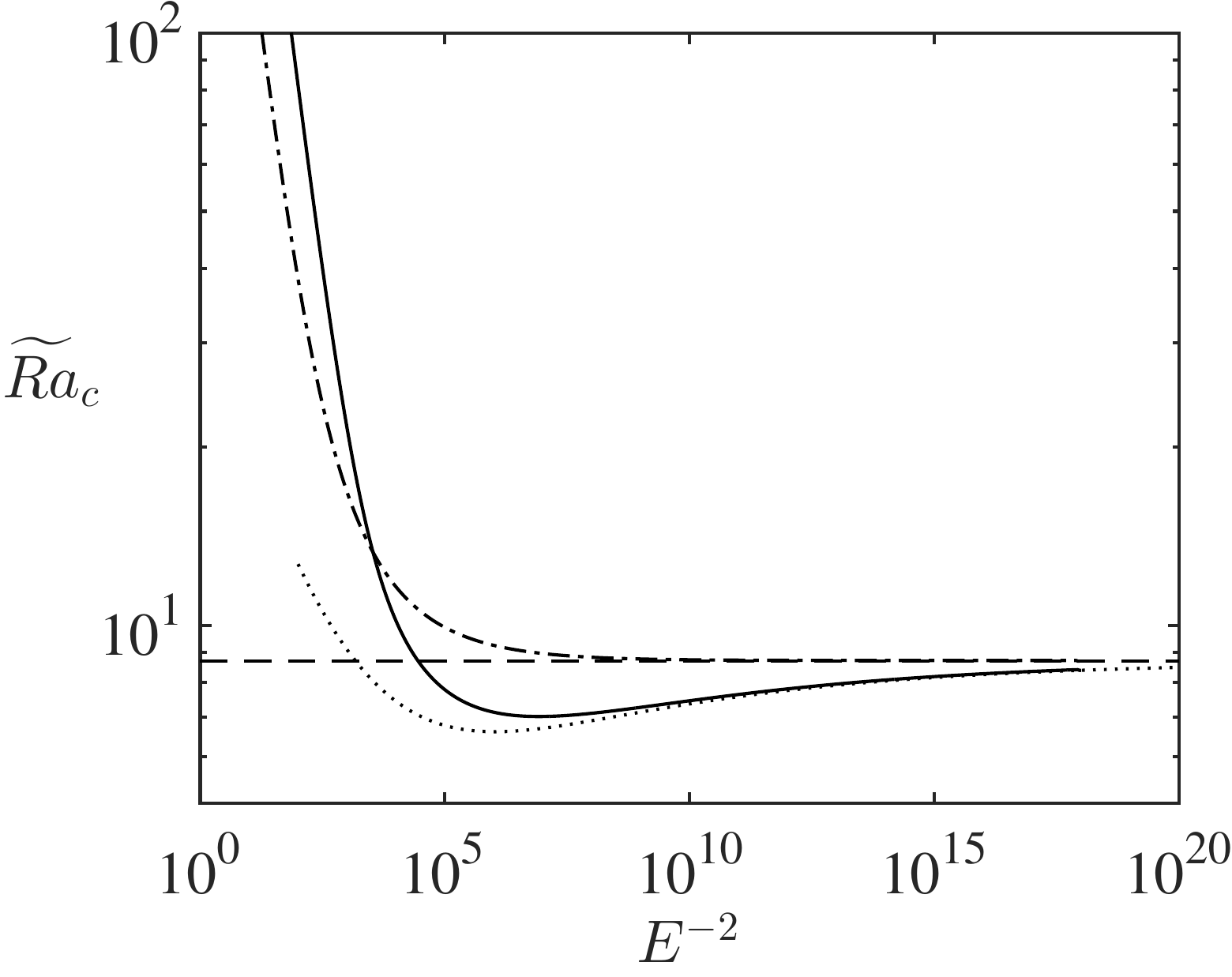}
     \hspace{0.5cm}
      \includegraphics[height=4.5cm]{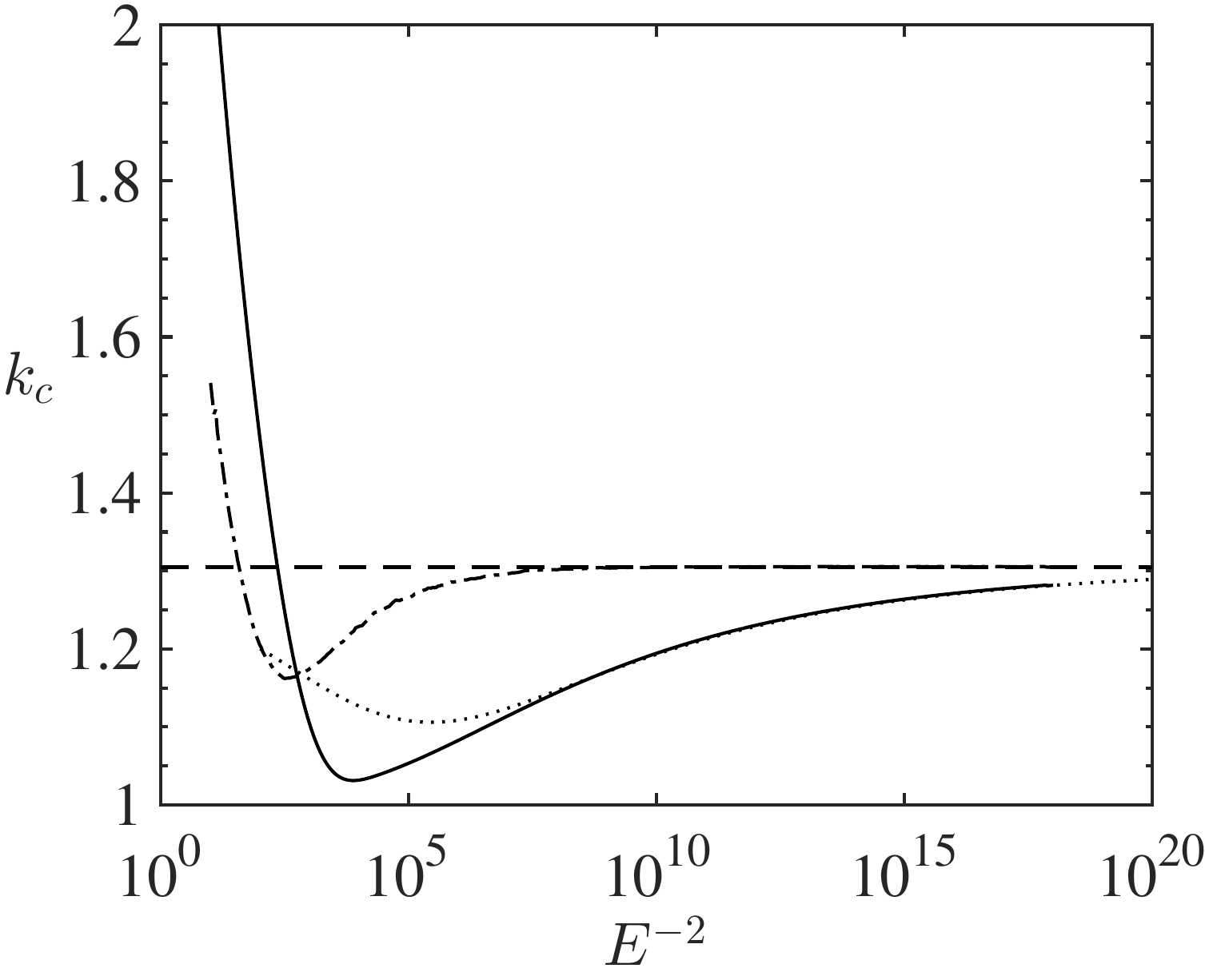}
  \end{center}
  \caption{\small{The effect of Ekman pumping on the onset of steady convection. (a) Critical reduced Rayleigh number $\widetilde{Ra}$ vs $E^{-2}$ and (b) critical reduced wavenumber $k_c$ vs $E^-{2}$.
  %Asymptotically compensated marginal stability results are presented for (a) $\widetilde{Ra}_c$ vs $E^{-2}$, and  (b) $k_{c}$ vs $E^{-2}$. 
  Shown are results obtained from the unapproximated Navier-Stokes equations for no-slip boundaries (solid) and stress-free boundaries (dashed-dotted) for $E^{-2}\le 10^{18}$. Asymptotic approximations are obtained from the reduced NH-QGE with 
 (i) impenetrable stress-free boundary conditions (dashed) where  $(\widetilde{Ra}_c,k_{c})=(8.6956,1.3048)$, 
and (ii) Ekman pumping boundary conditions (dotted). The linear effect of Ekman pumping is quantified by the interval between dashed and dotted lines at a given $E^{-2}$.}}
\label{fig:linear}
\end{figure}

A comparison of the linear stability results between the reduced composite model and the incompressible Navier-Stokes equations is illustrated in the 
$\widetilde{Ra}$--$E^{-2}$ diagram of Figure~\ref{fig:linear} and the data given in Table~\ref{table:critonset}. All numerical results were obtained by solving the two-point boundary value eigenproblem using an iterative Newton-Raphson-Kantorovich (NRK) scheme \citep{BOOK_Henrici, cS82} with $1025$ spatial grid points placed at the  Chebyshev-Gauss-Lobatto points. All spatial derivatives were computed with fourth order  finite differences.  Irrespective of the mechanical boundary condition selected, the onset of steady convection corresponds to minimal values $(\widetilde{Ra}_c, \widetilde{k}_{c} )=(8.6956,1.3048)$ as $E\rightarrow0$ (dashed line).  As shown in Figure~\ref{fig:linear},  for stress-free boundaries excellent convergence to the asymptotic values is observed when $E^{-2}\gtrsim 10^{8}$, or equivalently $E\lesssim 10^{-4}$ (dashed-dotted lines). For the no-slip case, where Ekman pumping is present, one may observe that the unapproximated values (solid lines) differ from the asymptotic values by an $\mathcal{O}(1)$ amount. Most strikingly, the asymptotic convergence is very slow, and differences are still visible at $E^{-2}\approx 10^{20}$. The inclusion of Ekman pumping in the reduced model and the associated corrections rectifies this difference and asymptotically accurate values of $\widetilde{Ra}$ and ${k}_\perp $ are recovered (cf. solid and dotted lines). 
 
Clearly, the reduced model has an enhanced capability of reaching lower $E$ with equivalent computational resources.
The results of the composite model are in quantitative agreement with the asymptotic result of \cite{wH71} who find the following analytic result for marginal stability
\Beq
\widetilde{Ra}_{m} = \lb \frac{\pi^2}{k^2_\perp} + k^4_\perp\rb -  2\sqrt{2}\frac{\pi^2}{k^4_\perp} \epsilon^{1/2}+  6 \frac{\pi^2}{k^6_\perp} \epsilon,
\qquad \epsilon=E^{1/3}.
\Eeq
The higher order asymptotic corrections respectively capture the influence of Ekman pumping (also reported by \cite{pN65}) and thermal regularization within the middle layer.

\begin{table}%[htdp]
\begin{center}
\begin{tabular}{|c|c|c|c|c|}

\hline% \\ \\
Ekman & CNH-QGE & & NS & \\
\hline
$E$ & $\widetilde{Ra}_c$ & ${k}_{c}$& $\widetilde{Ra}_c$&${k}_{c}$ \\ 
 \hline %\\
$10^{-6}\;$ & $7.7594$ & $1.2289$ & 7.7832 &1.2294\\ 
$10^{-7}\;$ & $8.0491$ & $1.2537$ & 8.0572 &1.2539\\
$10^{-8}\;$ & $8.2522$ & $1.2703$ & 8.2550 &1.2704\\
$10^{-10}$ & $8.4888$ & $1.2890$ & -- & --\\
$10^{-12}$ & $8.5995$ & $1.2975$ & --&--\\
$10^{-14}$ & $8.6510$ & $1.3014$ &-- &--\\
$10^{-15}$ & $8.6652$ & $1.3025$ &-- &--\\
$10^{-16}$ & $8.6749$ & $1.3032$ &-- &--\\
$10^{-\infty}$ & $8.6956$ & $1.3048$ &-- &-- \\
\hline
\end{tabular}
\end{center}
\caption{\small{Minimal critical onset value reduced Rayleigh number $\widetilde{Ra}_c$ and wavenumber $k_{c}$  as  a function of the Ekman number $E$ for the CNH-QGE and the unapproximated Boussinesq equations in the presence of no-slip boundary conditions.}}
\label{table:critonset}
\end{table}%

The quantitative impact of the inclusion of Ekman pumping in the asymptotically reduced equations can be assessed by computing fully nonlinear single-mode (or single horizontal wavenumber) solutions. Following \cite{kj98}, we pose for steady state solutions with $\sigma > 0.67$ the ansatz
 \beginar
\left( \Psi^{(c)},  \zeta^{(c)},  W^{(c)},  \Theta^{(c)}  \right)
=
 \sigma^{-1}  \left(\tilde{\Psi} (Z),  \tilde{\zeta} (Z),  \tilde{W}(Z),  \sigma \tilde{\Theta}(Z) \right) h(x,y),
\endar
where $h(x,y)$ is  a real-valued function satisfying the planform equation
\Beq
\nabla^2_\perp h = -k_\perp^2 h,
\Eeq
with normalization $\overline{h^2}=1$.
These planforms include rolls ($h= \sqrt{2} \cos k_\perp x$), squares ($h= \cos k_\perp x  + \cos k_\perp y$), hexagons ($h=\sqrt{2/3} (\cos k_\perp x + \cos(\frac{1}{2} k_\perp ( x + \sqrt{3}y)) + \cos(\frac{1}{2} k_\perp ( x - \sqrt{3}y))$), regular triangles  ($h=\sqrt{2/3} (\sin k_\perp x + \sin(\frac{1}{2} k_\perp ( x + \sqrt{3}y)) + \sin(\frac{1}{2} k_\perp ( x - \sqrt{3}y))$), and the patchwork quilt ($h= (\cos(\frac{1}{2} k_\perp ( x + \sqrt{3}y)) + \cos(\frac{1}{2} k_\perp ( x - \sqrt{3}y))$).  For such patterns horizontal advection vanishes, for instance,
\Beq
\nabla^\perp\Psi^{(c)}\cdot \nabla_\perp \zeta^{(c)} =  \sigma^{-2} \tilde{\Psi}^2_0 (Z) \nabla^\perp h \cdot \nabla_\perp \nabla^2_\perp h =
 -k_\perp ^2  \sigma^{-2}  \tilde{\Psi}^2_0 (Z)  \nabla^\perp h \cdot \nabla_\perp h \equiv 0.
\Eeq
The following $\sigma$-independent system of ordinary differential equations for the vertical structure is then obtained:
%\textcolor{red}{I Added Eq 5.7. I also made Eq 5.8 show Nu explicitly.}
\beginar
\label{eqn:sm1}
&\lb \pd{ZZ} - k_\perp^6 \rb  \tilde{W} + k^4_\perp {\widetilde{Ra}} \tilde{\Theta} = 0, \\
\label{eqn:sm2}
&\lb\epsilon^2 \pd{ZZ} - k_\perp^2 \rb  \tilde{\Theta} -   \tilde{W} \pd{Z} \overline{\Theta}^{(c)} 
%- \underline{\epsilon \overline{h^3}  \pd{Z}\lb  \tilde{W}\tilde{\Theta}\rb } 
 =0, \\
 & \label{eqn:sm3} \pd{Z}  \tilde{W} - k_\perp^2 \tilde{\zeta} = 0, \\ 
\label{eqn:sm4}
&- \pd{Z} \overline{\Theta}^{(c)} + \tilde{W}\tilde{\Theta}  = Nu,
\endar
together with the boundary conditions
\Beq
\tilde{W}(0) = \displaystyle{\frac{\epsilon^{1/2}}{\sqrt{2}}}\tilde{\zeta}(0),\quad
\tilde{W}(1) = -\displaystyle{\frac{\epsilon^{1/2}}{\sqrt{2}}} \tilde{\zeta}(1),\label{eqn:pump}
\Eeq
\Beq
\overline{\Theta}^{(c)}(0) = 1,\qquad
\overline{\Theta}^{(c)}(0) = 0, \qquad
\tilde{\Theta}(0) =\tilde{\Theta}(1) = 0.
\label{eqn:bctemp}
\Eeq
%With regard to nonlinear vertical advection (underlined term in (\ref{eqn:sm2})), we note that $\overline{h^3}=0$ for all planforms except hexagons where $\overline{h^3}=\sqrt{2/3}$. However, even for this latter case, we find \textsl{a posteriori} that the contribution from nonlinear vertical advection remains numerically negligible.
%%(KJ. However, I cannot prove this mathematically yet).

The above single-mode system represents a nonlinear two-point boundary value problem which we solve by successive over-relaxation on a discretized one-dimensional mesh.  An iterative NRK scheme is used with $\mathcal{O}(10^{-10})$ accuracy in the $L^2$ norm of the energy functional $E (Z)=( \tilde{W}^2 + \vert \nabla_\perp \tilde{\Psi}\vert^2)/2 $. The control parameters of the problem are the scaled Rayleigh number $\widetilde{Ra}$, the horizontal wavenumber $k_\perp$, and  the parameter $\epsilon=E^{1/3}$  measuring the strength of Ekman pumping.

In the absence of Ekman pumping ($\epsilon=0$) the only remaining nonlinearity in the reduced equations is the vertical divergence of the horizontally averaged convective flux appearing in (\ref{eqn:sm4}), which is incapable of generating energy exchanges between horizontal wavenumbers. \cite{JK_PoF_1999} have shown that single-mode solutions are in fact exact solutions to the reduced system. 
%Given the subdominance of vertical advection (\ref{eqn:sm2}) for all planforms (including hexagons), this result also holds true in the presence of Ekman pumping, i.e., when $\epsilon\ne0$.

\begin{figure}
  \begin{center}
      \includegraphics[height=7cm]{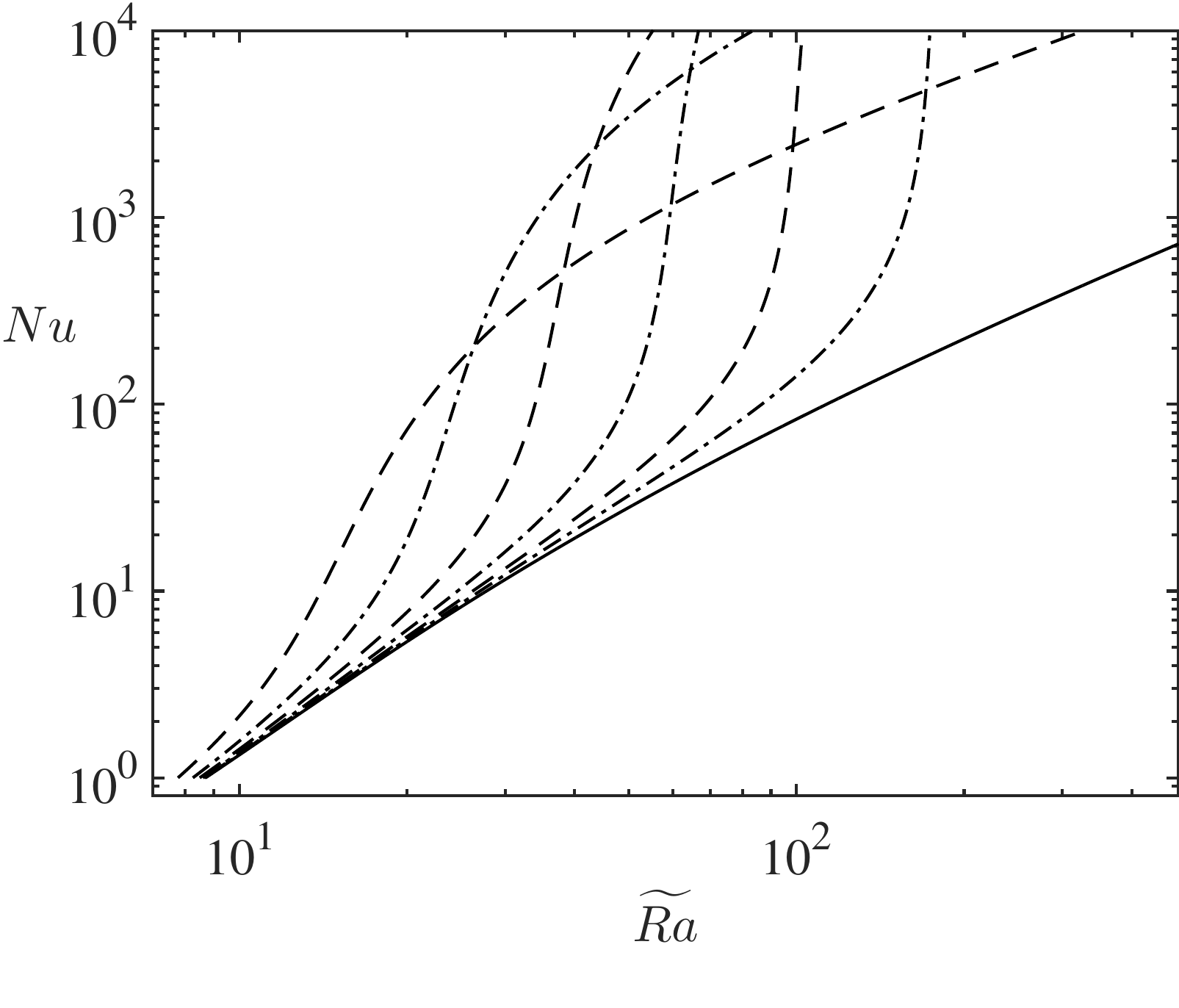}
  \end{center}
  \caption{\small{The Nusselt number $Nu$ vs $\widetilde{Ra}$ corresponding to the critical wavenumbers $k_\perp=k_{c}$ in the presence of Ekman pumping for single-mode solutions. The solid curve shows the results for the case of no pumping $\epsilon=0$. The remaining curves illustrate the enhancement due to Ekman pumping for, from right to left, $E=\epsilon^3=(10^{-16}, 10^{-14}, 10^{-12}, 10^{-10}, 10^{-8}, 10^{-6})$. See Table~\ref{table:critonset} for $k_{c}$.}}
\label{fig:Nuss-Ra-Ek}
\end{figure}

\begin{figure}
  \begin{center}
   \subfloat[]{
      \includegraphics[height=5.4cm]{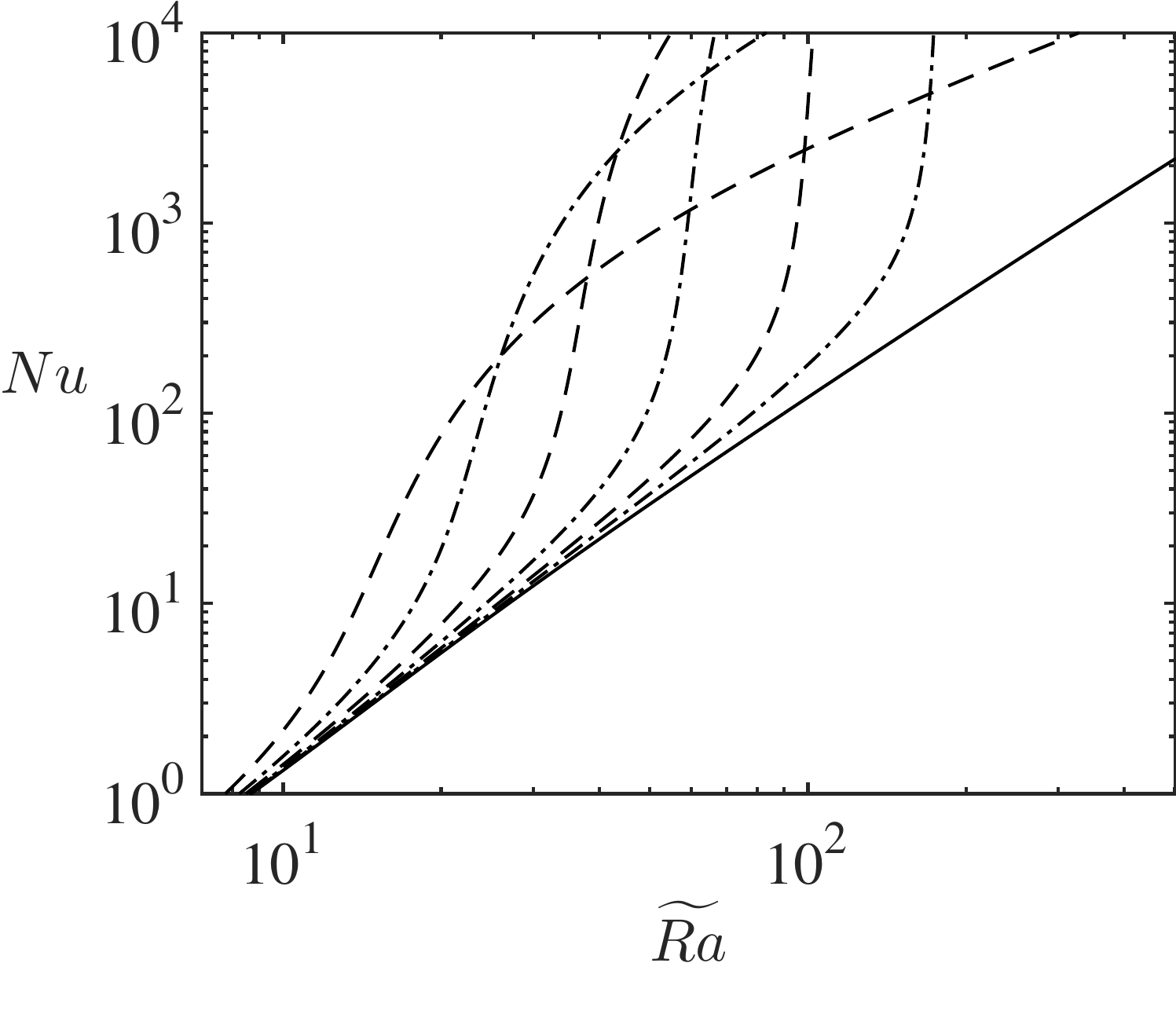}}
      \qquad
    \subfloat[]{
      \includegraphics[height=5.4cm]{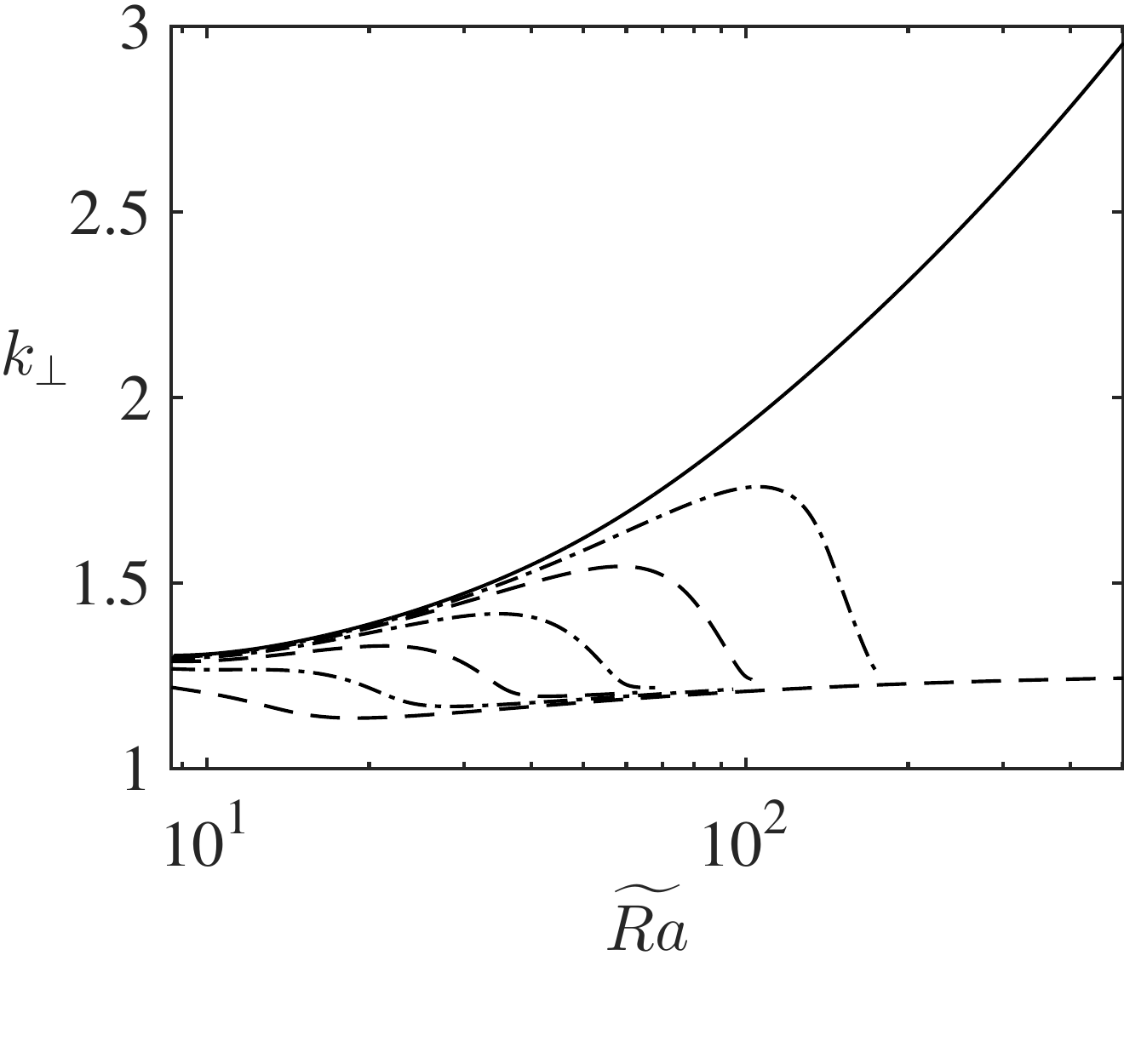}}
  \end{center}
   \caption{\small{(a) The maximal Nusselt number $Nu$ vs $\widetilde{Ra}$, and (b) the corresponding wavenumber $k_\perp$ in the presence of Ekman pumping for single-mode solutions. The solid curve shows the results for the stress-free case with no pumping ($\epsilon=0$).  The remaining curves illustrate the enhancement due to Ekman pumping for, from right to left, 
$E=\epsilon^3=(10^{-16}, 10^{-14}, 10^{-12}, 10^{-10}, 10^{-8}, 10^{-6})$. \textcolor{black}{The stress free case provides an upper bound for the horizontal wavenumber that maximizes heat transport in the single-mode theory.} }}
\label{fig:MaxRa}
\end{figure}

Results for the $Nu$-$\widetilde{Ra}$ relation from the fully nonlinear single-mode theory 
are presented in Figures~\ref{fig:Nuss-Ra-Ek} and \ref{fig:MaxRa}. Given no a priori means for selecting the wavenumber $k_\perp$, it is held fixed at the critical value obtained at linear onset  (Table~\ref{table:critonset})  in Figure~\ref{fig:Nuss-Ra-Ek}. 
In Figure~\ref{fig:MaxRa}, the wavenumber that maximizes $Nu$ at fixed $\widetilde{Ra}$ is selected.
% illustrates the maximal $Nu$ obtained by varying $k_\perp$ at fixing $\widetilde{Ra}$ such that $d Nu/d k_\perp =0$.  
In each case comparisons with the stress-free $Nu$-$\widetilde{Ra}$ curve corresponding to $\epsilon=0$ in Eq.~(\ref{eqn:pump}) (solid line) reveal strong departures once the predicted threshold $\widetilde{Ra}_{thres}\sim E^{-1/9}$ is reached. 

Estimates of this threshold are given in Table~\ref{table:rtrans}. 
\begin{table}%[htdp]
\begin{center}
\begin{tabular}{|c|c|}
\hline% \\
Ekman, $E$ & $\widetilde{Ra}_{thres}\sim E^{-1/9}$ 
\\ 
 \hline %\\
$10^{-6}\;$ & $\ 4.6415$\\ 
%$10^{-7}\;$ & $8.0491$\\
$10^{-8}\;$ & $\ 7.7426$\\
$10^{-10}$ & $12.9155$\\
$10^{-12}$ & $21.5443$\\
$10^{-14}$ & $35.9381$\\
%$10^{-15}$ & $8.6652$\\
$10^{-16}$ & $59.9484$\\
$10^{-\infty}$ & $\infty$\\
\hline
\end{tabular}
\end{center}
\caption{\small{Estimates for the transitional Rayleigh number $\widetilde{Ra}_{thres}\sim E^{-1/9}$ as a function of $E$. 
%Note that $\widetilde{Ra}_{c} = 8.6956$.
}}
\label{table:rtrans}
\end{table}%
For $E\gtrsim 10^{-9}$ we observe that this departure occurs immediately at onset. As $\widetilde{Ra}$ increases, the  $Nu$-$\widetilde{Ra}$ curves exhibit a transition region of strong monotonic increase with a positive curvature. This trend continues until a zero curvature point is reached.  In the range $Nu\le 10^4$, this occurs for $E\ge 10^{-14}$; smaller values of $E$ have yet to attain their zero curvature points. Beyond the zero curvature point a monotonically increasing curve of negative curvature is observed before the asymptotic branch is reached. Interestingly, Figure~\ref{fig:MaxRa}(b) indicates the transition region for maximal $Nu$ is bracketed by two limiting values of $k_\perp$ -- a monotonically increasing branch prior to transition and a saturated branch with $k_\perp=1.2434$ very close to the critical onset value $k_\perp=1.3048$. The latter provides evidence that all $Nu$-$\widetilde{Ra}$ curves with $\epsilon>0$ are topologically similar.

\textcolor{black}{It is worth providing more precise reasons for the heat transport scaling transition. In eqs.~(\ref{eqn:sm1})--(\ref{eqn:bctemp}) the small parameter $\epsilon$ makes two physically distinct appearances. This first is in the Ekman pumping boundary conditions, Eq.~(\ref{eqn:pump}). The second is in the vertical dissipation of thermal fluctuations, Eq.~(\ref{eqn:sm2}). Ostensibly, one might suppose that the latter is much less significant than the former: $\epsilon^2$ versus $\sqrt{\epsilon}$, respectively.  We can therefore entertain dropping the $\epsilon^2$ and retaining only the $\sqrt{\epsilon}$. We already know that the heat transport shows significant enhancement when $\widetilde{Ra} \sim \epsilon^{-1/3}$. What part of this results from pumping versus the thermal wind balance immediately adjacent to the Ekman boundary layer? The answer is that without thermal dissipation the Ekman pumping not only enhances the heat transport, but also causes it to diverge to infinity. However, thermal dissipation acts to arrest the divergence resulting in a finite albeit much enhanced value of $Nu$ over that with no Ekman pumping. } 

\begin{figure}
  \begin{center}
    \subfloat[]{
      \includegraphics[height=5.75cm]{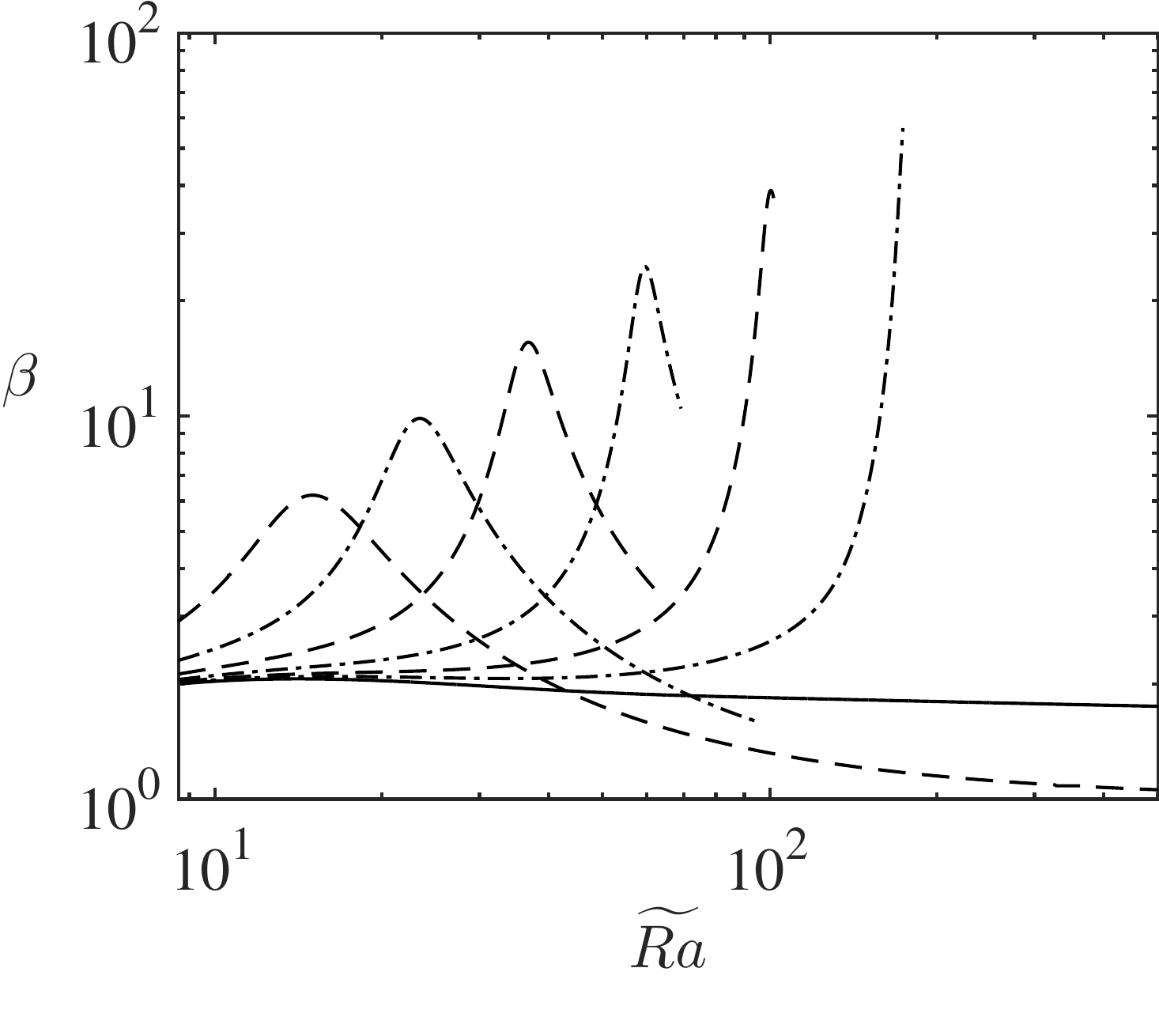}}
      \quad
    \subfloat[]{
      \includegraphics[height=5.75cm]{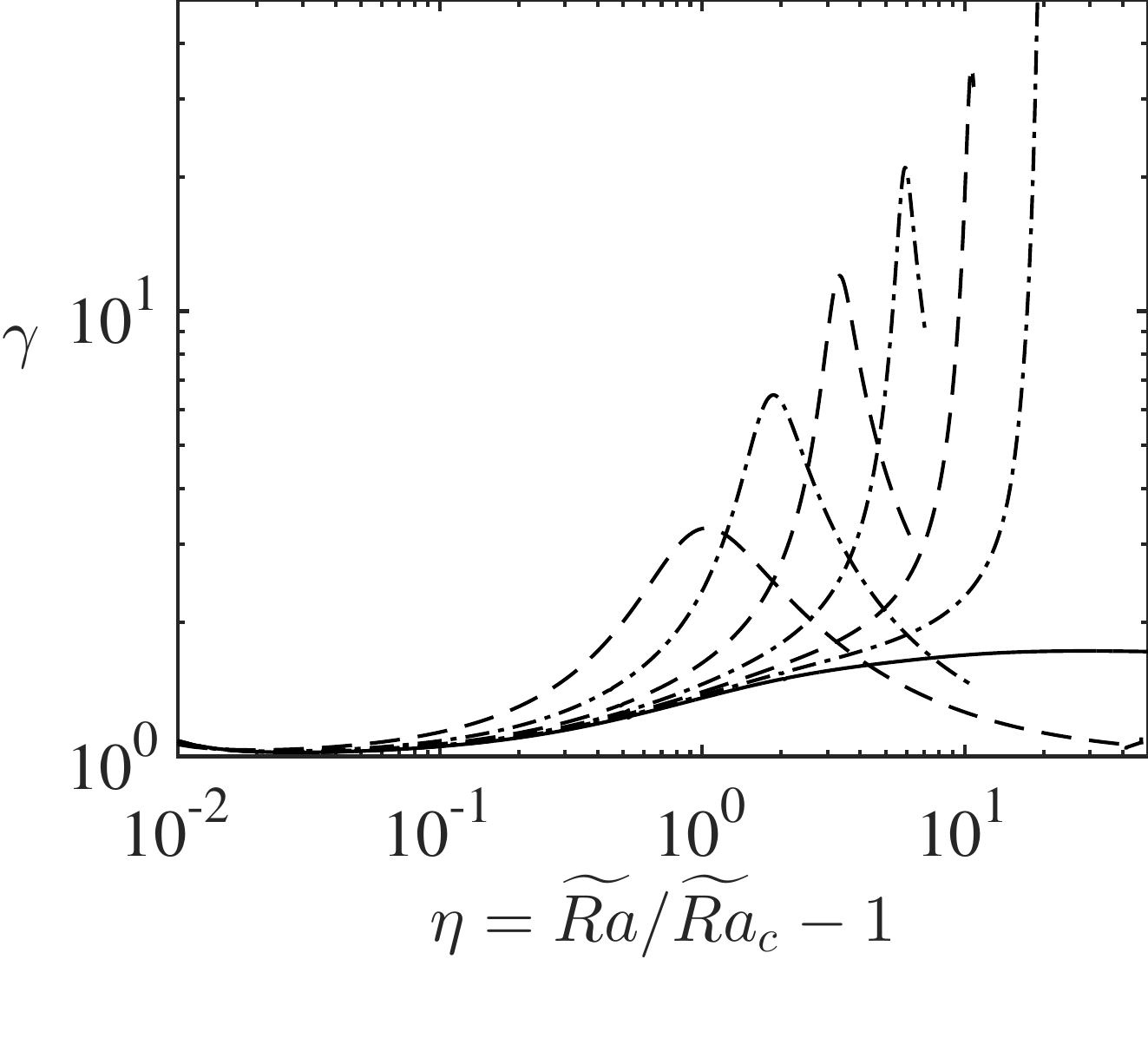}}      
  \end{center}
  \caption{\small{(a) Instantaneous heat transport exponent 
  (a)  $\beta =d\log_{10} Nu /d \log_{10} \widetilde{Ra}$ vs   $\widetilde{Ra}$.
  (b) $\gamma =d\log_{10} Nu /d \log_{10} \eta$ vs   $\eta=\widetilde{Ra}/\widetilde{Ra}_c-1$.
  The solid curve shows the results for the case of no pumping ($\epsilon=0$ in Eq.~(\ref{eqn:pump})). The remaining curves illustrate the enhancement due to Ekman pumping for, from right to left, $E=\epsilon^3=(10^{-16}, 10^{-14}, 10^{-12}, 10^{-10}, 10^{-8}, 10^{-6})$}.}
\label{fig:NussExp}
\end{figure}

It is evident from Figures~\ref{fig:Nuss-Ra-Ek} and \ref{fig:MaxRa}(a) that at fixed $\widetilde{Ra}$ the maximal heat transport is achieved for an intermediate value of $\epsilon>0$. Moreover as $\epsilon\rightarrow0$ the transition from the stress-free curve corresponding to $\epsilon =0$ in Eq.~(\ref{eqn:pump}) exhibits increasingly steep slopes. This is quantified in Figure~\ref{fig:NussExp}, where the instantaneous power exponents 
\beginar
&
\beta = \displaystyle{\frac{d\log_{10} Nu }{d \log_{10} \widetilde{Ra}}}&\quad\mbox{for}\quad Nu\propto\widetilde{Ra}^\beta \\
&
\label{eqn:super}
\gamma= \displaystyle{\frac{d\log_{10} (Nu-1)}{d \log_{10} \eta}}&\quad\mbox{for}\quad Nu\propto\eta^\gamma, 
\quad\eta\equiv\frac{\widetilde{Ra}}{\widetilde{Ra}_c}-1
\endar
%$\beta = d\log_{10} Nu /d \log_{10} \widetilde{Ra}$  for $Nu$-$\widetilde{Ra}^\beta$ and $\gamma= d\log_{10} (Nu-1) /d \log_{10} (\widetilde{Ra}/\widetilde{Ra}_c-1)$ for  $Nu$-$(\widetilde{Ra}/\widetilde{Ra}_c-1)^\gamma$ 
are plotted as a function of $\widetilde{Ra}$.  Owing to the inability of laboratory experiments and DNS to probe deeply into the high $\widetilde{Ra}$, low~$(E,Ro)$ regime it has been suggested \citep{rE15} that a more pertinent measure for the heat transport in this regime is one more closely related to weakly nonlinear theory, i.e., the supercriticality $\eta$ and the associated exponent $\gamma$ in Eq.~(\ref{eqn:super}) and Figure~\ref{fig:NussExp}(b). It can be seen in Figure~\ref{fig:NussExp} that both exponents display similar qualitative characteristics. 
 
For stress-free boundaries (corresponding to $\epsilon=0$ in Eq.~(\ref{eqn:pump})) we observe that $\beta=1.7466$ and $\gamma=1.7171$ as $\widetilde{Ra}\rightarrow\infty$ which is close to that obtained in the full simulations of the NH-QGE \citep{mS06} where $\beta=2.1$. We note that the expected weakly nonlinear result $\gamma =1$ is captured near onset at $\widetilde{Ra}=\widetilde{Ra}_c$ \citep{aB94,jD01,kJ12}. \textcolor{black}{In contrast, away from onset and the weakly nonlinear regime, Figure~\ref{fig:NussExp}(a) indicates that $\beta\approx 2$, revealing only a slight difference from the strongly nonlinear value $\beta=2.1$ determined from DNS.}

For no-slip boundaries with Ekman pumping ($\epsilon\ne0$), we observe that $\beta\ge 2$, indicating a tendency for Ekman pumping to increase heat transport. This trend is also observed in the supercriticality exponent $\gamma$. The maximal exponents all occur at the zero curvature point in the $Nu$-$\widetilde{Ra}$ curves (Figures~\ref{fig:Nuss-Ra-Ek} and \ref{fig:MaxRa}(a)) and this value trends to $\infty$ as $\epsilon\rightarrow0$ where the zero curvature point becomes inflectional. These observations within the single-mode setting provide an explanation of the measured increase in the heat transport exponent $\beta^{NS}_{rot}$ in the scaling relation $Nu\propto\widetilde{Ra}^\beta$ (see Figure~\ref{fig:NuRa}) and suggest that both laboratory experiments and DNS have yet to probe the saturated asymptotic state in the presence of Ekman pumping. In Figure~\ref{fig:NussExp}, we see that the single-mode branches asymptote to saturated exponents $\beta,\gamma \approx 1$ as $\widetilde{Ra}\sim\eta\rightarrow\infty$ which are significantly below the stress-free result of $\beta,\gamma \approx 2$ and close to that produced by weakly nonlinear theory. Thus after an initial range of enhancement in the heat transport it appears that Ekman pumping diminishes heat transport efficiency as measured by the exponents.  \textcolor{black}{This fortuitous result is consistent with the claim by \citet{Ecke2015} that the $Nu$-$\eta$ relation should be interpreted from a weakly nonlinear standpoint where $\eta$ is interpreted as $\mathcal{O}(1)$.}

%Due to the inability of laboratory experiments and DNS to probe deeply into the low $E$-low $Ro$ regime it has been suggested (Ecke 2015) that 
%a more pertinent measure for the exponent is one more closely related to weakly nonlinear theory, i.e., $\beta = d\log_{10} (Nu -1) /d \log_{10} \eta$
%where $\eta = \widetilde{Ra}/\widetilde{Ra}_c -1$ denotes supercriticality (see Figure~\ref{fig:NussExp} (b)). Interestingly, we observe that the saturated branch due to Ekman pumping asymptotes to $\beta_\eta=1.04$ which is also the exponent predicted by weakly nonlinear theory (REFS). For the stress-free case it appears that the supercriticality exponent $\beta_\eta = 1.7171$ remains the $Nu-\widetilde{Ra}$ exponent $\beta=1.7466$. 

\begin{figure}
\begin{center}
 \begin{minipage}{1.0\linewidth}
\hspace{-0em}
   \includegraphics[height=0.45\textwidth]{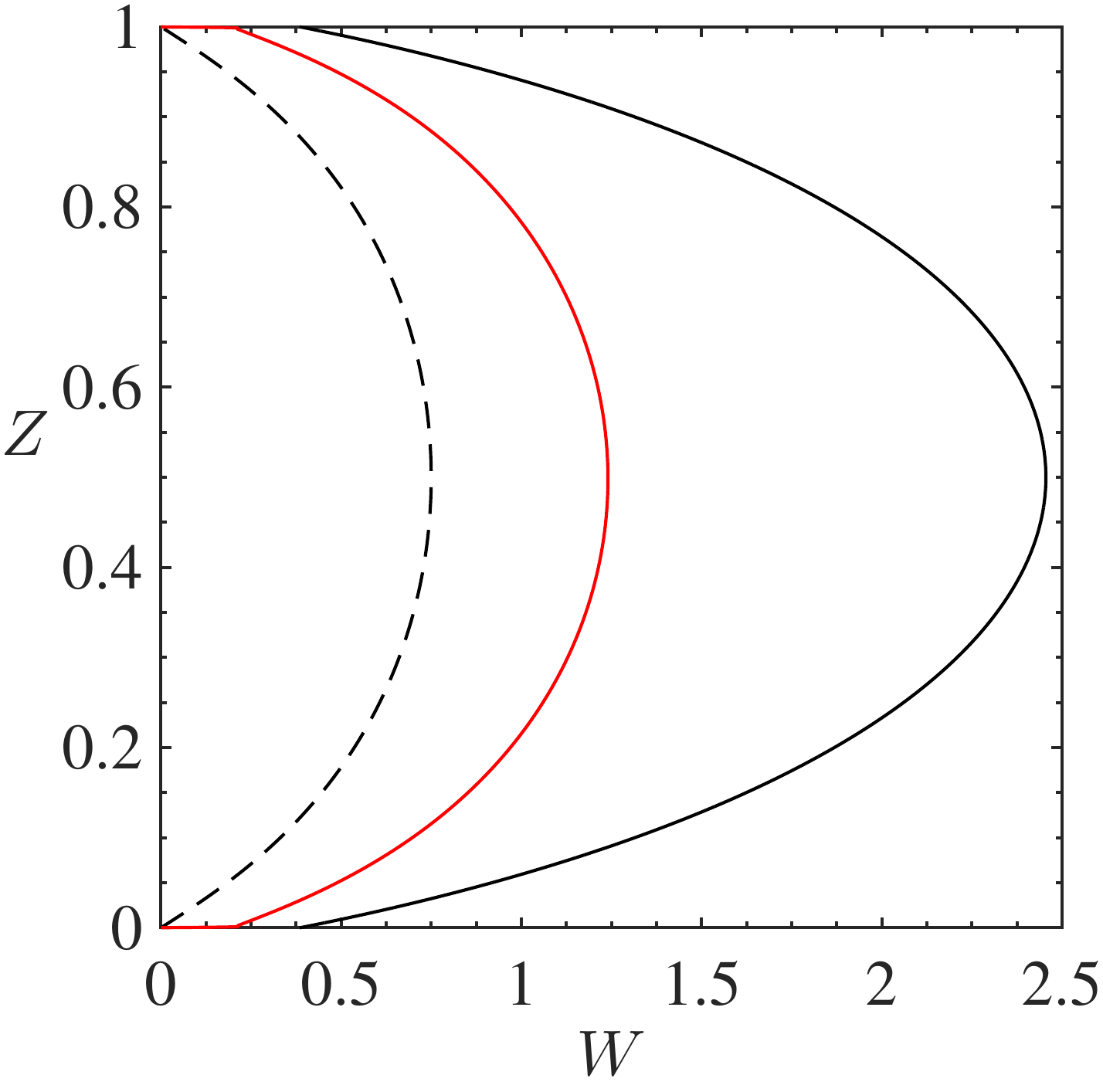}
 \hspace{+1em}
   \includegraphics[height=0.45\textwidth]{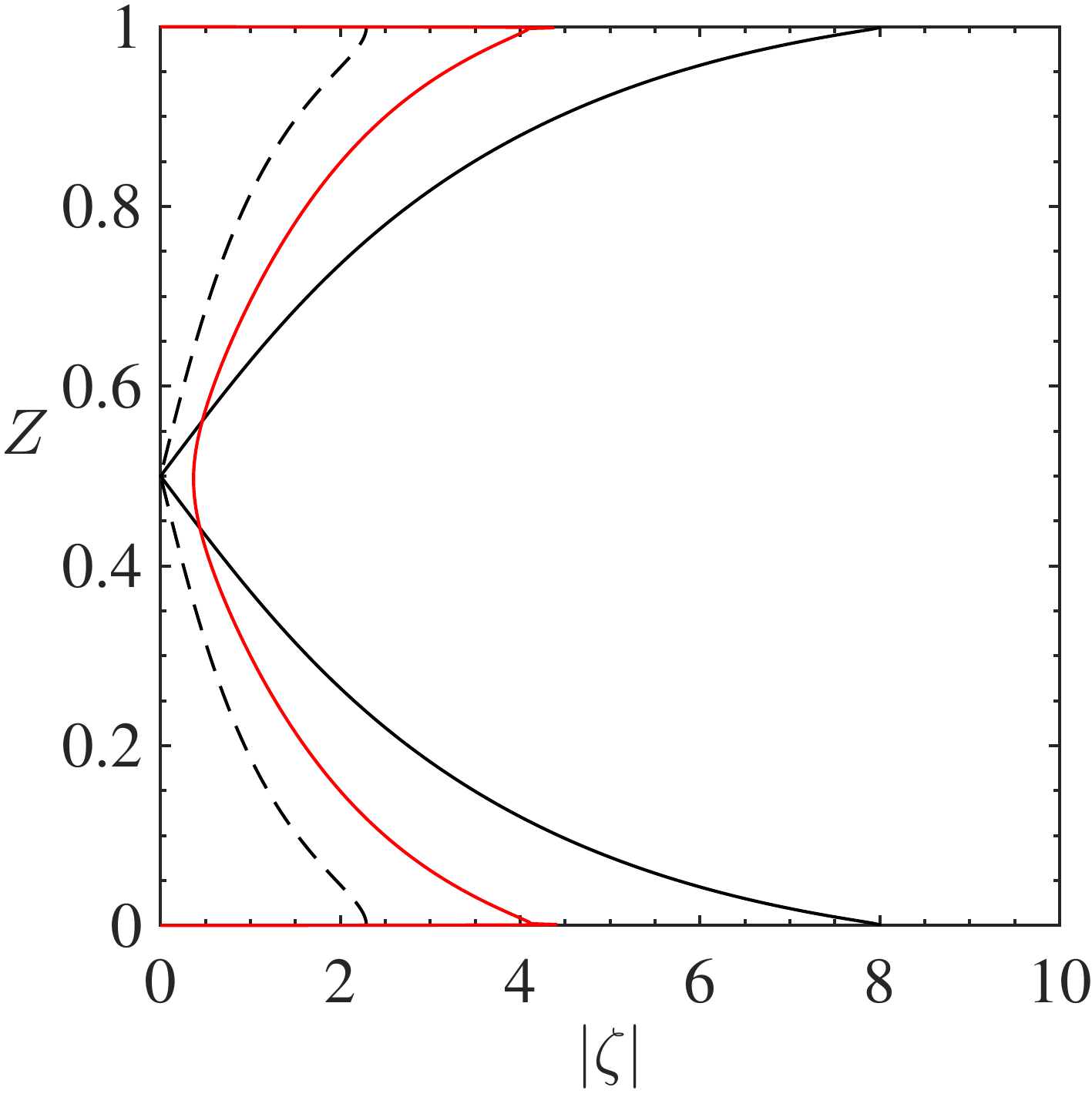} 
  \end{minipage}
   \begin{minipage}{1.0\linewidth}
\hspace{-0em}
   \includegraphics[height=0.45\textwidth]{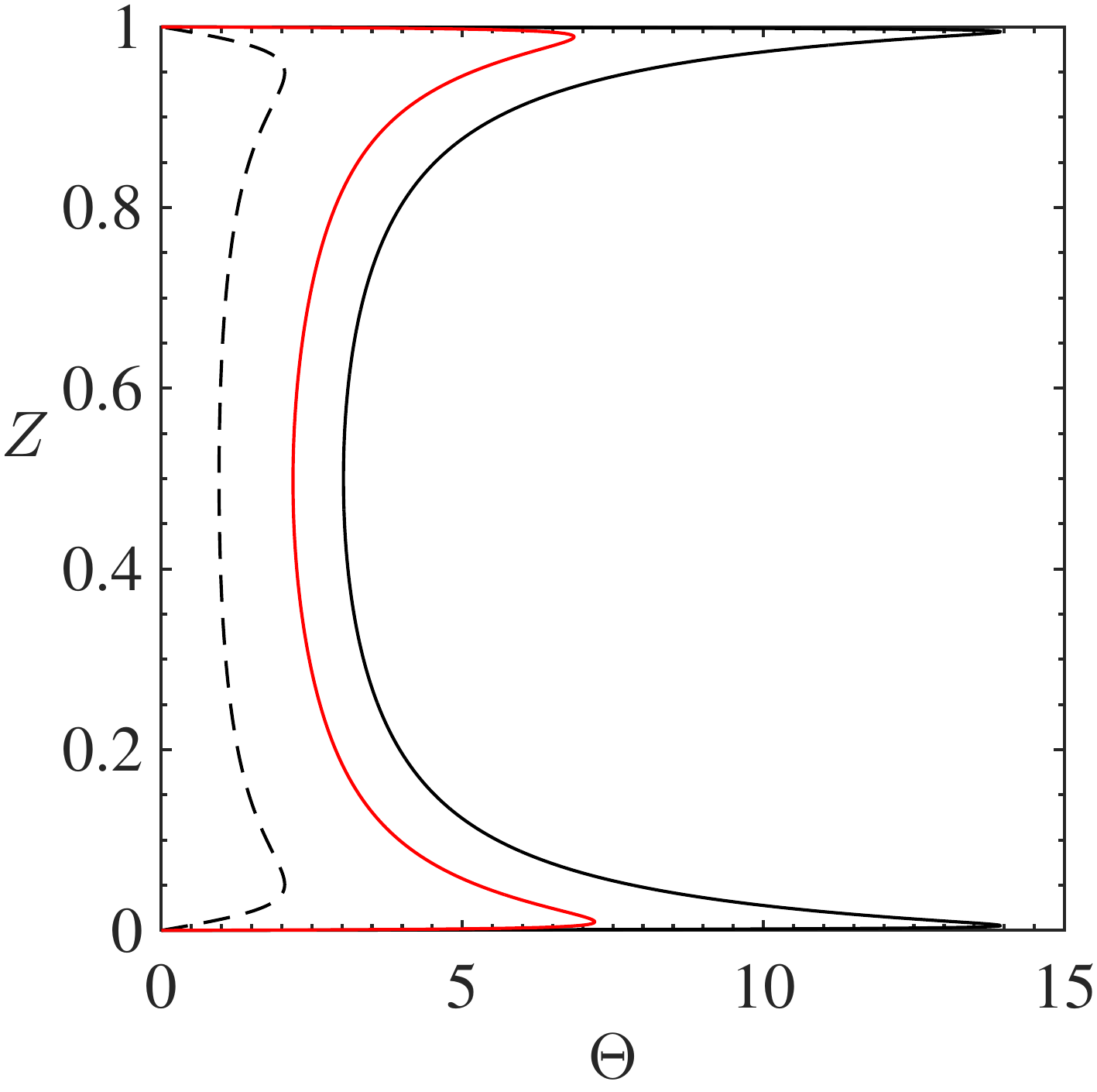}
 \hspace{+1em}
   \includegraphics[height=0.45\textwidth]{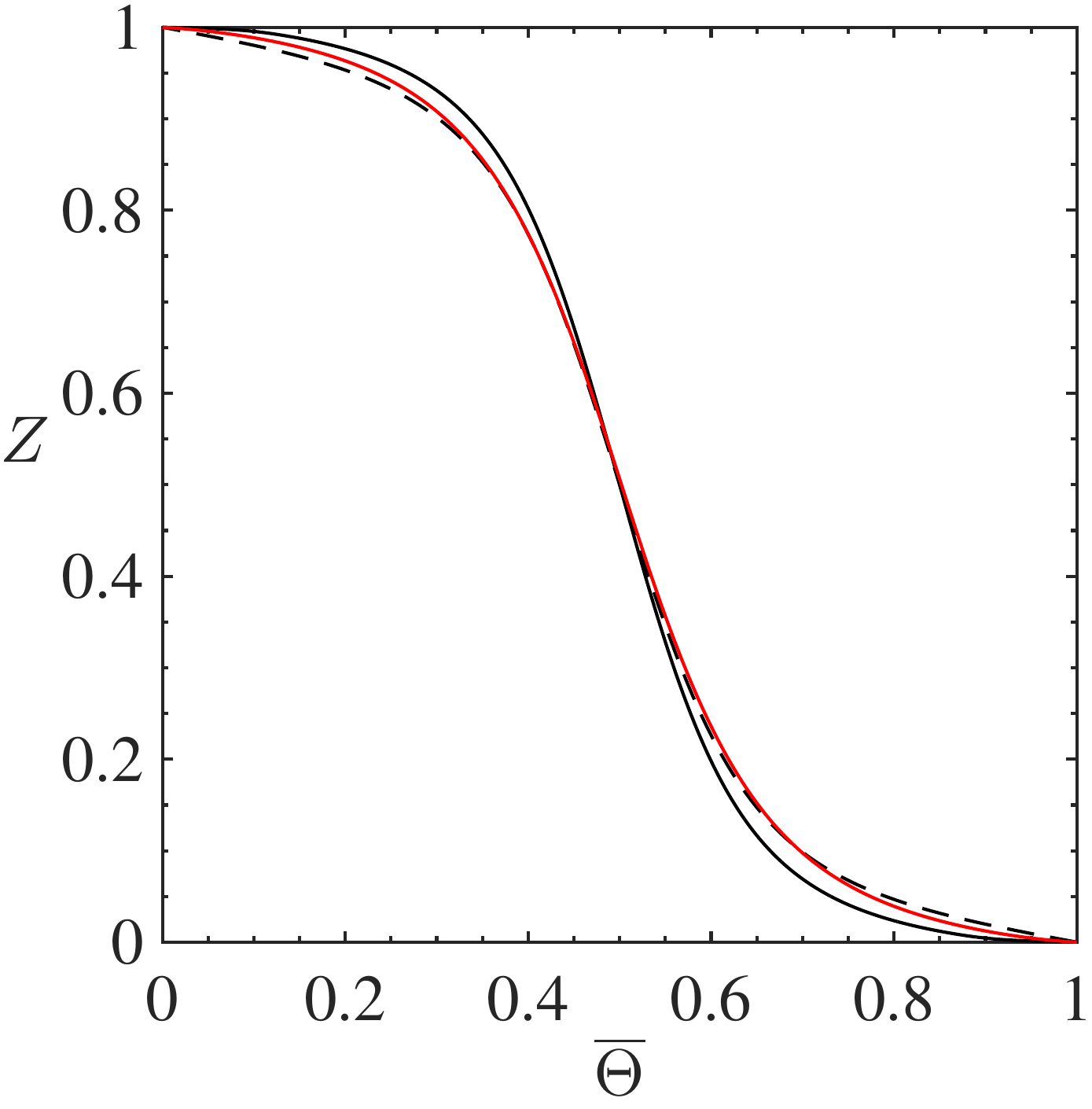} 
  \end{minipage}
\caption{\small{Vertical structure profiles at $\widetilde{Ra}=20$, $E=10^{-7}$, $\sigma=7$ obtained from the reduced NH-QGE with stress-free boundary conditions (black, dashed line), reduced CNH-QGE (black, solid line) and DNS (red, solid line). (a) RMS vertical velocity, (b) RMS vertical vorticity, (c) RMS temperature, (d) mean temperature. The stress-free and no-slip single-mode results respectively constitute lower and upper bounds on the DNS results.}}
\label{fig:Eigprofile}
\end{center}
\end{figure}

The fidelity of the CNH-QGE and single-mode solutions is further demonstrated in the vertical structure profiles of Figure~\ref{fig:Eigprofile}. Comparison with DNS at fixed $\widetilde{Ra}=20$, $E=10^{-7}$, $\sigma=7$ clearly shows that a periodic array of convection cells described by a single-mode solution overestimates the DNS amplitudes. This fact is known from stress-free investigations \citep{mS06,sS14}.  However,  excellent agreement is found in the topology of the vertical profiles. The stress-free and no-slip single-mode results \textcolor{black}{respectively constitute lower and upper bounds on the DNS results.} This is also borne out in an explicit comparison of the heat transport at $Pr=7, E=10^{-7}$ (Figure~\ref{fig:Nuss-Ra-Ek2}).

%vigor of convection in the regime of validity of the CNH-QGE ($\widetilde{Ra}\lesssim E^{1/3}\approx215$ at $E=10^{-7}$), as borne out in an explicit %comparison of the heat transport at $Pr=7, E=10^{-7}$ (Figure~\ref{fig:Nuss-Ra-Ek2}). 
%This regime includes a subset of the laboratory results.

\begin{figure}
  \begin{center}
    \includegraphics[height=8cm]{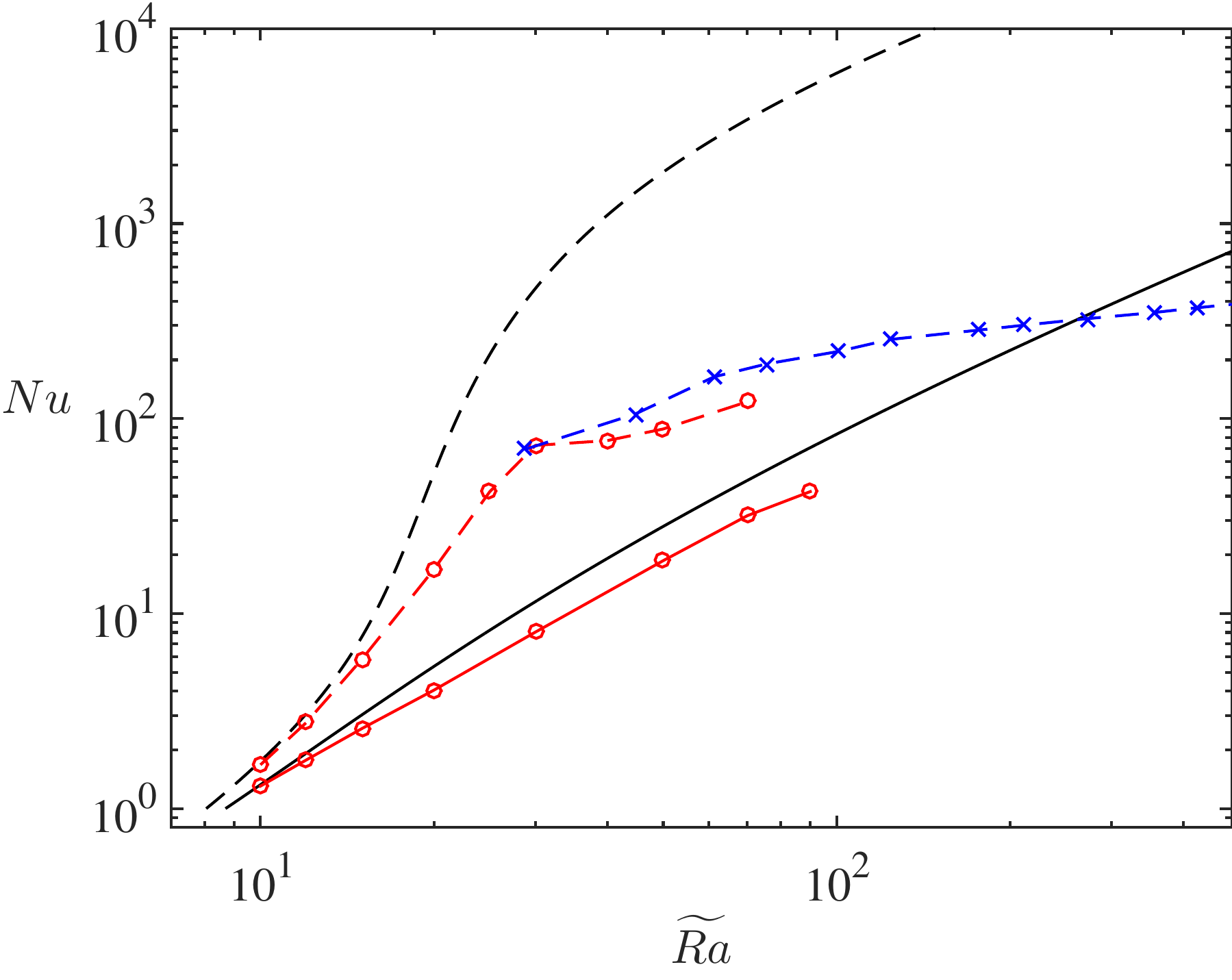}
  \end{center}
  \caption{\small{The Nusselt number $Nu$ vs $\widetilde{Ra}$ at $\sigma=7$ and $E=10^{-7}$:
  single-mode data for stress-free boundaries at $\epsilon=0$ (solid black curve);  single-mode data for no-slip boundaries (black dashed curve);  DNS data for stress-free boundaries (red solid curve); DNS data for no-slip boundaries (red dashed curve); laboratory data (blue dashed curve).  The single-mode solutions provide an upper bound to the DNS Nusselt number.}} 
\label{fig:Nuss-Ra-Ek2}
\end{figure}

\section{Conclusion}

Discrepancies between recent synergistic investigations of rotating thermal convection performed through laboratory experiments \citep{jS15}, DNS \citep{sS14}, and reduced models \citep{kJ12} have drawn attention to the non-trivial impact of Ekman pumping on the efficiency of heat transport. Laboratory experiments and DNS are unable to access the geophysically and astrophysically relevant high $Ra$--low $(Ro,E)$ parameter space.
In the present study, this difficulty is overcome by performing a detailed asymptotic analysis in the limit $(Ro,E)\rightarrow0$ and extending the previously developed NonHydrostatic-QuasiGeostrophic Equations to incorporate the effects of Ekman pumping. The analysis reveals the existence of three distinct fluid regions each characterized by a different dominant physical balance: a geostrophically balanced \textcolor{black}{bulk} where fluid motions are predominately aligned with the axis of rotation, Ekman layers adjacent to the bounding plates where viscous stresses attenuate the interior geostrophic velocity field, and intermediate thermal wind layers driven by Ekman pumping. A classical Ekman pumping parameterization $W^{(o)}=\pm E^{1/6}\zeta^{(o)}_0/\sqrt{2}$ is utilized to alleviate the need for spatially resolving the Ekman boundary layers and a  reduced model, the CNH-QGE, is constructed using the method of composite expansions \citep{nayfeh2008perturbation}. The model bears all the hallmarks of its stress-free counterpart where horizontal advection of momentum and heat, linear vortex stretching through the Coriolis force, and the vertical advection of the local mean temperature dominate the buoyancy driven flow \citep{mS06,kJ12}.  However, once  a critical threshold $\widetilde{Ra} = \mathcal{O}(E^{-1/9})$
%, or equivalently  ${Ra} = \mathcal{O}(E^{-13/9})$, 
is  reached, Ekman pumping provides a substantial source of buoyancy production in the vicinity of the bounding plates  and the system transitions
from asymptotically weak Ekman pumping to $\mathcal{O}(1)$ Ekman pumping. 

The physical explanation for this phenomenon is the ascendence to dominance of a new source 
of buoyancy production in the thermal wind layer through the vertical advection of the mean temperature by Ekman pumping.
%, $w_E \pd{Z} \overline{\Theta}\sim E^{1/6} \zeta \pd{Z} \overline{\Theta}$. 
This occurs through the intensification of the mean temperature gradient 
%$\pd{Z} \overline{\Theta}$ 
and the vortical motions
%$\zeta \gtrsim E^{-1/6}$ 
that drive vertical Ekman transport, 
%based on boundary layer intensification of vortical motions owing to an increase in the pumping velocities from $w^*_E=\mathcal{O}(E^{1/6} U)$ to $w^*_E=\mathcal{O}(U)$,  thereby introducing a new source 
%of buoyancy production through the vertical advection of the mean temperature that is comparable to that produced in the geostrophic interior.
%, i.e., $\pd{t} \Theta^{\prime(c)} \sim E^{1/6} \zeta^{(c)}\pd{Z} \overline{\Theta}^{(c)}= U \pd{Z} \overline{\Theta}^{(c)}$. 
\textcolor{black}{resulting in convective fluxes generated by
Ekman transport that are comparable to those generated in the bulk.}
% i.e.,   $\overline{E^{1/6}\zeta^{(c)} \Theta^{\prime(c)}}\sim \overline{ U\Theta^{\prime(c)}}$. 
Thus $\mathcal{O}(1)$ changes to the Nusselt number are produced when compared to rapidly rotating convection in the presence of stress-free boundaries.

Single-mode theory \textcolor{black}{for cellular patterns was} used to quantify the dramatic enhancement in the Nusselt number as measured in the $Nu\propto \widetilde{Ra}^\beta$ or $\Nu-1\propto (\widetilde{Ra}/\widetilde{Ra}-1)^\gamma$ scaling relations. \textcolor{black}{Similar single-mode theory applied to isolated, radially symmetric profiles of convective Taylor columns \citep{iG10} yields qualitatively similar results. This theory provides} an upper bound on the laboratory and DNS results which have reported heat transport exponents $\beta^{NS}_{rot}$ that appear to increase with decreasing $E$ (see Figure~\ref{fig:NuRa}). Results from single-mode theory suggest that the current experiments are probing the steeper transition region before the asymptotic regime is reached.  Indeed, as $E\rightarrow0$, this region exhibits an ever-steepening scaling (see Figures~\ref{fig:Nuss-Ra-Ek} and \ref{fig:MaxRa}). Measurements of the instantaneous heat transport exponents show that once the asymptotic regime is reached the Ekman pumping reduces the exponents to values below those observed in the stress-free case. Owing to the challenges in probing the high $Ra$--low $(Ro,E)$ regime these investigations are unable to reach the upper parts of the branch which are of geophysical and astrophysical interest. In addition, we found that at finite $E\ll1$ {(and within the regime of validity of the theory)}, the $Nu$-$Ra$ heat transport law achieved by DNS and in the laboratory is bounded from below by single-mode theory for stress-free boundaries and bounded from above by single-mode theory for no-slip boundaries at the specified $E$. In this regard, the results of single-mode theory have considerable utility. However, a more detailed numerical investigation of the CNH-QGE will further our understanding of the rotationally constrained regime. Specifically, motivated by laboratory experiments, our analysis has focused primarily on water for which $\sigma=7$. For smaller Prandtl numbers, $\sigma \lesssim 1$, \textcolor{black}{geostrophic turbulence is known to be triggered at much lower reduced Rayleigh numbers \citep{kJ12b,dN14}. Assuming that the physics behind heat transport enhancement is robust, the quantitative effect of Ekman pumping in this regime and associated scaling laws can now be explored via simulations of the CNH-QGE derived here.} 
%\textcolor{red}{We note that \citet{Ecke2015} suggests that given the moderate supercriticality $\eta$ of current DNS and laboratory experiments the $Nu$-$\widetilde{Ra}$ should interpreted from the standpoint of weakly nonlinear theory but with $\eta=\mathcal{O}(1)$. The strongly nonlinear results of Figure~\ref{fig:NussExp} suggest that this analogy may be fortuitous.}

Given that the range of reduced Rayleigh numbers $\widetilde{Ra}$ for which the  CNH-QGE are valid can span five decades in geophysical and astrophysical settings where $E=\mathcal{O}(10^{-15})$, the flow morphologies of low Prandtl number rotationally constrained convection are likely to be as rich as in nonrotating convection. As illustrated by \cite{kJ12b} and \cite{sS14} multiple heat transport scaling regimes are therefore likely to exist.  

{\bf Acknowledgements}

{This work was supported by the National Science Foundation under grants EAR \#1320991 (MAC, KJ and JMA), EAR CSEDI \#1067944 (KJ and JMA) and DMS-1317596 (EK). \textcolor{black}{GV acknowledges funding from the Australian Research Council, Project No. DE140101960. The authors wish to acknowledge the hospitality of the UCLA IPAM long program on the Mathematics of Turbulence where some of this work was done as well as important conversations with Greg Chini, Robert Ecke, James McWilliams, David Nieves and Meredith Plumley.}
}

\newpage
\section{Appendix}
\label{sec:appx}
As deduced in subsection~\ref{sec:eksig} in the regime prior to the enhancement of heat transport by Ekman pumping, i.e., where
$\zeta_0^{(o)}(0) < \mathcal{O}(\epsilon^{-1/2})$, an asymptotic theory may be developed based solely on higher order corrections to the NH-QGE. Such corrections may be considered to be the nonlinear extension of the linear work of \cite{pN65} and \cite{wH71}.

%on the higher order pumping, namely
%\Beq
%W_{1/2}^{(o)}(0) =\frac{1}{\sqrt{2}} \zeta_0^{(o)}(0), \quad
%W_{1/2}^{(o)}(1) =-\frac{1}{\sqrt{2}} \zeta_0^{(o)}(1).
%\Eeq
On proceeding to ${\cal O}(\epsilon^{1/2})$, the mean dynamics are described by
\beginar
\label{eqn:mean_redcr}
%\left.
%\begin{array}{c}
  \pd{Z} \overline{P}^{(o)}_{1/2}  &=&\displaystyle{\frac{\widetilde{Ra}}{\sigma}} \overline{\Theta}^{(o)}_{1/2}, \\ 
 \pd{\tau}\overline{\Theta}^{(o)}_{1/2} + 
\pd{Z} \lb \overline{{\overline{ W^{(o)}_0  \Theta^{\prime(o)}_{3/2} +W^{(o)}_{1/2}  \Theta^{\prime(o)}_{1}  }}}^\mathcal{T}\rb &=& \displaystyle{\frac{1}{\sigma}} \pd{ZZ}\overline{\Theta}^{(o)}_{1/2}, \\ 
\overline{\Theta}_{1/2}^{(o)}(0) = 0,\  \ 
\overline{\Theta}_{1/2}^{(o)}(1) &=& 0.
%\end{array}
%\right\}
\endar
Following the perturbation analysis of section~\ref{sec:outer} for the fluctuations to the next order gives 
\Beq
\label{eqn:Geoc}
\mathcal{L}_{geo}
\lb
\begin{array}{c}
\eub^{\prime(o)}_{1/2} \\ P^{\prime(o)}_{3/2}
\end{array}
\rb = \boldsymbol{RHS}.
\Eeq
Upon application of the solvability condition, we find that the corrections are geostrophically balanced with
\Beq
\eub^{\prime(o)}_{1/2}= \nabla^\perp \Psi^{(o)}_{1/2} + W^{(o)}_{1/2}\hr, \quad P^{\prime(o)}_{3/2}=\Psi^{(o)}_{1/2}
\Eeq
and evolve according to 
\beginar
\label{eqn:rpsicr}
D^{\perp}_{0t} \zeta^{(o)}_{1/2} + \eub^{(o)}_{1/2} \cdot\nabla_\perp \zeta^{(o)}_{0}  -\pd{Z}  W^{(o)}_{1/2} &=& \nabla_\perp^2 \zeta^{(o)}_{1/2}, 
\\ 
%\nn \\
\label{eqn:rwcr}
D^{\perp}_{0t} W^{(o)}_{1/2} + \eub^{(o)}_{1/2} \cdot\nabla_\perp W^{(o)}_{0}  + \pd{Z} \Psi^{(o)}_{1/2}   &=&\frac{\widetilde{Ra}}{\sigma} \Theta_{3/2}^{\prime(o)}  + \nabla_\perp^2 W^{(o)}_{1/2}, 
\\ 
%\nn \\
\label{eqn:rtempfcr}
D^{\perp}_{0t}\Theta_{3/2}^{\prime(o)}  + \eub^{(o)}_{1/2} \cdot\nabla_\perp\Theta_{1}^{\prime(o)}   + W^{(o)}_0 \pd{Z} \overline{\Theta}^{(o)}_{1/2}  
+\underline{ W^{(o)}_{1/2} \pd{Z} \overline{\Theta}^{(o)}_{0}  }&=& \frac{1}{\sigma} \nabla^2_\perp   \Theta_{3/2}^{\prime(o)}.
\endar
Importantly, it is evident from the underlined term that \textcolor{black}{the temperature fluctuations at the boundaries remain nonzero, $\Theta_{3/2}^{\prime(o)}(0),\Theta_{3/2}^{\prime(o)}(1)\ne0$, implying that a middle layer correction is required}. On developing the asymptotics for the middle region, the corrections are given by the following set of equations  %$\hr\times\eub^{\prime(m)}_{3/2} = -\nabla \Psi^{(m)}_{3/2}$:
\beginar
%\left.
%\begin{array}{c}
\hr\times\eub^{\prime(m)}_{3/2} = -\nabla_\perp \Psi^{(m)}_{3/2}, \\ %\\
\pd{z} \Psi^{(m)}_{3/2}   =\displaystyle{\frac{\widetilde{Ra}}{\sigma}} \Theta^{\prime(m)}_{3/2},  \\ %\\
\label{eqn:allmiddlefcr}
   D^{\perp}_{0t} \Theta^{\prime(m)}_{3/2} 
%   + W^{(o)}_0  \pd{z} \overline{\Theta}^{(m)}_1  
%+  \pd{z} \lb   W^{(o)}_0 \Theta^{\prime(m)}_1 - \overline{W^{(o)}_0 \Theta^{\prime(m)}_1} \rb
    =\displaystyle{ \frac{1}{\sigma}} \nabla^2   \Theta^{\prime(m)}_{3/2}, \\% \\
    \Theta^{\prime(m)}_{3/2} (Z_b)+\Theta^{\prime(o)}_{3/2}(Z_b)=0, \quad \Theta^{\prime(m)}_{3/2} (\infty)=0,
%\end{array}
%\right\}
\hspace{2em}
\endar
where $Z_b=0$ or $1$.
The dynamics are in thermal wind balance and evolve according to an advection-diffusion equation. Notable in 
(\ref{eqn:allmiddlefcr}) is the absence of vertical advection of the mean temperature indicating that buoyancy production 
by Ekman pumping is negligible.  

We extend the scaling analysis of subsection \ref{sec:rathres} for the CTC regime to the case where horizontal advection in the reduced dynamics is subdominant, and set
\beginar
\label{eqn:expcr0}
& W^{(o)}_{1/2} = \widetilde{Ra}^{\tilde w} \widehat{W}^{(o)}_{1/2} ,\
\Psi^{(o)}_{1/2} = \widetilde{Ra}^{\tilde \psi} \widehat{\Psi}^{(o)}_{1/2} ,\
\zeta^{(o)}_{1/2} = \widetilde{Ra}^{\tilde \zeta} \widehat{\zeta}^{(o)}_{1/2} ,\\
&\Theta^{\prime(o)}_{3/2} = \widetilde{Ra}^{\tilde \theta} \widehat{\Theta}^{(o)}_{3/2},\
\pd{Z} \overline{\Theta}^{(o)}_{1/2} =  \widetilde{Ra}^{\tilde {dt}} \widehat{\pd{Z} \overline{\Theta}}^{(o)}_{1/2}.\nn
%Nu =  \widetilde{Ra}^{\hat{\beta}} \widehat{Nu}. \nn
\endar
In the bulk, the algebraic equations satisfied by the exponents are identical to those found in the NH-QGE (see Eq.~(\ref{eqn:exp1})), namely, 
\Beq
\label{eqn:expcr}
{\tilde w}={\tilde \psi} ={\tilde \zeta} = \frac{ \hat{\beta}+1}{2},\quad
{\tilde \theta}=\frac{ \hat{\beta}-1}{2},\quad
\tilde {dt} =-1.
\Eeq
Inspection of the two highest terms of the asymptotic series for the fluid variables establishes that series remains uniform in all variables, e.g., 
%{\bf where does the factor $\epsilon^{1/2}$ in the second eq. below come from -- it is not in (7.13)? see also (7.17) and below.
%KEITH: Yes, equation 3.78 and 7.13 indicates that the first to terms in the asymptotic series remain uniform. 7,15b is simply stating that the second term in 7.15a is always smaller than the first. Is there a better way to express this.}
\Beq
%w \approx \widetilde{Ra}^\frac{\hat{\beta}+1}{2} \lb \widehat{W}^{(o)}_{0}+\epsilon^{1/2} \widehat{W}^{(o)}_{1/2}\rb
w \approx \widetilde{Ra}^\frac{\hat{\beta}+1}{2} \lb \widehat{W}^{(o)}_{0}+\epsilon^{1/2} \widehat{W}^{(o)}_{1/2}+\cdots \rb, \quad\mbox{s.t.}\quad
\epsilon^{1/2} W^{(o)}_{1/2} = o\lb W^{(o)}_{0}\rb.
%\theta^\prime \approx \widetilde{Ra}^\frac{\hat{\beta}-1}{2} \lb \widehat{\Theta}^{(o)}_{1}+\epsilon^{1/2} \widehat{\Theta}^{(o)}_{3/2}\rb,
\Eeq
%such that $\epsilon^{1/2} W^{(o)}_{1/2} = o\lb W^{(o)}_{0}\rb$. 
Similar expressions hold for the other variables identified in (\ref{eqn:expcr}). Indeed, this finding always holds for the NH-QGE in the region of asymptotic validity for impenetrable stress-free boundary conditions.

The validity in the presence of impenetrable no-slip boundary conditions is thus a central question of interest.
In the thermal boundary layer, where vertical fluid motions persist due to Ekman pumping, the exponents satisfy instead 
\Beq
\label{eqn:expcr2}
{\tilde w}= \frac{ \hat{\beta}+1}{2},\quad
{\tilde \psi} ={\tilde \zeta} = {\hat{\beta}+1},\quad
\tilde {dt} ={\tilde \theta}=\frac{ \hat {3K}+1}{2}.
\Eeq
Comparison with the leading order exponents (Eq. (\ref{eqn:exp2})) shows that the vortical and thermal corrections increase at a much greater rate than in the stress-free case, suggesting an eventual loss of uniformity in the asymptotic expansion. Since
\beginar
&\zeta \approx \widetilde{Ra}^\frac{\hat{\beta}+1}{2} \lb \widehat{\zeta}^{(o)}_{0}+\epsilon^{1/2}\widetilde{Ra}^\frac{\hat{\beta}+1}{2} \widehat{\zeta}^{(o)}_{1/2}\rb, \quad
\Psi \approx \widetilde{Ra}^\frac{\hat{\beta}+1}{2} \lb \widehat{\Psi}^{(o)}_{0}+\epsilon^{1/2}\widetilde{Ra}^\frac{\hat{\beta}+1}{2} \widehat{\Psi}^{(o)}_{1/2}\rb, 
\\
&\theta^\prime \approx  \widetilde{Ra}^{\hat{\beta}} \lb \widehat{\Theta}^{(o)}_{1}+\epsilon^{1/2} \widetilde{Ra}^\frac{\hat{\beta}+1}{2}  \widehat{\Theta}^{(o)}_{3/2}\rb, \quad
\pd{Z} \overline{\Theta}^{(o)} \approx  \widetilde{Ra}^{\hat{\beta}} \lb\widehat{\pd{Z} \overline{\Theta}}^{(o)}_{0}+\epsilon^{1/2} \widetilde{Ra}^\frac{\hat{\beta}+1}{2}  \widehat{\pd{Z} \overline{\Theta}}^{(o)}_{1/2}\rb\nn
\endar
the loss of uniformity occurs when $\epsilon^{1/2} \widetilde{Ra}^\frac{\hat{\beta}+1}{2} =\mathcal{O}(1)$. A more detailed boundary layer
analysis (not presented here) reveals a particularly grave situation where all orders in the asymptotic sequence exhibit non-uniformity and become 
as large as the leading order prediction. This situation is resolved by promoting the effect of Ekman pumping to leading order, as done in this paper.
The higher order system (\ref{eqn:rpsicr})-(\ref{eqn:rtempfcr}) is contained within the CNH-QGE when the effects of Ekman pumping are subdominant.

\bibliographystyle{jfm}
\bibliography{Dynamobib}

\end{document}